\documentclass{article}
\usepackage[utf8]{inputenc}
\usepackage{graphicx}
\usepackage[portrait, margin=0.8in]{geometry}
\usepackage{hyperref}
\usepackage{authblk}
\usepackage{array}
\usepackage{todonotes}
\usepackage{amsmath,amsthm,amssymb}
\usepackage{braket}
\usepackage{enumitem}
\usepackage{mathabx}
\newcommand{\Jibm}{J_\mathit{IBMQ}}
\newcommand{\hibm}{h_\mathit{IBMQ}}
\newcommand{\Tibm}{T_\mathit{IBMQ}}
\newcommand{\Jdwave}{J_\mathit{QA}}

\newcommand{\Tdwave}{T_\mathit{QA}}
\usepackage[sorting = none, backend = bibtex, style=numeric-comp]{biblatex}

\usepackage{booktabs,multirow}

\title{Simulating Heavy-Hex Transverse Field Ising Model Magnetization Dynamics Using Programmable Quantum Annealers}

\author[1]{Elijah Pelofske\thanks{Email: epelofske@lanl.gov}}
\author[1]{Andreas Bärtschi}
\author[1]{Stephan Eidenbenz}
\affil[1]{Los Alamos National Laboratory}

\date{\vspace{-6ex}}

\begin{document}
\maketitle

\begin{abstract}
Recently, a Hamiltonian dynamics simulation was performed of a ferromagnetic 2D transverse field Ising model with a connectivity graph native to the $127$ qubit heavy-hex IBM Quantum architecture using ZNE quantum error mitigation. We demonstrate that one of the observables in this Trotterized Hamiltonian dynamics simulation, namely magnetization, can be efficiently simulated on current superconducting qubit-based programmable quantum annealing computers. We show this simulation using two distinct methods: reverse quantum annealing and h-gain state encoding. Each of the methods use anneal schedules with pauses at varying anneal fractions, small programmed coupler weights, and fast quenches to read out the qubit states. This simulation is possible because the $127$ qubit heavy-hex connectivity graph can be natively embedded onto the D-Wave Pegasus quantum annealers hardware graph and because there exists a direct equivalence between the energy scales of the two types of quantum computers. We derive equivalent anneal pauses in order to simulate the Trotterized quantum circuit dynamics for varying Rx rotations $\theta_h \in (0, \frac{\pi}{2}]$, using quantum annealing processors. Multiple disjoint instances of the Ising model of interest can be embedded onto the D-Wave Pegasus hardware graph, allowing for parallel quantum annealing. We report equivalent magnetization dynamics of the $127$ qubit heavy-hex Ising model using quantum annealing for time steps of $20, 50, 100$ up to $10,000$, which we find are consistent with exact classical $27$ qubit heavy-hex Trotterized circuit magnetization dynamics (with low-Trotter step size), and we observe reasonable, albeit noisy, agreement with the existing simulations for single site magnetization at $20$ Trotter steps for some parameters. Because of the size of the D-Wave Pegasus hardware graph, we were also able to embed and execute equivalent magnetization dynamics simulations on a $384$ node heavy-hex graph. Notably, the quantum annealers are able to simulate equivalent magnetization dynamics for thousands of time steps, significantly out of the computational reach of the digital quantum computers on which the original Hamiltonian dynamics simulations were performed. 
\end{abstract}

\section{Introduction}
\label{section:introduction}

Quantum computers are approaching the regime where exact classical verification of their computations is difficult, or intractable \cite{arute2019quantum, zhong2020quantum, madsen2022quantum, Wu_2021}. This is an interesting regime for NISQ \cite{Preskill_2018} technology to be in, and it has sparked increasing attention on good classical methods for approximating, or even exactly computing, observable measurements from large quantum systems \cite{liu2022validating, villalonga2022efficient, doi:10.1137/050644756}. Recently, Ref.~\cite{kim2023evidence} studied the dynamics of 2D Transverse field Ising model, 
\begin{equation}
    H = -J \sum_{(i, j)} Z_i Z_j + h \sum_i X_i
    \label{equation:problem_Hamiltonian}
\end{equation}
using a Trotterized circuit method on a $127$ qubit Eagle processor IBM Quantum device \texttt{ibm\_kyiv} with a heavy-hex \cite{PhysRevX.10.011022} hardware graph. This computation made use of the quantum error mitigation algorithm Zero Noise Extrapolation (ZNE) post processing \cite{PhysRevA.102.012426, Kandala_2019, Giurgica_Tiron_2020, Temme_2017, LaRose2022mitiqsoftware, PhysRevA.105.042406, tran2023locality} in order to extend the computational capabilities for simulating observable measures, such as magnetization. This study was motivated by the Ising model being of fundamental interest in quantum many-body physics simulations. The hardware experiments also made use of random Pauli twirling \cite{PhysRevA.94.052325, Cai_2019} to further reduce errors in the computation. This computation was of a sufficient size that exact and brute force verification methods could not be applied, but certain classical approximation methods, namely tensor networks, could be applied for the shorter depth Trotterized circuits. Following this study, several groups then identified various fast approximate classical simulation methods for these Hamiltonian dynamics \cite{kechedzhi2023effective, tindall2023efficient, begušić2023fast, liao2023simulation, begušić2023fast_2, rudolph2023classical, shao2023simulating, patra2023efficient, tindall2024confinement}. Refs.~\cite{anand2023classical, 2308.01339} examined other aspects of these simulations as well. This Hamiltonian simulation that Ref.~\cite{kim2023evidence} studied is a 2-dimensional transverse field Ising model \cite{pineda2014two}, which has the same form as quantum annealing with a transverse field driving Hamiltonian \cite{Kadowaki_1998, morita2008mathematical, das2008colloquium};
\begin{equation}
     H = - \frac{A(s)}{2}  \sum_i^n X_i + \frac{B(s)} {2} \left( \sum_{i}^n h_i Z_i + \sum_{i < j}^n J_{ij} Z_i Z_j \right)
     %H = - \frac{A(s)}{2} \Big( \sum_i^n \sigma^x_i \Big) + \frac{B(s)} {2} \Big( \sum_{i}^n h_i \sigma_i^z + \sum_{i < j}^n J_{ij} \sigma_i^z \sigma_j^z \Big) %% Rewriting the Pauli Sigmas to be consistent across Hamiltonians.
    \label{equation:QA_Hamiltonian}
\end{equation}

D-Wave Quantum Processing Units (QPU's) implement this transverse field driving quantum annealing using superconducting flux qubits, on a fixed hardware graph~\cite{Harris_2010, harris2018phase, king2021scaling, PhysRevX.4.021041, PhysRevA.92.062328, johnson2011quantum, PhysRevA.91.042314, PhysRevX.8.031016, boixo2016computational, King_2022, King_2023}. Therefore, current D-Wave quantum annealing devices can be programmed to accept Hamiltonians of the form Eq.~\eqref{equation:problem_Hamiltonian} by setting the relevant quadratic terms to be ferromagnetic ($J < 0$) and set $h_i=0$. There is a sign difference between the two transverse field Hamiltonians (Eq.~\eqref{equation:problem_Hamiltonian} vs Eq.~\eqref{equation:QA_Hamiltonian}), however in this case the eigenvalues of both systems (regardless of the sign of the transverse field Hamiltonian) are the same. Using a varied anneal schedule the ratio of the problem Hamiltonian and the transverse field Hamiltonian can be attenuated. 

Quantum annealing has been used to study the properties of various large-scale spin systems in the transverse field Ising model, in particular frustrated spin systems \cite{narasimhan2023simulating, Lopez_Bezanilla_2023, Park_2022, PhysRevLett.131.166701, King_2021}. In this study, we demonstrate that the measure of magnetization of the Hamiltonian studied in ref. \cite{kim2023evidence} can be very efficiently simulated (with respect to total compute time) using current D-Wave quantum annealing hardware, which implements a quantum Ferromagnetic Ising spin system in a transverse field.  The other physical observables measured in these studies are not currently possible to measure using D-Wave quantum annealing hardware, because the hardware does not allow for universal qubit rotations to put qubits into the required basis to measure non computational basis (Pauli Z-basis) states. This computation that D-Wave quantum annealers are capable of is not a digital computation of the Trotterized Hamiltonian simulation circuit of Ref.~\cite{kim2023evidence}, instead quantum annealing can directly perform a simulation of the system as an analog computation. We will derive the equivalent quantum annealing parameters to the Trotterized Hamiltonian simulation dynamics of Ref.~\cite{kim2023evidence}, and then present results from these simulations using two D-Wave quantum annealers. We report mean lattice magnetization results, and show that these samples can then be used to compute higher order Z observables. The restriction of Z-basis observables is specific to the current D-Wave hardware, which presents us from simulation arbitrary Pauli-string observables from this Hamiltonian simulation. However, it should be noted that in general analog quantum computation could in principle perform these types of Hamiltonian dynamics computations and, depending on the hardware, be able to access other Pauli basis. This study focusing on the Z observables is strictly a result of the hardware capabilities of the current D-Wave devices.

\begin{figure*}[t!]
    \centering
    \includegraphics[width=0.66\textwidth]{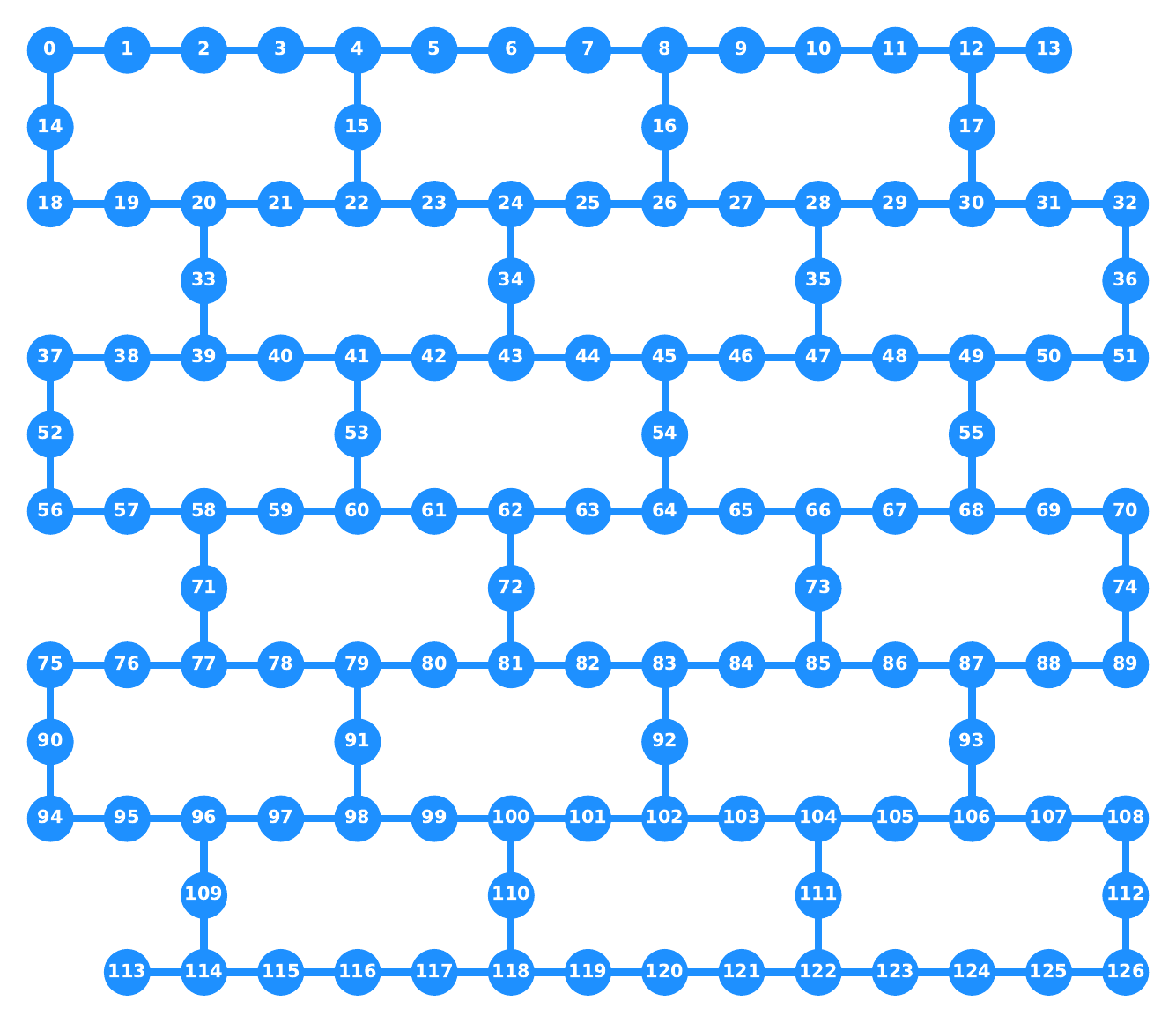}
    \caption{$127$ qubit heavy-hex hardware graph, where edges represent two qubit gates such as CNOT or ECR. This hardware graph is shared by \texttt{ibm\_washington}, \texttt{ibm\_sherbrooke}, \texttt{ibm\_brisbane}, \texttt{ibm\_nazca}, \texttt{ibm\_kyiv}, \texttt{ibm\_cusco}, not accounting for specific non functioning CNOT or ECR gates. }
    \label{fig:hardware_graph}
\end{figure*}

Next we briefly summarize the state of the classical simulation approaches for this Ising model Hamiltonian dynamics problem to provide context for how the following reported quantum annealing results compare. Refs. \cite{begušić2023fast_2, begušić2023fast, rudolph2023classical, shao2023simulating, liao2023simulation, patra2023efficient, kechedzhi2023effective} each used different approaches to replicate the various observables from the original $127$ qubit experiment for both $5$ and $20$ Trotter steps - specifically notably for the qubit $62$ single site mean magnetization at $20$ Trotter steps. Ref. \cite{tindall2023efficient} extended the tensor network simulations of site $62$ magnetization to infinite-sized heavy-hex lattices for up to $50$ Trotter steps, but only for $\theta_h=0.1$ up to $\theta_h=0.9$, at greater $\theta_h$ values this simulation could only be extended to approximately $25$ Trotter steps. Ref. \cite{patra2023efficient} reported efficient simulation of average lattice magnetization, and single site magnetization, observables for up to a $1121$ qubit heavy-hex graph for $37$ Trotter steps using tensor network methods for $\theta_h = 1.0$. Ref. \cite{kechedzhi2023effective} used light-cone truncation to simulate mean ($127$ qubit) lattice magnetization for $5$ Trotter steps, and site $62$ magnetization for up to $120$ Trotter steps for a fixed $\theta_h = \frac{18 \pi}{64}$.

%%%%%%%%%%%%%%%%%%%%%%%%%%%%%%%%%%%
%%%%%%%%%%%%%%%%%%%%%%%%%%%%%%%%%%%
%%%%%%%%%%%%%%%%%%%%%%%%%%%%%%%%%%%
\section{Methods}
\label{section:methods}

Section \ref{section:methods_Derivation_of_QA_parameters} shows the derivation of the equivalent annealing parameters that we use in order to perform simulations equivalent to the Trotterized circuit experiments. Importantly, the Trotterized circuit experiments \cite{kim2023evidence} were performed on a ``kicked'' Ferromagnetic Ising model in a transverse field, meaning that the state in which the evolution begins is in the spin up state $\uparrow$. D-Wave quantum annealers allow users to program quantum Ising spin systems in a transverse field, but the initial spin up state does require more advanced control features. In Section \ref{section:methods_RA_approach} and Section \ref{section:methods_h_gain_approach} we outline two different approaches for simulating magnetization dynamics of a Ferromagnetic Ising spin systems in a transverse field using a programmable quantum annealer. These two methods we will refer to as reverse annealing, and h-gain state encoding. Section \ref{section:methods_parallel_quantum_annealing} describes the use of tiled parallel quantum annealing, and Section \ref{section:methods_parameter_setting} describes the experimental settings that are used.

%%%%%%%%%%%%%%%%%%%%%%%%%%%%%%%%%%%%%%%%%%%%%%%%%%%%
%%%%%%%%%%%%%%%%%%%%%%%%%%%%%%%%%%%%%%%%%%%%%%%%%%%%
%%%%%%%%%%%%%%%%%%%%%%%%%%%%%%%%%%%%%%%%%%%%%%%%%%%%
\subsection{Derivation of Equivalent D-Wave Quantum Annealer Parameters for the IBM Quantum Trotterized Experiment}
\label{section:methods_Derivation_of_QA_parameters}

Ref.~\cite{kim2023evidence}, considers Hamiltonian simulation of a Transverse Field Ferromagnetic Ising Hamiltonian on its Heavy-hex topology of the form 
\begin{align}
    H   &   = -\Jibm \sum_{\langle i,j \rangle}{Z_i Z_j} + \hibm \sum_i{X_i}    %\\
        &   = -\frac{\pi}{4} \sum_{\langle i,j \rangle}{Z_i Z_j} + \frac{\theta_h}{2} \sum_i{X_i} \label{eq:HIBM}
\end{align}
for a total time of $\Tibm = N \delta_t = N$, where
\begin{itemize}[noitemsep]
    \item $\theta_h \in [0, \tfrac{\pi}{2}]$ varies between 0 and $\tfrac{\pi}{2} =: J_{IBM}$,
    \item $N$ denotes simultaneously the number of Trotter steps and the evolution time (interpreting $\delta_t :=1$),
    \item all parameters are unitless.
\end{itemize}
Ref.~\cite{kim2023evidence} then implemented a first order Trotter approach with a circuit of the form
\begin{align}
    e^{-i N\delta_t H}  &   = \prod_{p=1}^N \left( \prod_{\langle i,j\rangle}{R_{Z_iZ_j}(-\tfrac{\pi}{2})} \prod_i{R_x(\theta_h)} \right)
\end{align}
which fits the IBM hardware particularly nicely because $R_{Z_iZ_j}(-\tfrac{\pi}{2})$ can be implemented with only 1 CNOT and thus $\prod_{\langle i,j\rangle}R_{Z_iZ_j}(-\tfrac{\pi}{2})$ can be implemented in CNOT depth 3 due to a 3-edge-coloring of the heavy-hex graph~\cite{kim2023evidence,pelofske2023qavsqaoa,pelofske2023short}.

The goal is to implement simulation of the same original Hamiltonian on D-Wave quantum annealers. The heavy-hex graph is a strict subgraph of the D-Wave chip Pegasus graphs \cite{pelofske2023short, pelofske2023qavsqaoa}. We simplify the problem by considering only anneal schedules with fast quenches and an anneal pause at a specified anneal fraction $s$, which (neglecting the quenches) applies the following Hamiltonian (from \cite{D_Wave_qpu_annealing_time}):

\begin{align}
    H(s)      & = \frac{B(s)}{2} \left(\sum_i h_i\; Z_i + \sum_{\langle i,j\rangle} J_{i,j}\; Z_iZ_j\right)-\frac{A(s)}{2}  \sum_i X_i  %\\
              & = \frac{B(s)}{2} \cdot \Jdwave \sum_{\langle i,j\rangle} Z_iZ_j-\frac{A(s)}{2}  \sum_i X_i  
\label{eq:HDwave}
\end{align}

for some time $\Tdwave$, where the functions $A(s),\ B(s)$ are device-specific for the systems \texttt{Advantage\_system4.1} and \texttt{Advantage\_system6.2} \cite{D_Wave_physical_properties}, and where
\begin{itemize}[noitemsep]
    \item  we can program $\Jdwave$ to whatever value we want, as long as it is within the device specifications,
    \item  $s \in [0,1]$ is the anneal fraction, interpolating between the Ising term and the Transverse Field,
    \item  the ratio $A(s)/B(s)$ is monotonically decreasing in $s$ from $\gg 0$ at $s=0$ to $0$ at $s=1$,
    \item  $A(s)/h,\ B(s)/h$ are device-specific functions given in GHz (with Planck constant $h$),
    \item  $\Tdwave$ is given in $\mu s$.
\end{itemize}

\begin{figure*}[p!]
    \centering
    \includegraphics[width=0.90\textwidth]{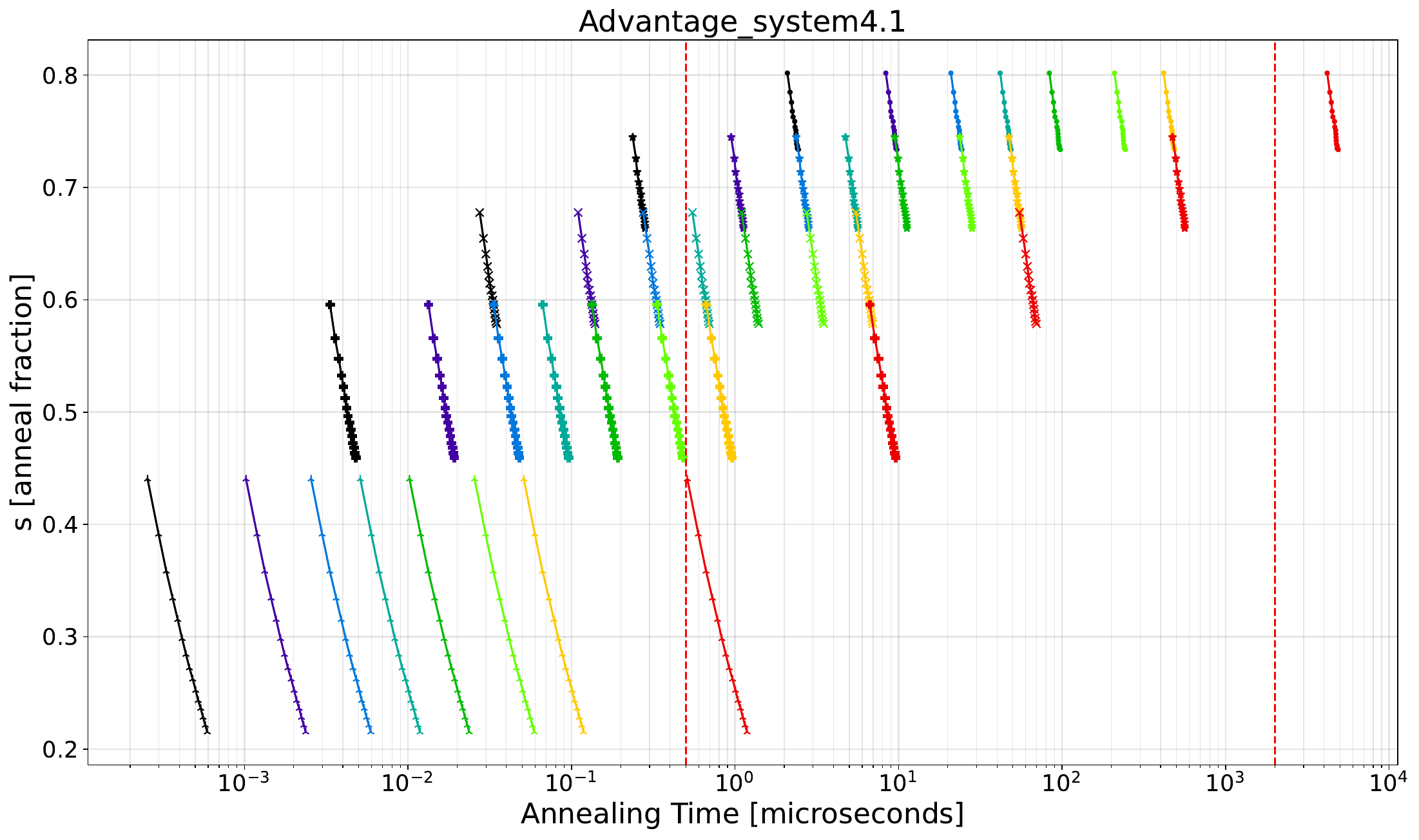}\\[1ex]%
    \includegraphics[width=0.90\textwidth]{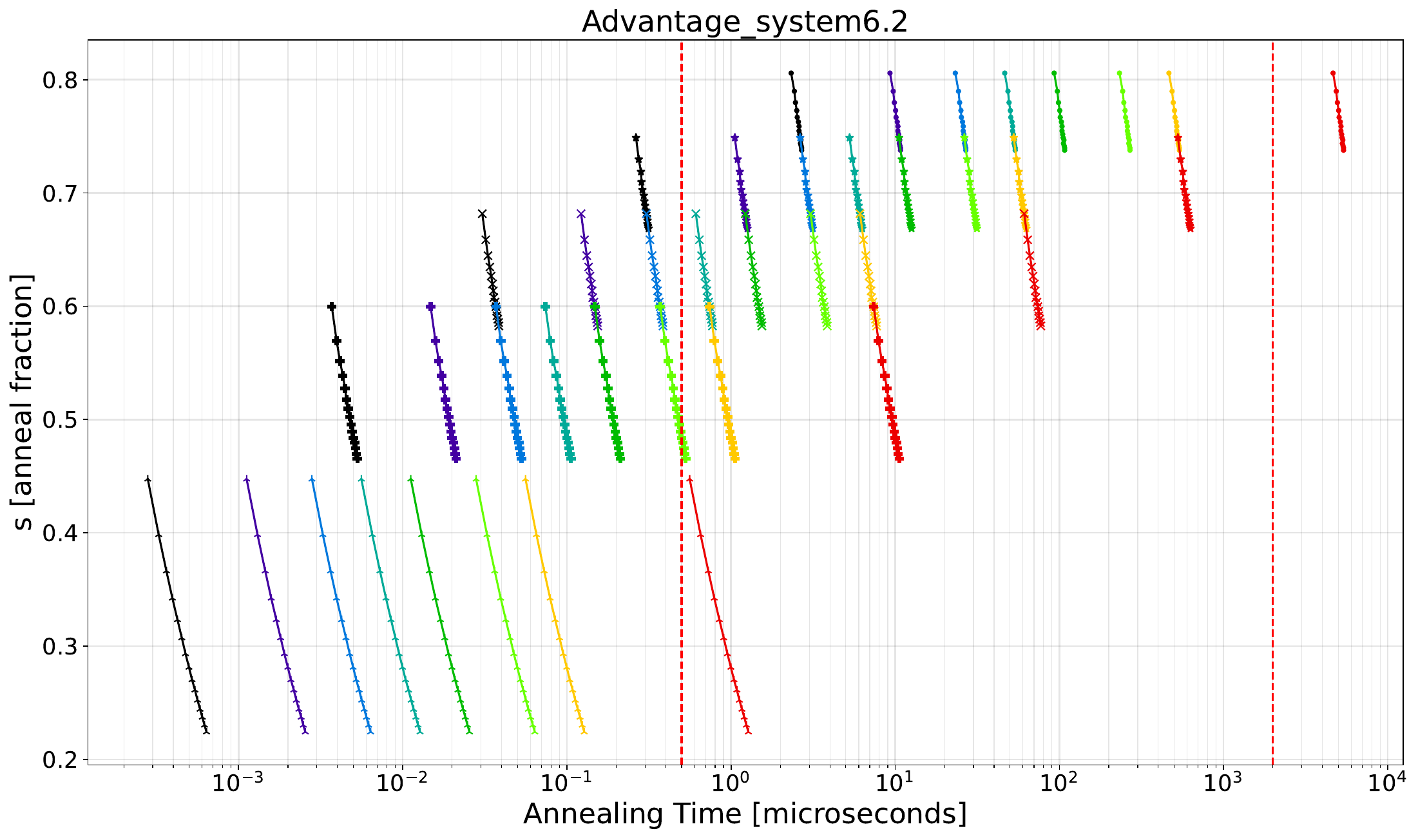}\\[1ex]%
    \includegraphics[width=1.0\textwidth]{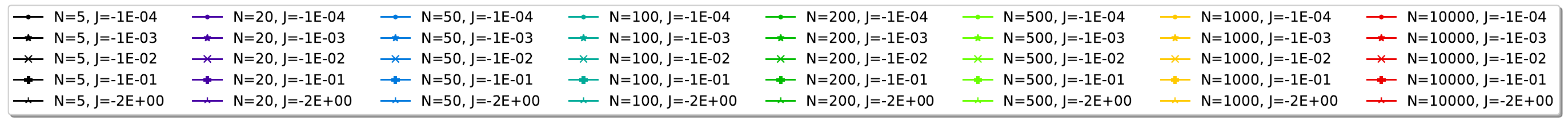}
    \caption{Anneal schedules for performing the equivalent Trotterized Ferromagnetic Transverse Field Ising Model (TFIM) simulation in Ref.~\cite{kim2023evidence} for $15$ linearly spaced values of $\theta_h \in [0, \frac{\pi}{2}]$, for different values of $N$ (time steps, grouped by color), and programmed Ferromagnetic coupler strengths ($J$, grouped by symbol). The anneal schedules are defined in terms of anneal times (specifically, these are \emph{pause times} neglecting ramp times) and the anneal fractions at which those pauses would occur. Vertical dashed red lines denote the minimum and maximum allowed annealing times for each device. x-axis has a log-scale. Note that the smallest coupling strengths shown here, $J=-0.0001$ is outside of the programmable coefficient range on these two D-Wave quantum annealers. }
    \label{fig:equivalent_anneal_parameters_to_IBMQ_Hamiltonian}
\end{figure*}

Our goal is \emph{for any given} $\theta_h$ and Trotter step number $N$ (= $\Tibm$) to set $\Jdwave,\ s,\ \Tdwave$ such that 
we evolve equivalent Hamiltonians on the right-hand sides of Equations~\eqref{eq:HIBM} and~\eqref{eq:HDwave} for an equivalent number of time steps;
we plot our parameter setting results in Figure~\ref{fig:equivalent_anneal_parameters_to_IBMQ_Hamiltonian}.
To this end, we need to have
\begin{align}
    \frac{\Tdwave}{\hbar} \left( \frac{B(s) \Jdwave}{2} \sum_{\langle i,j \rangle} Z_i Z_j {\ +\ } \frac{A(s)}{2} \sum_i X_i \right)  
    & = \frac{2\pi\cdot\Tdwave}{h} \left( \frac{B(s) \Jdwave}{2} \sum_{\langle i,j \rangle} Z_i Z_j {\ +\ } \frac{A(s)}{2} \sum_i X_i \right)  \\
    \stackrel{!}{=} \Tibm \left( -\Jibm \sum_{\langle i,j \rangle}{Z_i Z_j} + \hibm \sum_i{X_i}\right)                
    & = \qquad N \quad \left( -\frac{\pi}{4} \sum_{\langle i,j \rangle}{Z_i Z_j} + \frac{\theta_h}{2} \sum_i{X_i}\right)
\end{align}

We are interested in programming different ferromagnetic coupling strengths on the D-Wave QPUs. The first step we take is that we choose the coupler coefficient $\Jdwave$. Then, for each $\theta_h$ step, we select the nearest discretized anneal fraction $s$, from the calibration data sheets available for each D-Wave device \cite{D_Wave_physical_properties} (see also Fig.~\ref{fig:DWave_calibration_data_plots}, Appendix~\ref{section:appendix_DWave_calibration_data}), such that Equation~\eqref{equation:select_s} is satisfied, which sets the correct ratio between the Ising term and the Transverse Field:
\begin{equation}
    \frac{A(s)}{B(s) \Jdwave} = -\frac{2\theta_h}{\pi}
    \label{equation:select_s}
\end{equation}
(Because $s$ is measured in discrete steps for the machine calibration, we choose the $s$ which minimizes the different between the left hand side and right hand side of the equation.)

Next, we exactly compute $\Tdwave$ such that Equations~\eqref{equation:compute_DWave_time} are satisfied, which sets the correct evolution time.
Because $A(s)/h$ and $B(s)/h$ are calibrated in discrete steps, the two equations will give annealing times that are not exactly equal, but agree up to the 10s of nanoseconds range (at least when $\theta_h > 0$). Therefore, in practice we take the mean of the two computed $\Tdwave$ times from the two equations in~\eqref{equation:compute_DWave_time}. This process of selecting equivalent $\Tdwave$ and $s$ values for each $\theta_h$ is then repeated for the range of $\theta_h$ angles we wish to simulate using quantum annealing. Importantly, there is a factor of $1000$ mismatch between $\mu s$ and GHz -- so $\Tdwave$ in Eq.~\eqref{equation:compute_DWave_time} is in nanoseconds. 

\begin{equation}
                \begin{cases}
                    2 \Tdwave \cdot 2 \pi \cdot B(s)/h \cdot \Jdwave  & = - N \cdot \pi,         \\
                    \ \ \Tdwave \cdot 2 \pi \cdot A(s)/h                & = \ \ N \cdot \theta_h,
                \end{cases}
                \label{equation:compute_DWave_time}
\end{equation}

There are a number of degrees of freedom we have when selecting the simulation parameters to execute the equivalent quantum annealing Hamiltonian dynamics simulation. In particular, we can choose different $\Jdwave$ values, which result in different (but theoretically equivalent) annealing times and annealing pauses. Figure~\ref{fig:equivalent_anneal_parameters_to_IBMQ_Hamiltonian} shows what some of these equivalent annealing schedules are for different values of $N$ and $\Jdwave$. Note that there are other ways of deriving equivalent quantum annealing parameters for these simulations -- for example we could also allow the ferromagnetic coupling strength $\Jdwave$ to be varied instead of being fixed. For the purposes of ensuring that we are within the D-Wave machine constraints however (in particular, coefficient precision), we opted for fixing $\Jdwave$ and then varying the anneal schedule parameters. There are D-Wave system constraints that are important here. The first are simply the minimum and maximum annealing times of 0.5$\mu s$ and 2000$\mu s$, respectively, allowed on the devices (see also Table~\ref{tab:hardware_summary}), and these are marked by dashed vertical red lines in Figure~\ref{fig:equivalent_anneal_parameters_to_IBMQ_Hamiltonian}.

%%%%%%%%%%%%%%%%%%%%%%%%%%%%%%%%%%%%%%%%%%%%%%%%%%%%
%%%%%%%%%%%%%%%%%%%%%%%%%%%%%%%%%%%%%%%%%%%%%%%%%%%%
%%%%%%%%%%%%%%%%%%%%%%%%%%%%%%%%%%%%%%%%%%%%%%%%%%%%
\subsection{Ising Model Embeddings: Parallel Quantum Annealing}
\label{section:methods_parallel_quantum_annealing}

\begin{figure*}[th!]
    \centering
    \includegraphics[width=0.99\textwidth]{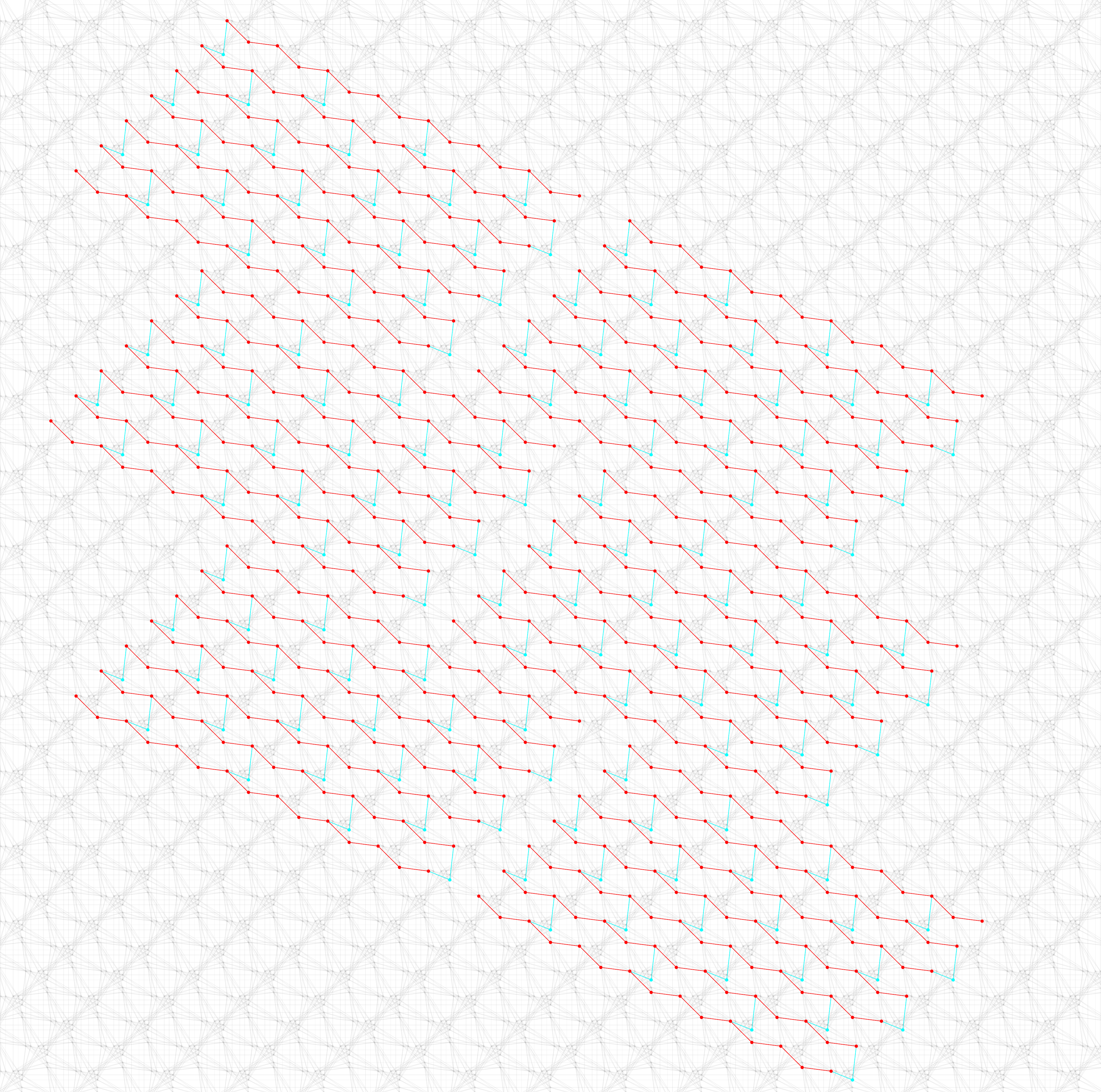}
    \caption{The $127$ qubit heavy-hex graph embedded onto a logical Pegasus graph (light grey), tiled a total of $6$ times allowing for parallel quantum annealing. The red qubits and couplers on the Pegasus graph encode the horizontal qubit lines on the heavy-hex lattice (shown in Figure~\ref{fig:hardware_graph}) and cyan qubits and couplers encode the vertical heavy-hex qubits. Note that in these embeddings there are complete horizontal qubit lines which are not actually used in the embedding since they are not in the $127$ qubit heavy-hex hardware graph. }
    \label{fig:Pegasus_embedding}
\end{figure*}

Figure~\ref{fig:hardware_graph} shows the $127$ qubit heavy-hex lattice \cite{PhysRevX.10.011022}, over which the Hamiltonian in Eq.~\eqref{equation:problem_Hamiltonian} is defined. In Refs.~\cite{pelofske2023qavsqaoa, pelofske2023short}, native embeddings of $127$ qubit heavy-hex lattices onto $P_{16}$ Pegasus graphs \cite{boothby2020nextgeneration, dattani2019pegasus, boothby2021architectural} were created, so we can use these tiled embeddings to embed the Ising model in question onto the current D-Wave Pegasus graph devices. Specifically, we can embed the Ising model of interest onto both the two D-Wave devices \texttt{Advantage\_system6.2} and \texttt{Advantage\_system4.1}. Figure~\ref{fig:Pegasus_embedding} defines the native embeddings of the $127$ qubit heavy hex graph onto a logical $P_{16}$ Pegasus graph. Both of the D-Wave devices have missing qubits and couplers from the logical Pegasus $P_{16}$ lattice, which means that not all $6$ disjoint instances can be embedded onto the hardware graph. On \texttt{Advantage\_system4.1}, $3$ independent instances can be embedded without encountering missing hardware from the logical lattice, and on \texttt{Advantage\_system6.2}, $4$ independent instances can be embedded, thereby allowing for parallel quantum annealing \cite{Pelofske_2022, PhysRevA.91.042314, Pelofske_2023}. Parallel quantum annealing increases the number of samples obtained per anneal-readout cycle, and also allows an averaging over different parts of the hardware which reduces the effects of local noise on fixed parts of the chip. Note that parallel quantum annealing is also known as \emph{tiling} \cite{D_Wave_tiling}.

%%%%%%%%%%%%%%%%%%%%%%%%%%%%%%%%%%%%%%%%%%%%%%%%%%%%
%%%%%%%%%%%%%%%%%%%%%%%%%%%%%%%%%%%%%%%%%%%%%%%%%%%%
%%%%%%%%%%%%%%%%%%%%%%%%%%%%%%%%%%%%%%%%%%%%%%%%%%%%
\subsection{Reverse Quantum Annealing Magnetization Dynamics Method}
\label{section:methods_RA_approach}

\begin{figure*}[t!]
    \centering
    \includegraphics[width=0.66\textwidth]{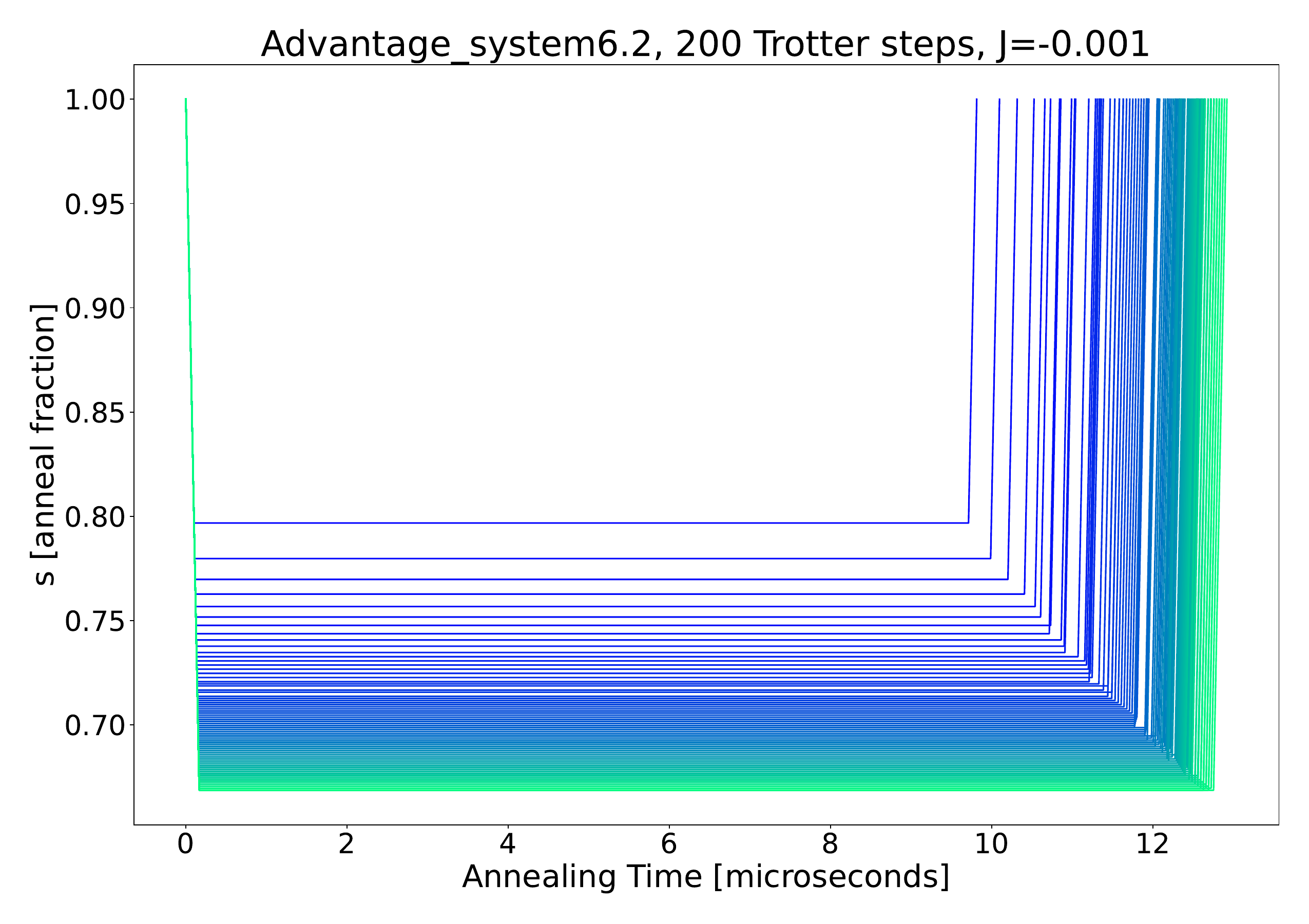}\\%
    \includegraphics[width=0.50\textwidth]{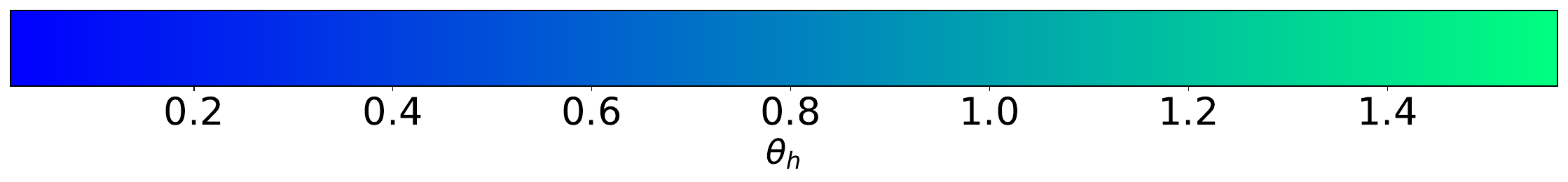}
    \caption{Example reverse quantum annealing schedules for $N=200$ Trotter steps, using $J=-0.001$, on \texttt{Advantage\_system6.2}. Each anneal schedule corresponds to varying $\theta_h$ angles (100 linearly spaced $\theta_h$ angles), color coded by the colorbar shown beneath the plot. }
    \label{fig:example_equivalent_QA_schedules_RA}
\end{figure*}

A clear way to prepare the spin up state, and then evolve the system for some time is using the reverse annealing feature of D-Wave quantum annealers \cite{D_Wave_solver_parameters}. This feature initializes the system at $t=0$ in a classical spin state (which is user-specified, and in this case is $+1$ for all active qubits), and then over the course of the anneal some amount of the transverse field term can be introduced into the anneal (defined by a user specified anneal schedule). The anneal fraction parameter $s$ defines this proportion of the transverse field Hamiltonian. If $s = 1$ then the state is entirely in the classical state, and if $s=0$ then the transverse field is the only Hamiltonian present at that point in time. In order to simulate the Hamiltonian magnetization as a function of applied transverse field (e.g. Rx rotations on all qubits), we can vary the anneal fraction $s$ at which we leave the anneal for most of the annealing time. The initial state must begin at $s=1$, and the anneal readout must occur at $s=1$ as well. The maximum rate at which the anneal fraction can be changed is dependent on the system parameters, and for the quantum annealers we use this rate is the inverse of the minimum annealing time, which is $0.5$ microseconds. The resulting annealing schedules are depicted in Figure~\ref{fig:example_equivalent_QA_schedules_RA}.
For all reverse quantum annealing simulations, \texttt{reinitialize\_state} is set to true, which re-prepares the specified initial state after each readout cycle.

%%%%%%%%%%%%%%%%%%%%%%%%%%%%%%%%%%%%%%%%%%%%%%%%%%%%
%%%%%%%%%%%%%%%%%%%%%%%%%%%%%%%%%%%%%%%%%%%%%%%%%%%%
%%%%%%%%%%%%%%%%%%%%%%%%%%%%%%%%%%%%%%%%%%%%%%%%%%%%
\subsection{H-gain State Encoding Magnetization Dynamics Method}
\label{section:methods_h_gain_approach}

D-Wave quantum annealers support a feature called the \emph{h-gain schedule} \cite{D_Wave_solver_parameters}, which allows users to program a time dependent schedule $g(t)$ in Eq.~\eqref{equation:QA_Hamiltonian_h_gain} that multiplies all of the linear terms (e.g. the classical Hamiltonian Z terms) by a factor, defined by $g(t)$ for each point in time during the anneal:
\begin{equation}
     H = - \frac{A(s)}{2} \sum_i^n X_i + \frac{B(s)} {2} \left( g(t) \sum_{i}^n h_i Z_i + \sum_{i < j}^n J_{ij} Z_i Z_j \right)
    \label{equation:QA_Hamiltonian_h_gain}
\end{equation}
This applied longitudinal field can be used to bias the state of the qubits, based on the programmed weights on the qubits. In particular this means that an initial state can be encoded into the anneal by programming the linear weights on the qubits and the h-gain schedule, we will refer to this method is \emph{h-gain state encoding} \cite{9259948, pelofske2023initial}. This longitudinal magnetic field control that the h-gain schedule offers has been used to study the magnetic properties of various spin systems \cite{harris2018phase, PRXQuantum.2.010349, PRXQuantum.2.030317, PhysRevResearch.5.013224} and properties of quantum annealing qubits \cite{PhysRevApplied.19.034053} on D-Wave quantum annealers. 

\begin{figure*}[t!]
    \centering
    \includegraphics[width=0.49\textwidth]{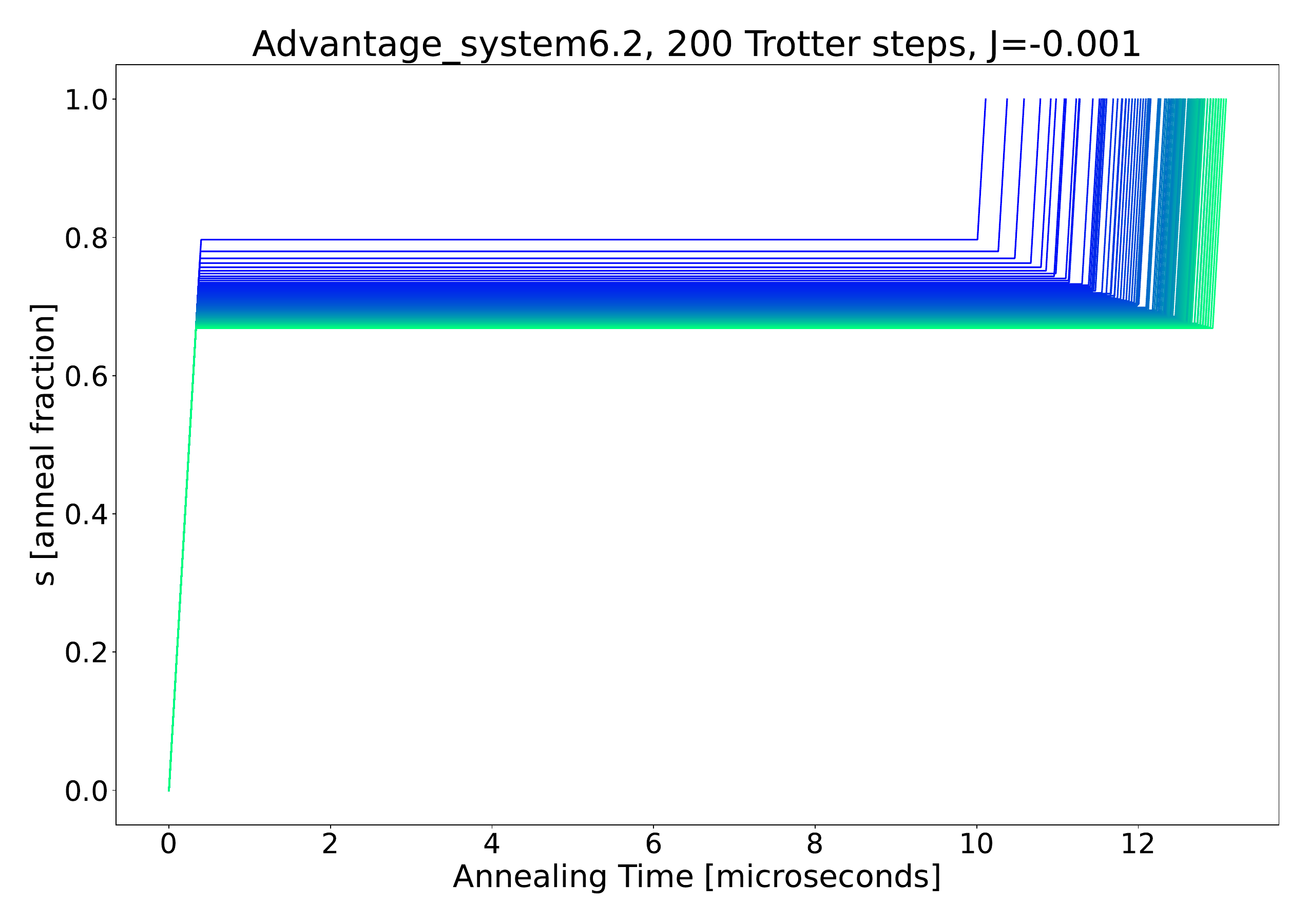}\hfill%
    \includegraphics[width=0.49\textwidth]{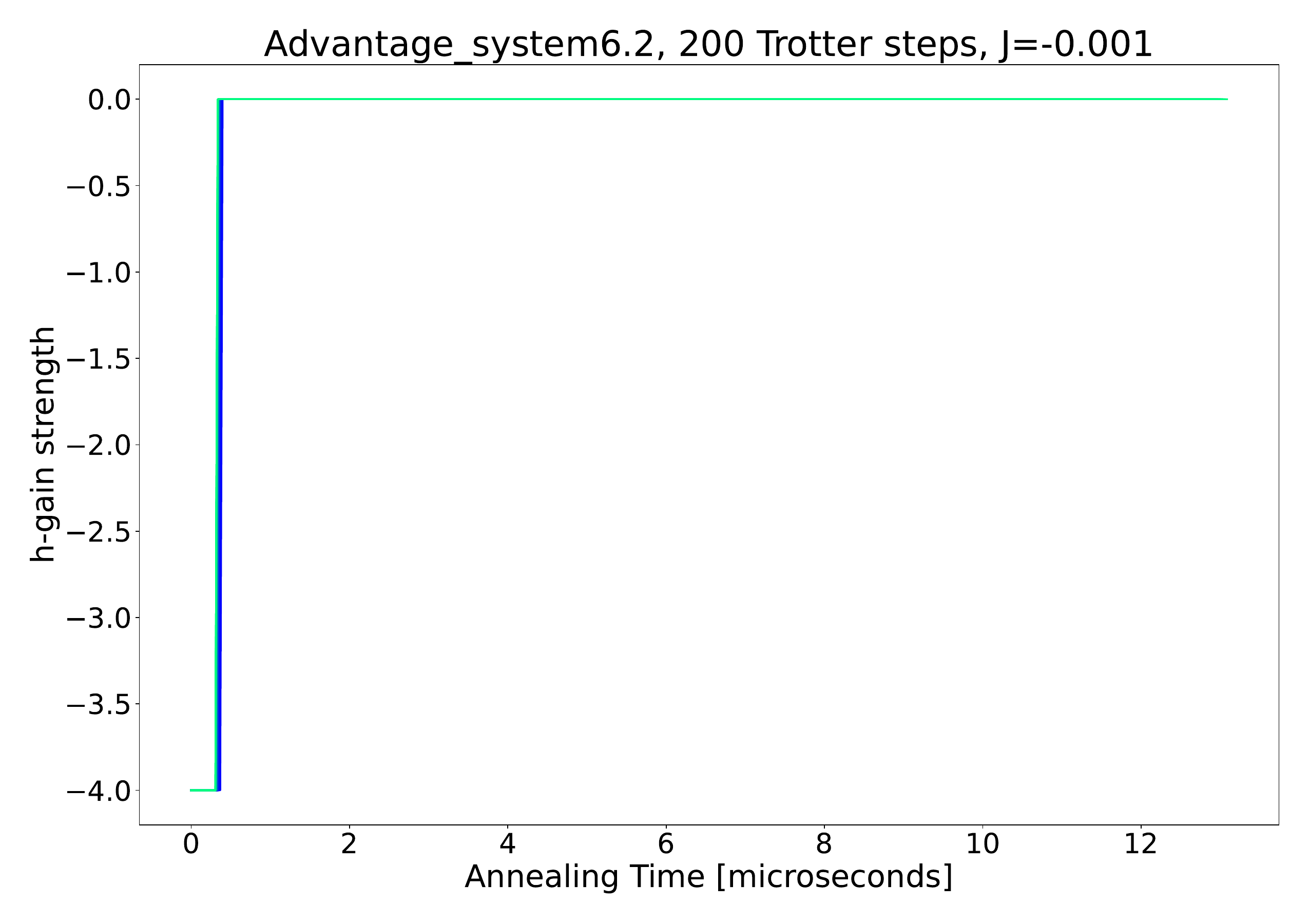}\\%
    \includegraphics[width=0.50\textwidth]{figures/fast_quench_schedules/legend.pdf}
    \caption{Example quantum annealing schedules (left), using h-gain schedule (right), for $N=200$ time (Trotter) steps, using $J=-0.001$, on \texttt{Advantage\_system6.2}. Each anneal schedule corresponds to varying $\theta_h$ angles, color coded by the colorbar shown beneath the plot. Left hand plot shows the varying forward quantum annealing schedules, and the right hand plot shows the corresponding h-gain field schedules. Notice that all of the h-gain schedules are nearly identical -- the only variance among them comes from the varying ramp durations during the initial portion of the forward anneal, during all of which the h-gain field is applied.}
    \label{fig:example_equivalent_QA_schedules_hgain}
\end{figure*}

In this case, we can program all of the active qubits to have a weight of $1$, and program an h-gain schedule (which is $g(t)$) to begin the anneal with a strong negative weight, and then at some point go to $0$ (the h-gain schedule being at $0$ corresponds to having no local weights on the Ising model, which is what the original un-modified Ising is). This protocol enforces that during the beginning of the anneal, all of the qubits are in a spin up $\uparrow$ state -- specifically the h-gain field is turned up to its maximum value, thus fully saturating the system to the spin up ground-state (e.g. the $B(s)$ term in Eq.~\eqref{equation:QA_Hamiltonian_h_gain}. Note that is specifically due to the qubit coefficients being set to a positive value, and the applied h-gain field being set to a negative value. In conjunction with an h-gain schedule, any annealing schedule can be programmed as well \cite{9259948, pelofske2023initial} -- specifically either reverse annealing, where the anneal begins at $s=1$, or forward quantum annealing where the anneal begins at $s=0$. For the current D-Wave quantum annealers there is not a strictly defined maximum h-gain schedule slope \cite{D_Wave_solver_parameters}, so we fix the time over which the h-gain schedule is reduced to a strength of $0$ to be $30$ nanoseconds (independent of the characteristics of the other anneal schedule). 

For the purposes of the magnetization dynamics we wish to compute, we use forward quantum annealing schedules, in conjunction with an h-gain schedule. The forward anneal schedule is set to ramp as fast as possible (the maximum slope is the inverse of the minimum annealing time of $0.5$ microseconds) to the required anneal fraction, then pausing for the required duration at that anneal fraction, and then ramping as fast as possible (under the maximum possible slope) to $s=1$ for qubit readout. The h-gain schedule is set to initialize at time $0$ in the maximum negative h-gain strength (for \texttt{Advantage\_system6.2} this is $-4$, but for \texttt{Advantage\_system4.1} this is $-3$). The h-gain schedule can have much sharper slopes compared to the anneal schedule \cite{D_Wave_solver_parameters}. Next, the h-gain schedule stays at the same strong applied field for a duration equal to $30$ nanoseconds less than the total anneal schedule ramp duration (which, as described previously is set to the maximum possible slope for the $s$ that is being ramped to). Next, the h-gain schedule decreases its strength over the next $30$ nanoseconds to an h-gain strength of $0$ and is left there for the rest of the anneal. Coefficient autoscaling is turned off for all QA simulations. This means that for both the qubit coefficients (which are always $1$), and for the coupler coefficients, the programmed energy scales are left as programmed instead of being normalized to within the maximum hardware range. 
The resulting schedules are depicted in Figure~\ref{fig:example_equivalent_QA_schedules_hgain}.

This h-gain state encoding method is expected to have very similar sources of error compared to the reverse annealing method - including primarily the quench at the beginning of the anneal and the quench at the end of the anneal before measurement. However, the h-gain state encoding method has an additional potential source of error which is that if the $s$ at which the pause occurs is very small (close to $0$), then the h-gain bias on the local field terms of the classical Hamiltonian will be negligible, meaning that the spin-up state would not be established. However, we expect this source of error to not be significant for the simulations used in this study because the h-gain field multiplier is very strong (e.g. $3$ or $4$) compared to the typical local field strength and the $s$ values reached at the end of the h-gain ramp (where the pause begins) are not near to $0$ as shown in Figure~\ref{fig:equivalent_anneal_parameters_to_IBMQ_Hamiltonian}.

The primary advantage of the h-gain state encoding method is that it minimizes the effects of the initial ramp that are encountered in the reverse annealing approach, since the spin up state $\uparrow$ is being biased towards whenever the h-gain field strength is less than $0$. In particular, during the forward anneal ramp the h-gain field is still being applied -- and it is only turned entirely off once the anneal schedule reaches the pause. However, this approach still experiences the effects of the fast quench up to $s=1$ during the readout stage.

%%%%%%%%%%%%%%%%%%%%%%%%%%%%%%%%%%%%%%%%%%%%%%%%%%%%
%%%%%%%%%%%%%%%%%%%%%%%%%%%%%%%%%%%%%%%%%%%%%%%%%%%%
%%%%%%%%%%%%%%%%%%%%%%%%%%%%%%%%%%%%%%%%%%%%%%%%%%%%
\subsection{Quantum Annealing Experimental Settings}
\label{section:methods_parameter_setting}

In order to perform the quantum annealing simulations that implement the equivalent Trotterized TFIM dynamics, we derive the equivalent annealing parameters (Section \ref{section:methods_Derivation_of_QA_parameters}) for $100$ linearly spaced angles $\theta_h \in (0, \frac{\pi}{2}]$ (not including $\theta_h = 0$). We then execute these annealing parameters, using anneal schedule ramps that are fast as possible, as outlined in Sections \ref{section:methods_RA_approach}, \ref{section:methods_h_gain_approach}. In particular, the pause duration that is derived (see Figure~\ref{fig:equivalent_anneal_parameters_to_IBMQ_Hamiltonian}) is set to the programmed D-Wave anneal pause time. Because of the ramps, the total annealing time is greater than just the pause duration, but this is necessary due to the constraints of the D-Wave machines. For each of the $100$ $\theta_h$ steps, we use a total of $10,000$ anneals, which are executed in sequences of $10$ device calls each using $1,000$ anneal-readout cycles. \texttt{readout\_thermalization} and \texttt{programming\_thermalization} times are both set to $0$ for all experiments. 

\begin{table*}[t!]
    \begin{center}
        \begin{tabular}{@{}llclll@{}}
            \toprule
            \ & & Annealing time & \multicolumn{3}{c}{Hardware numbers} \\
            \cmidrule(lr){3-3}
            \cmidrule(lr){4-6}
            D-Wave QPU Chip ID & Topology & $(\min,\ \max)$\hspace*{5pt} & Qubits & Couplers & Embeddable heavy-hex subgraphs \\ % Number of disjoint 127 qubit heavy-hex graphs embeddable on the chip \\
            \midrule[\heavyrulewidth]
            \texttt{Advantage\_system4.1} & Pegasus $P_{16}$ & $(0.5\mu s,\ 2000\mu s)$  & 5,627 & 40,279 & 3 disjoint 127 qubit graphs\\
            \midrule
            \texttt{Advantage\_system6.2} & Pegasus $P_{16}$ & $(0.5\mu s,\ 2000\mu s)$  & 5,614 & 40,106 & 4 disjoint 127 qubit graphs\\
            \phantom{\texttt{Advantage\_system}}\texttt{6.3} & & & & 40,105 & \\
            %\phantom{\texttt{Advantage\_system}}\texttt{6.3} & Pegasus $P_{16}$ & $(0.5\mu s,\ 2000\mu s)$  & 5,614 & 40,105 & 4 disjoint 127 qubits \\
            %\texttt{Advantage\_system6.3} & Pegasus $P_{16}$ & $(0.5\mu s,\ 2000\mu s)$  & 5,614 & 40,105 & 4 disjoint 127 qubits \\
            \bottomrule                
        \end{tabular}
    \end{center}
    \vspace{-2ex}
    \caption{D-Wave Quantum Annealing processor summary.\newline%
    %\texttt{Advantage\_system6.2} and \texttt{Advantage\_system6.3} are \emph{the same chip}, with a single coupler de-activated in \texttt{6.3}. 
    During the course of this study, the coupler between qubit 5580 and qubit 5595 on \texttt{Advantage\_system6.2} was de-activated, leading to a new hardware graph, and a new chip id labeled as \texttt{Advantage\_system6.3}, and subsequently the coupler between qubit 104 and 119 was de-activated leading to a new chip id \texttt{Advantage\_system6.4}. Therefore, some of the experiments we report as being on \texttt{Advantage\_system6.2} were executed on \texttt{Advantage\_system6.3} or \texttt{Advantage\_system6.4}. The calibrated energy scales remained the same since the chip was the same, so the results were left labeled as a single chip id. The de-activated coupler did not impact the parallel disjoint embeddings on the chip.
    }
    \label{tab:hardware_summary}
\end{table*}

For both reverse annealing and h-gain state encoding methods, we turn off the coefficient autoscaling in order to precisely control the programmed coefficient weights. For both methods the anneal schedule ramps are performed as fast as allowed by the device specifications. The goal of these fast quenches is to attempt to minimize the effect they have on the system, since ideally we would want to evolve the system at a specific $s$ for the desired duration. Although there are proposals for minimizing the effects of the quenches \cite{Pelofske_2022_inferring, Pelofske_2019}, we employ as fast as possible quenches in order to get the simulations to be as near to the intended annealing time as possible. This is also reasonable since the range of anneal fractions that are varied over are in a similar range (see Figure~\ref{fig:equivalent_anneal_parameters_to_IBMQ_Hamiltonian}). One experimental parameter choice that is important with the anneal schedules that necessarily include quenches is that a pause duration that takes on the order of the same time as the ramp durations will result in an evolution that is dominated by quenching, whereas we want the paused evolution time to dominate the effects of the anneal. Therefore, in general we expect the experimental parameters which require shorter annealing times in Figure~\ref{fig:equivalent_anneal_parameters_to_IBMQ_Hamiltonian} will result in simulations which have higher divergence from the ideal Hamiltonian simulation.

Figure~\ref{fig:example_equivalent_QA_schedules_RA} shows an example set of reverse quantum annealing schedules that implement this equivalent protocol to the Trotterized Hamiltonian dynamics circuits (with rapid schedule ramps). Figure~\ref{fig:example_equivalent_QA_schedules_hgain} shows the same thing for the same experimental parameters, but using the the h-gain state encoding method. 

There are a number of sources of noise and error in the current D-Wave quantum annealers \cite{9465651, pearson2019analog}. One of the ways to mitigate local hardware biases is to perform random spin reversal transforms on the Ising model (and the resulting samples are also spin reversed) \cite{king2014algorithm}, which is also known as gauge transforms. Reverse quantum annealing is not currently compatible with server side spin reversal transforms \cite{D_Wave_solver_parameters}, however server side spin reversal transforms are compatible with forward quantum annealing. Therefore, when using the h-gain state encoding method we utilize $100$ server-side spin reversal transforms per backend call of $1,000$ anneals. 

Table \ref{tab:hardware_summary} summarizes the high level details of the two D-Wave quantum annealers used to perform the simulations. %
%\footnote{During the course of this study, the coupler between qubit 5580 and qubit 5595 on \texttt{Advantage\_system6.2} was de-activated, leading to a new hardware graph, and a new chip id labeled as \texttt{Advantage\_system6.3}. Therefore, some of the experiments we report as being on \texttt{Advantage\_system6.2} were executed on \texttt{Advantage\_system6.3}. The calibrated energy scales remained the same since the chip was the same, so the experiments were left labeled as a single chip id. The de-activated coupler did not impact the parallel disjoint embeddings on the chip.}

For brevity in the remainder of the text, \texttt{Advantage\_system6.2} is occasionally abbreviated as \texttt{DW\_6.2}, and \texttt{Advantage\_system4.1} is abbreviated as \texttt{DW\_4.1}. The D-Wave quantum annealers have several important precision limitations on the programmed parameters users can select. First, the programmed annealing time on the D-Wave devices has a of resolution of $0.01$ microseconds \cite{D_Wave_annealing_time_resolution}. So any higher precision than $0.01$ microseconds will not result in a difference in the executed annealing time. Second, the effective Ising model coefficients that are programmed on the hardware also have a finite resolution due to coefficient quantization that must occur for the digital to analog programming \cite{Bunyk_2014, Johnson_2010}. Although we do not have the exact resolution limits for couplers we use on the two D-Wave quantum annealers that we will use, based on the D-Wave documentation \cite{D_Wave_ICE_error, D_Wave_physical_properties} we expect the quantization limit to be on average approximately $0.001$ to $0.0005$ for near-zero negative coefficients. Therefore, we expect simulations using these coupling strengths to not show any or very little state change compared to the initial uniform spin up state. 

We expect that another contributor to the error of these simulations is the fast quenching up to $s=1$ for qubit measurement. In particular, when the total time spent in quenches are a large proportion of the total simulation time. Because of this, and because of the coupling strength precision limits on the current D-Wave quantum annealers (see the anneal parameters for $N=5$ in Figure~\ref{fig:equivalent_anneal_parameters_to_IBMQ_Hamiltonian}), we do not perform equivalent QA simulations for $N=5$ Trotter steps. On future quantum annealers with slightly higher coupler precision (for instance, $J=-0.0001$), we think that meaningful equivalent $N=5$ simulations could be performed. 

The spin bath polarization effect \cite{lanting2020probing} causes anneals that are repeated in sequence (which reduces total QPU time) to be self correlated. We expect that stronger applied coefficients will result in stronger spin bath polarization effect, therefore we generally expect that weaker coupling coefficients result in less noisy and self correlated data. 
We do not perform low level calibrations such as correcting the anneal offsets, flux bias offsets, or quadratic coefficients to ensure that the simulation is performed as intended on the hardware (e.g. to calibrate for the effective coefficients and anneal schedules which are used) \cite{chern2023tutorial, PRXQuantum.1.020320, King_2022, King_2023, Barbosa_2021, PRXQuantum.2.030317}. Such calibrations would improve the simulation quality of the Hamiltonian, but we leave this open for future improved simulations. Additionally, the current D-Wave quantum annealers are not fully coherent at the annealing times that are available to users \cite{King_2022, King_2023, PhysRevLett.124.090502}. It is of interest to be able to perform these same simulations in a fully coherent analog quantum simulator in the future. 

When reporting the quantum annealing magnetization results in Section \ref{section:results}, we refer to $N$ (introduced in Section \ref{section:methods_Derivation_of_QA_parameters}), which in the IBM Quantum experiment \cite{kim2023evidence} is the number of Trotter steps, as the number of \emph{time steps}. The reason for this is because in the IBM Quantum experiment, the Trotterization for larger $N$ is not performing finer discretizations of the Hamiltonian simulation, but rather longer evolution times (the reason for this is because of the fixed RZZ gate angle being set to $-\frac{\pi}{2}$). Therefore, when we perform the equivalent Hamiltonian simulation using quantum annealing, $N$ is the number time steps, which is equivalent to the Trotter steps of Ref.~\cite{kim2023evidence}, but we do not use the phrase \emph{Trotter steps} in order to make it clear that the quantum annealing simulations are not performing finer discretized, Trotterized, simulation.

Note the connection between the structure of this Trotterized Hamiltonian dynamics circuit and the Quantum Approximate Optimization Algorithm \cite{QAOA, farhi2015quantum}, also known as the Quantum Alternating Operator Ansatz \cite{Hadfield_2019}. The Trotterized Hamiltonian dynamics circuit is essentially performing an amplitude amplification-style computation, with fixed angle phase separator and a fixed angle mixer. Each layer of hardware-compatible Rzz gates corresponds to a fixed-angle $-\frac{\pi}{2}$ phase separator, and a $\theta_h$ layer of single qubit Rx gates that correspond to the standard transverse field mixer with a fixed angle, which is analogous to hardware-compatible QAOA experiments that have been performed on IBM Quantum computers \cite{pelofske2023qavsqaoa, pelofske2023short} except that the order of the analogous phase separator and the mixer layers is switched compared to standard QAOA. This observation is a passing comment on the similarity between existing state of the art quantum computations, but does not inform the time dynamics simulation that this study is examining.

\begin{figure*}[t!]
    \centering
    \includegraphics[width=0.49\textwidth]{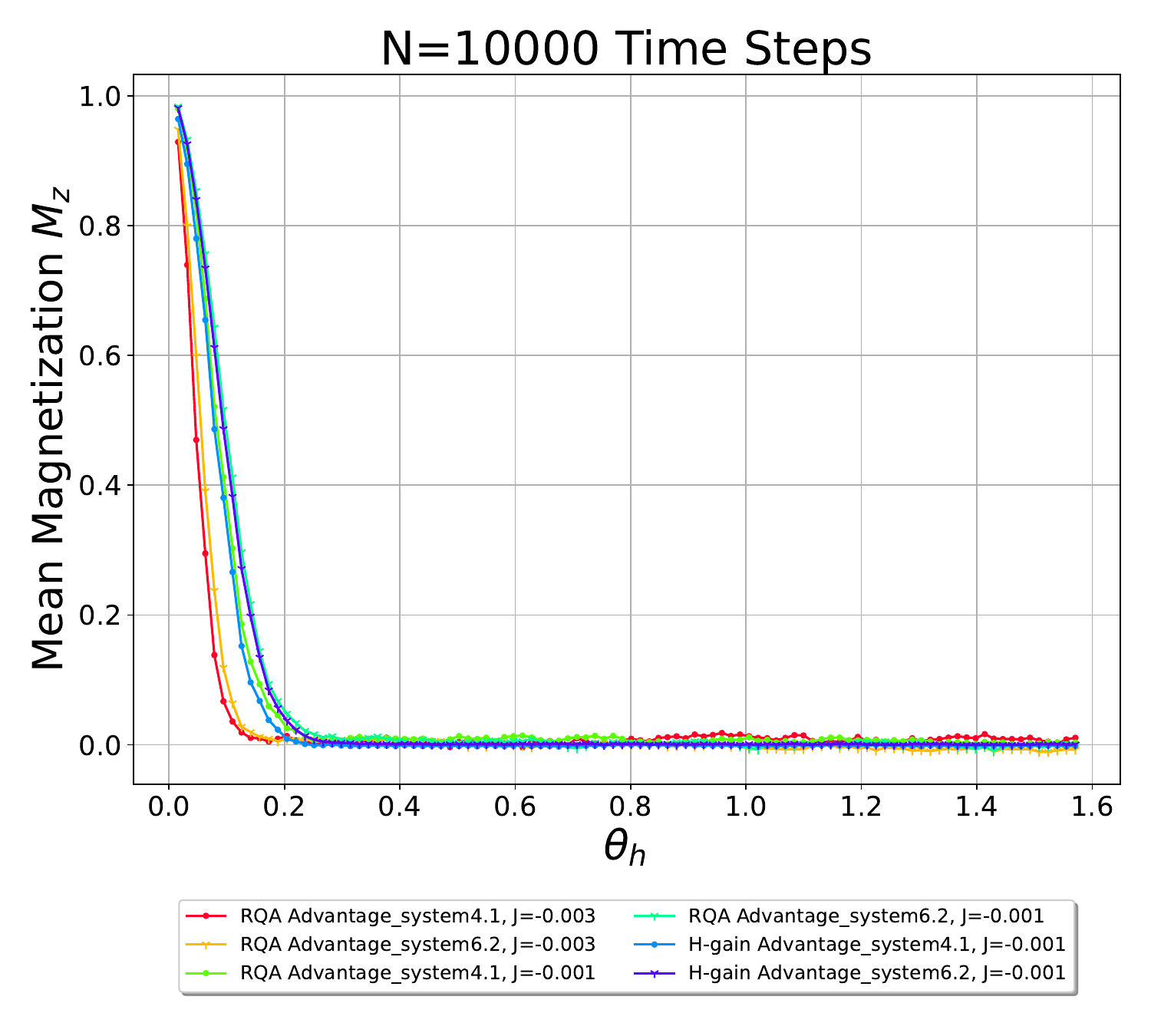}\hfill%
    \includegraphics[width=0.49\textwidth]{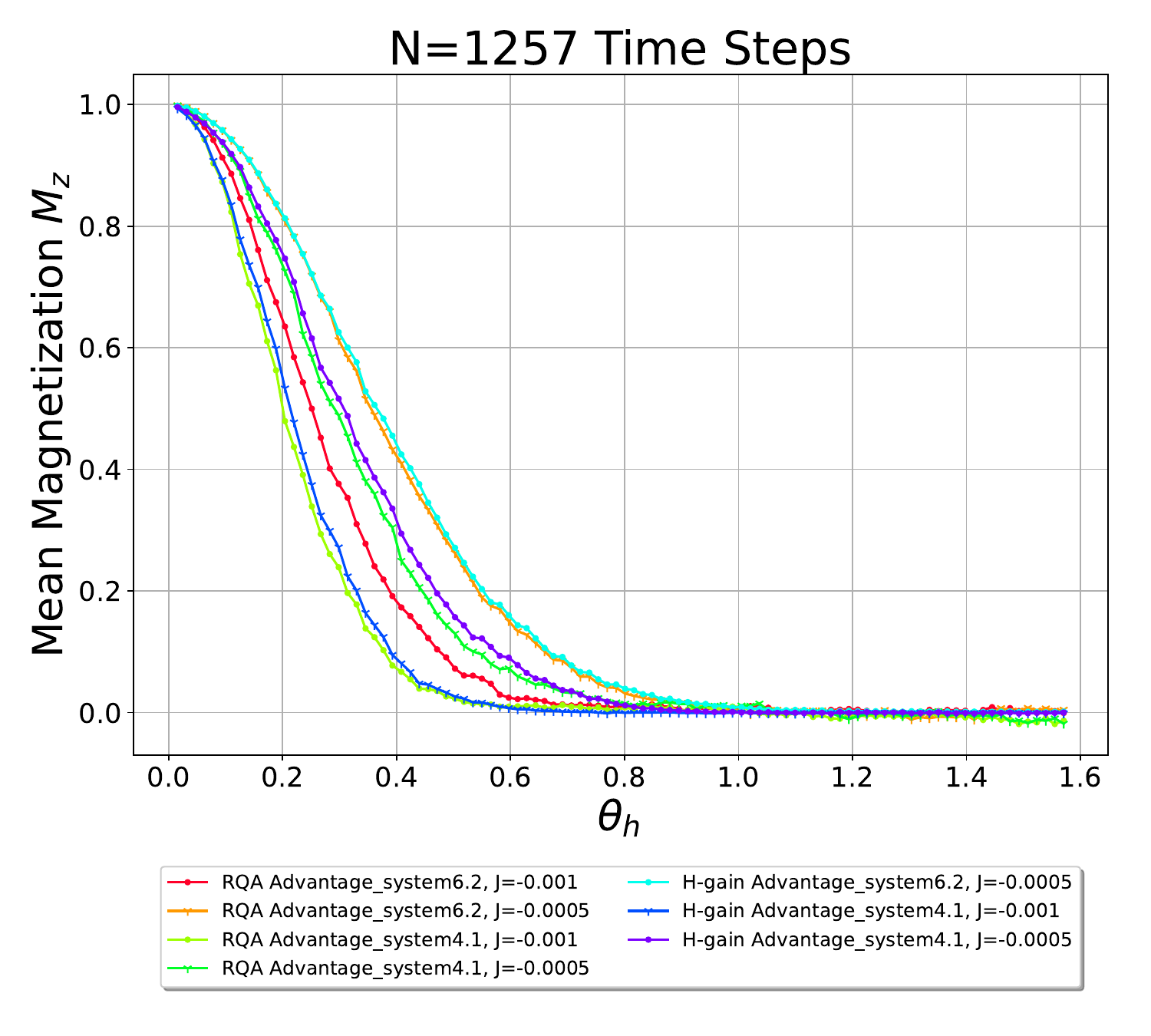}\\%
    \includegraphics[width=0.49\textwidth]{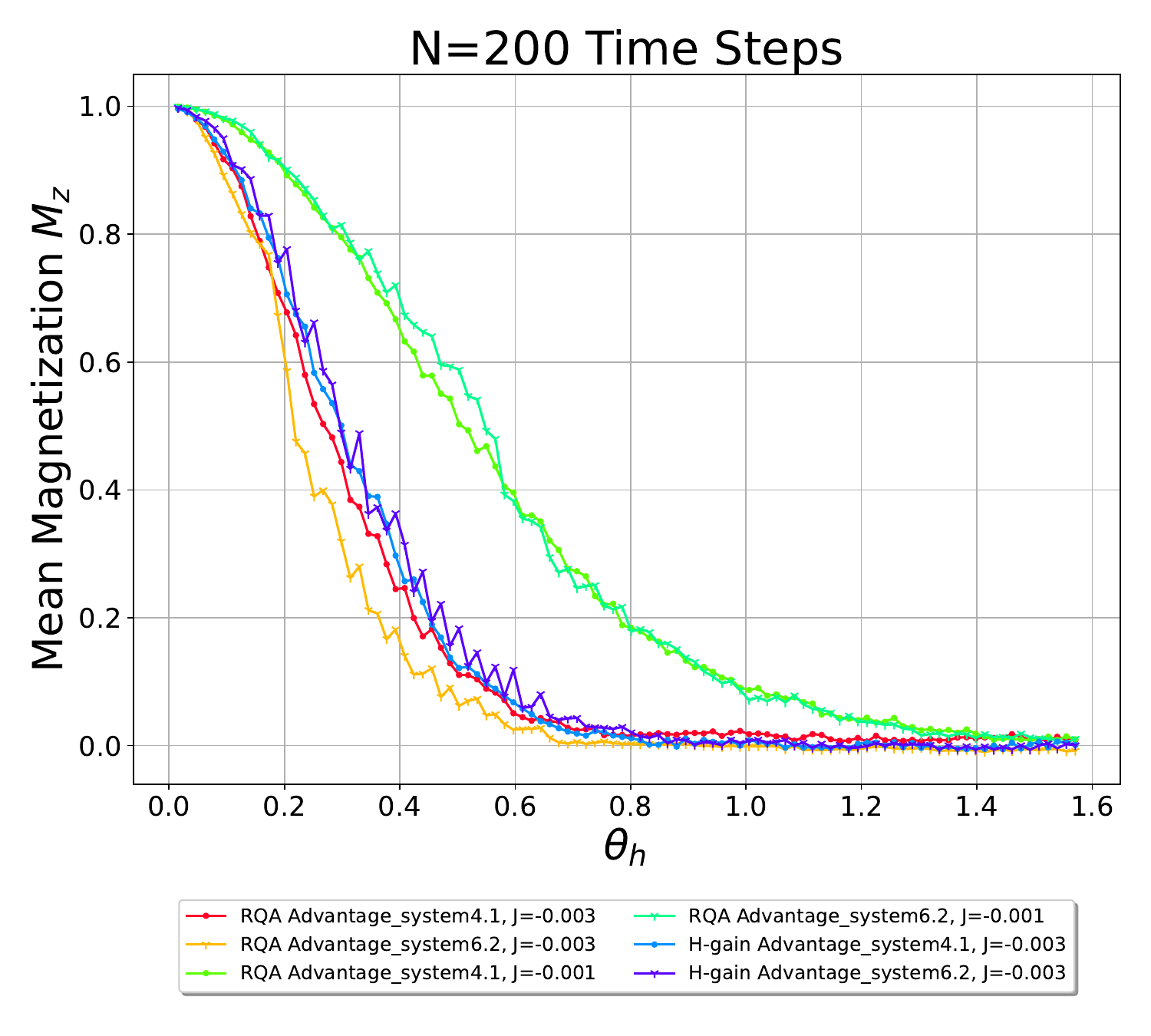}\hfill%
    \includegraphics[width=0.49\textwidth]{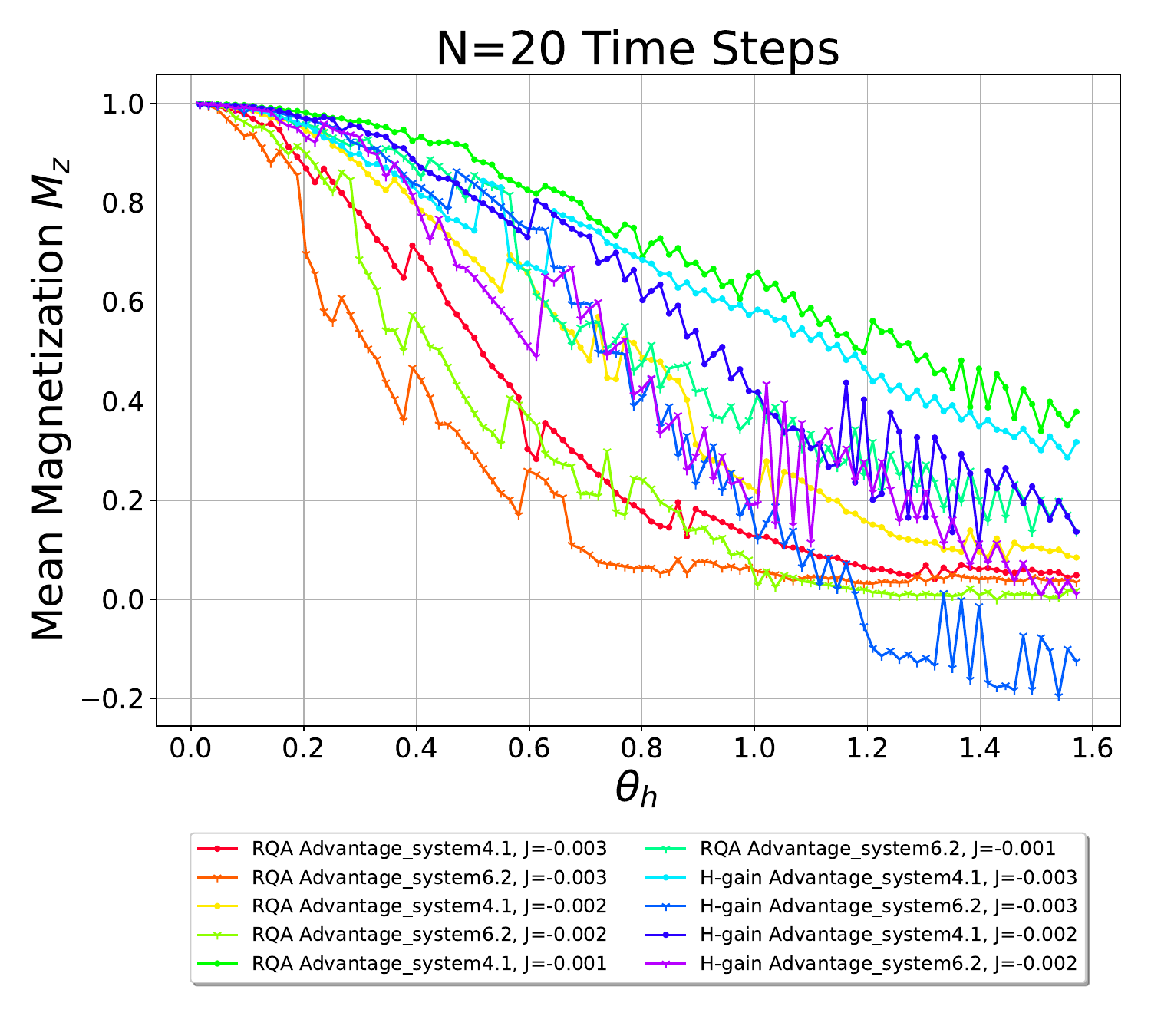}%
    \caption{Equivalent quantum annealing computed mean magnetization observable for different numbers of time (Trotter) steps $N$. These simulations were performed using two distinct methods of initializing the system in the spin up $\uparrow$ state; RQA denotes the \emph{reverse quantum annealing} method (Section \ref{section:methods_RA_approach}), and H-gain denotes the \emph{H-gain state encoding} approach (Section \ref{section:methods_h_gain_approach}). Note that for the different devices and $J$'s, the programmed quantum annealing schedules vary (see Figure~\ref{fig:equivalent_anneal_parameters_to_IBMQ_Hamiltonian}), but the equivalent $\theta_h$ values are plotted on the x-axis so that the experiments can be compared to each other. Here the physical observable being measured is the mean spin (e.g. magnetization) across the entire Ising model.  }
    \label{fig:equivalent_magnetization_on_DWave}
\end{figure*}

\begin{figure*}[t!]
    \centering
    \includegraphics[width=0.49\textwidth,trim={0 80 0 0},clip]{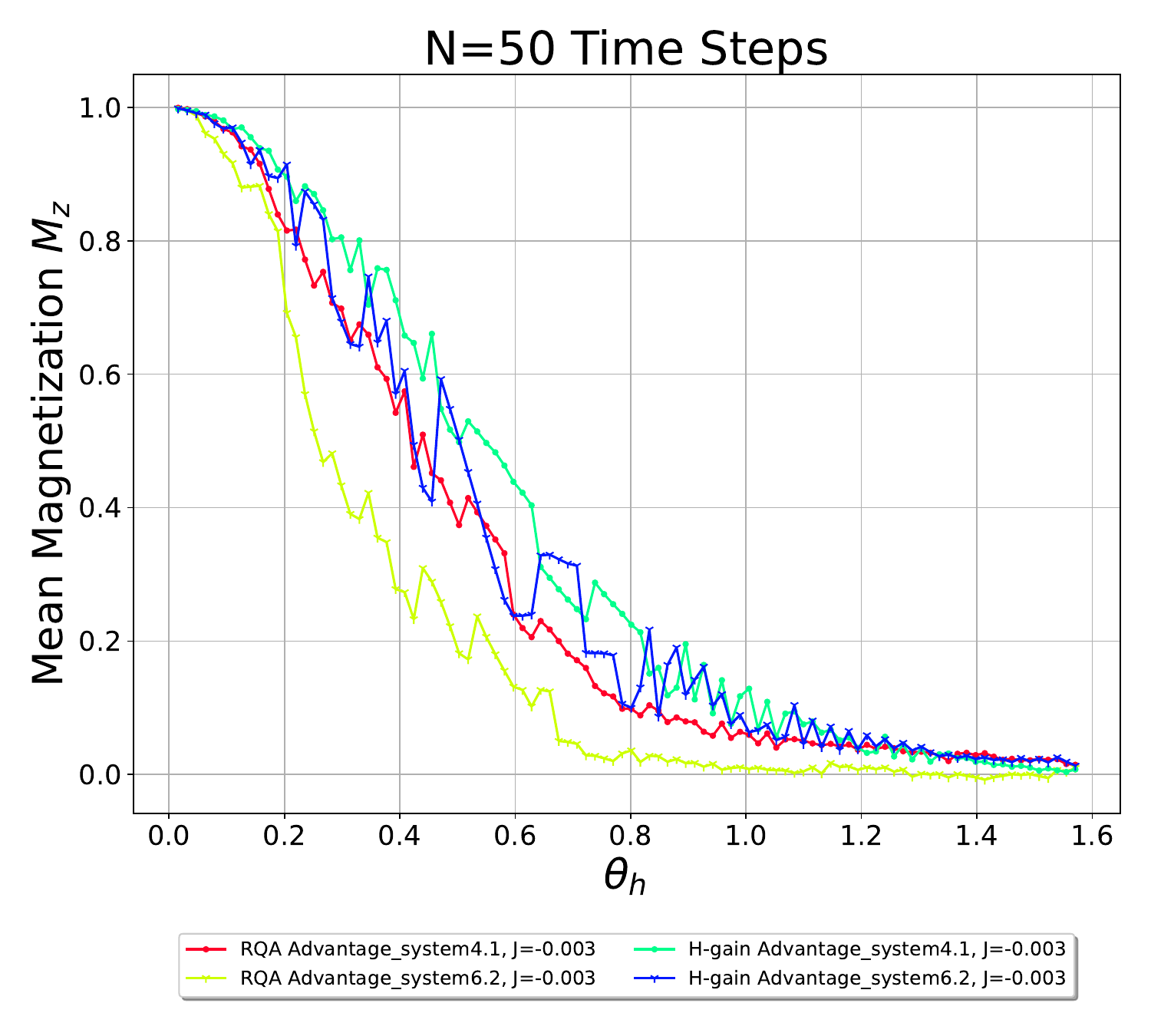}\hfill%
    \includegraphics[width=0.49\textwidth,trim={0 80 0 0},clip]{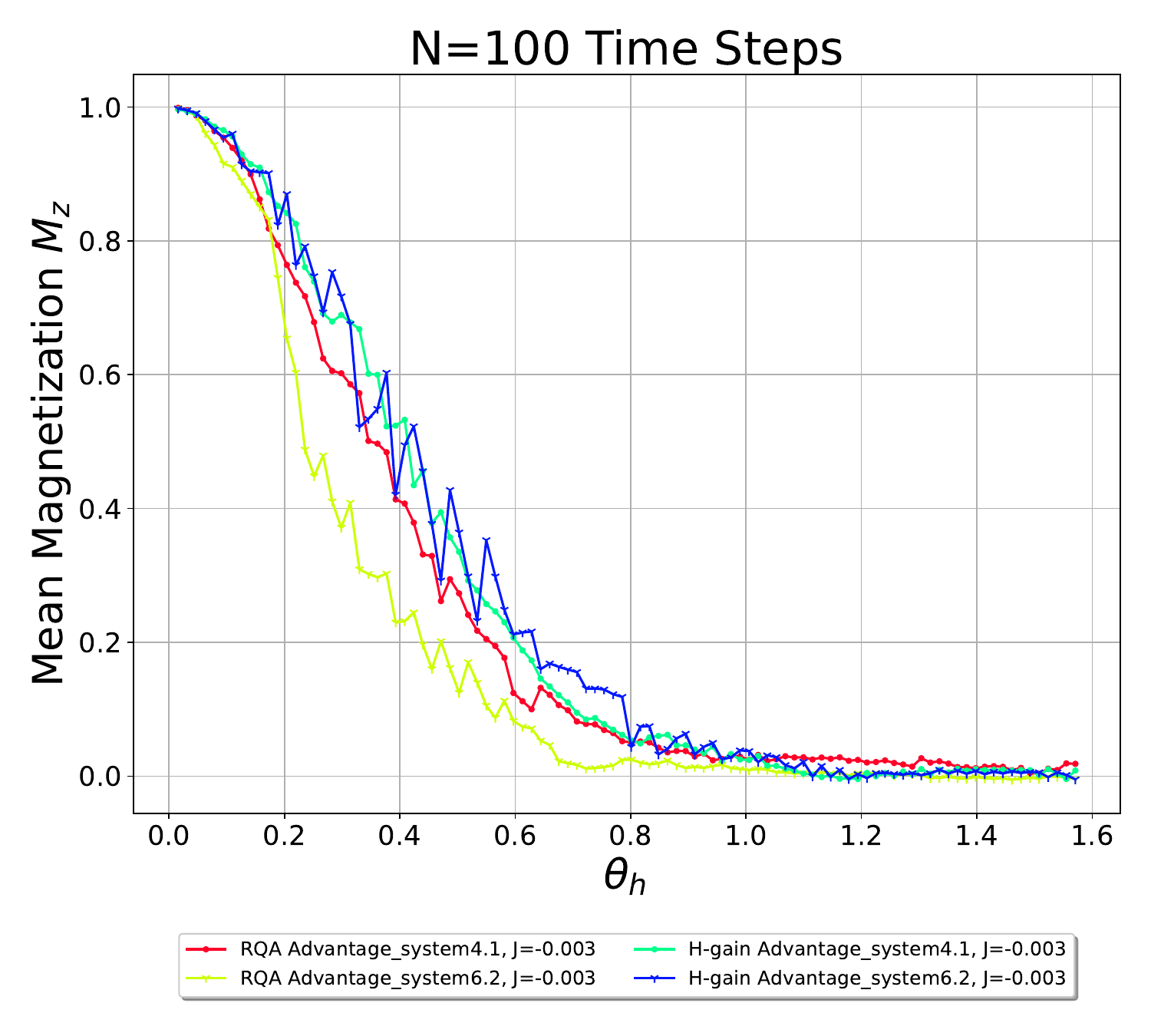}\\%
    \includegraphics[width=0.98\textwidth]{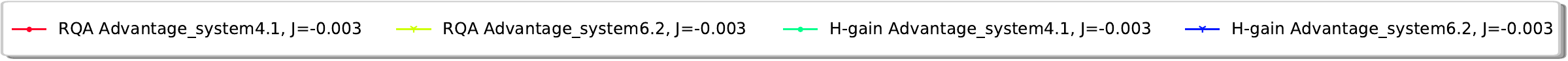}\\[2ex]%
    \includegraphics[width=0.32\textwidth,trim={0 130 0 0},clip]{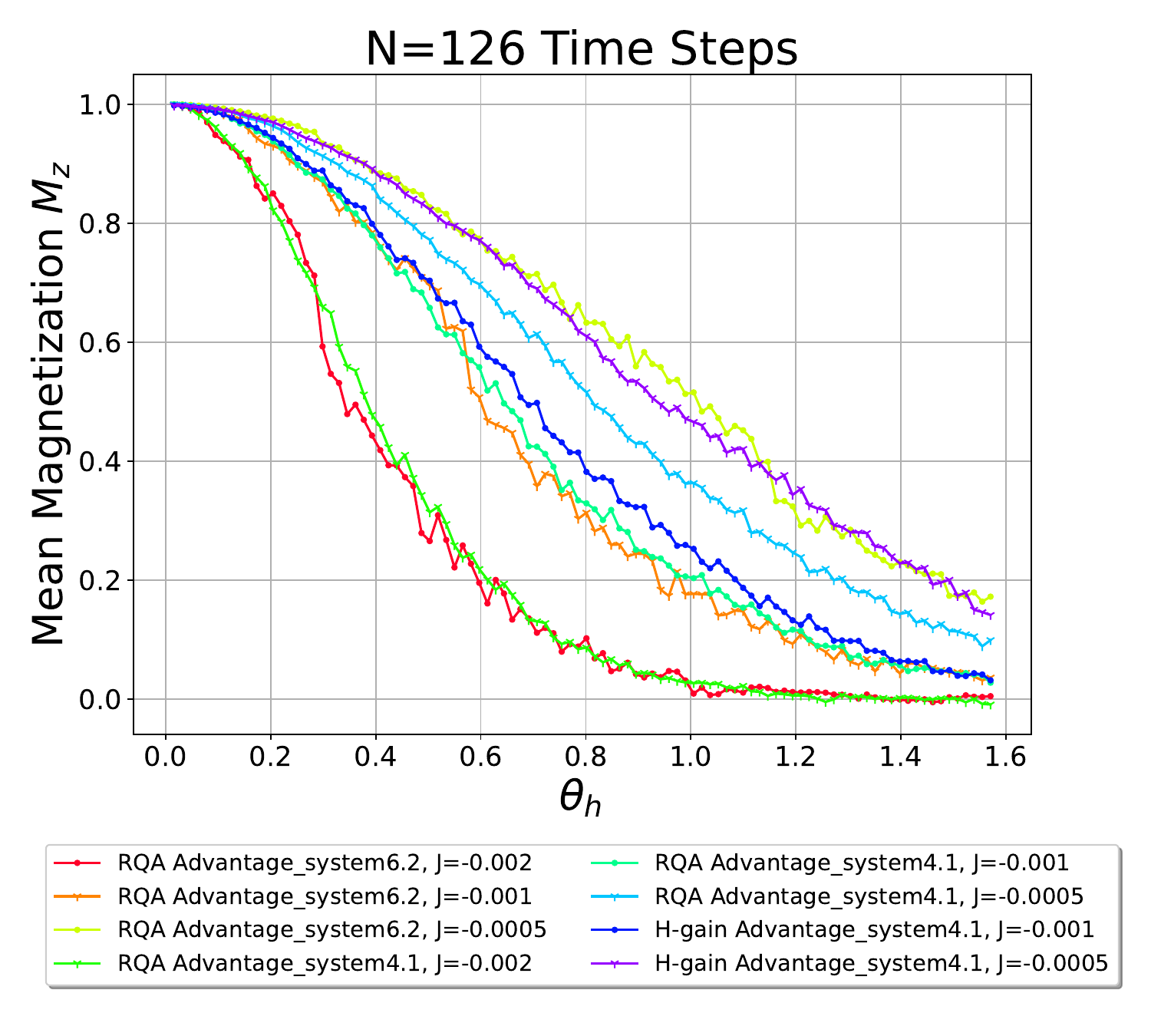}\hfill%
    \includegraphics[width=0.32\textwidth,trim={0 130 0 0},clip]{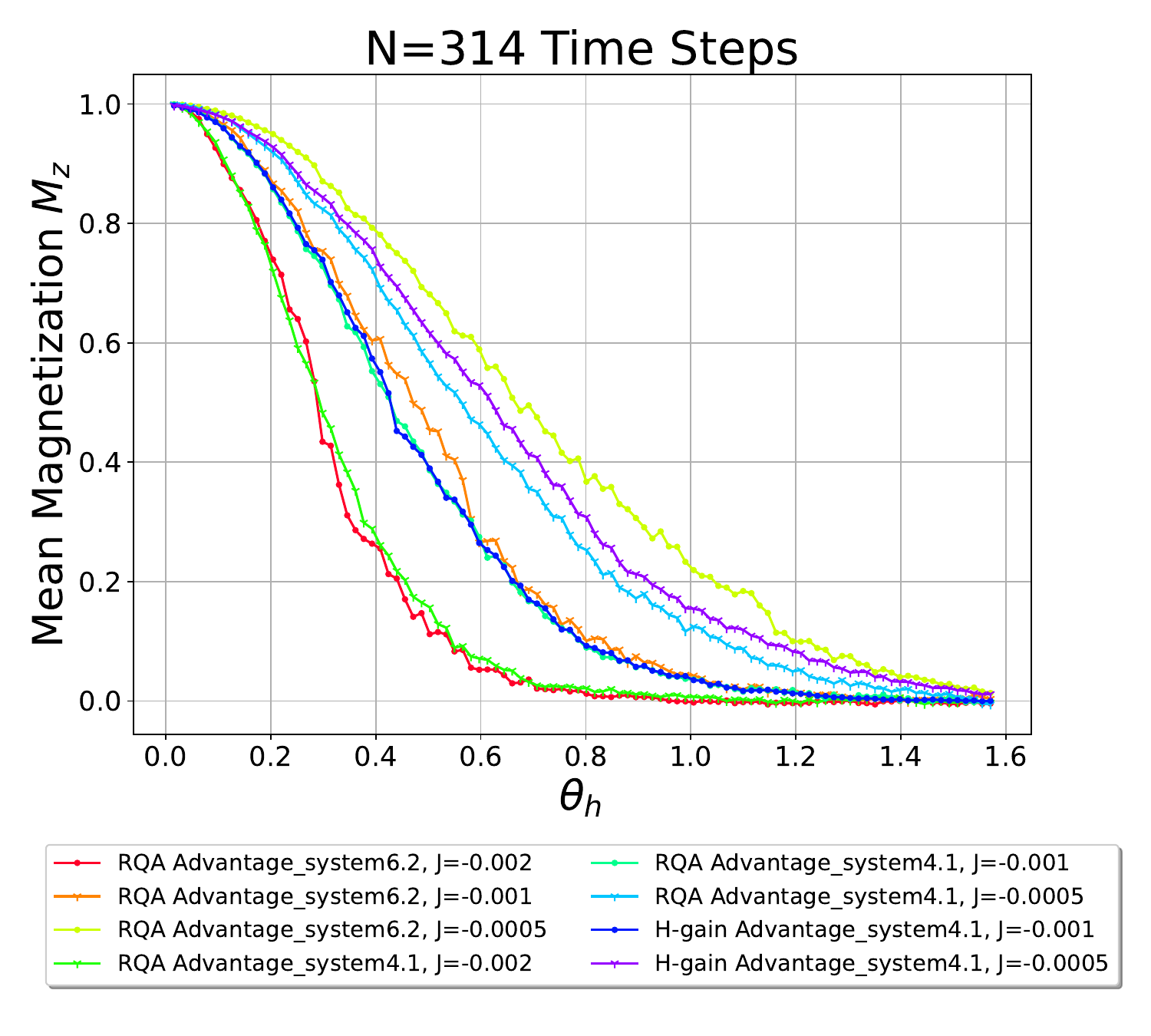}\hfill%
    \includegraphics[width=0.32\textwidth,trim={0 130 0 0},clip]{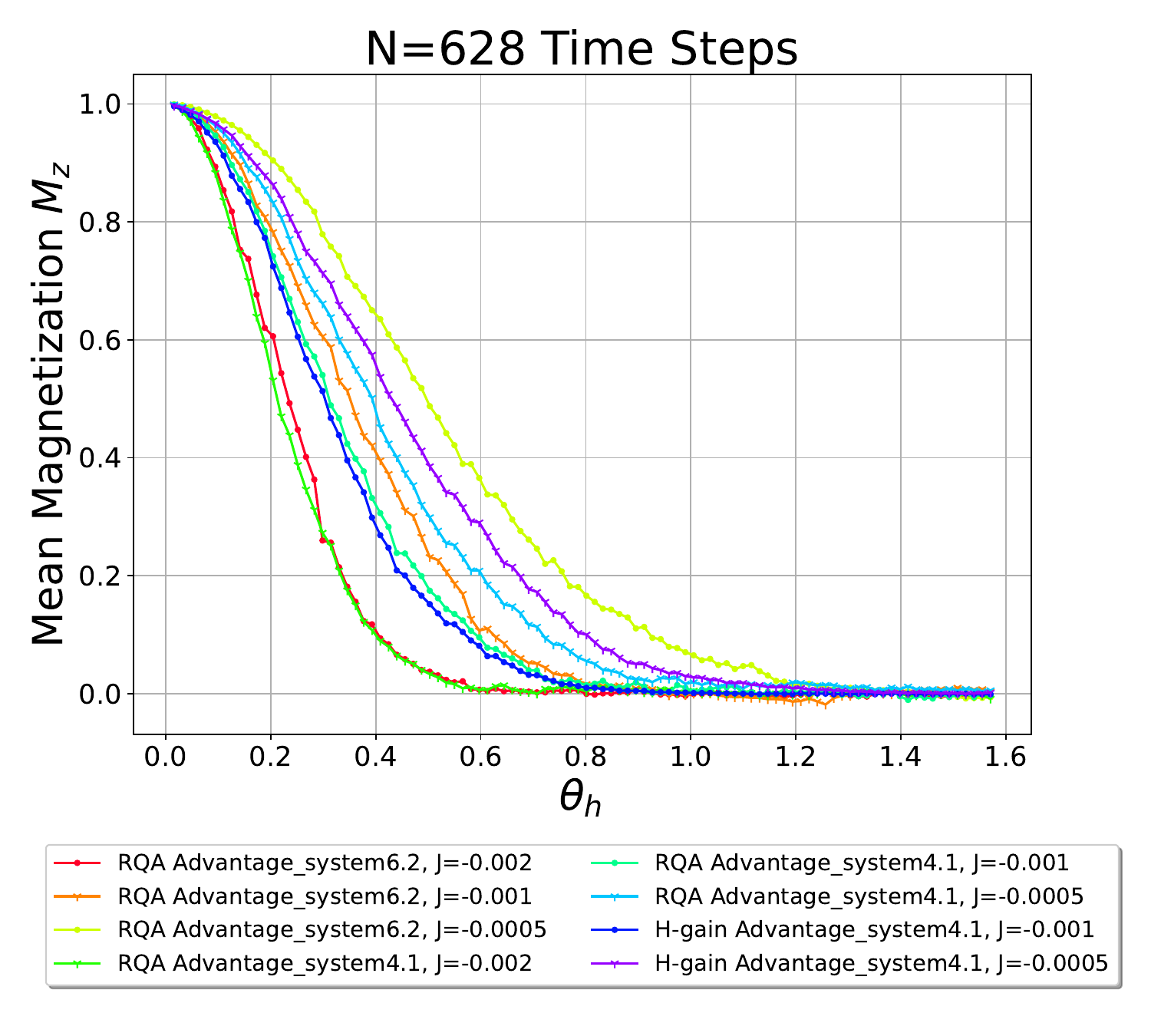}\\%
    \includegraphics[width=0.98\textwidth]{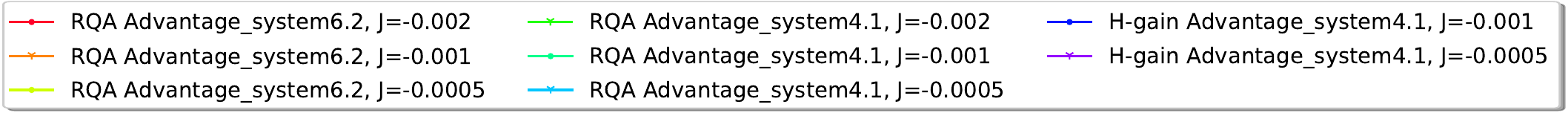}
    \caption{Continuing from Figure~\ref{fig:equivalent_magnetization_on_DWave}, these sub-plots present more equivalent quantum annealing mean lattice magnetization dynamics for varying number of time steps $N$, for varying $\theta_h$, using both RQA (reverse quantum annealing) and H-gain (H-gain state encoding) simulation approaches.}
    \label{fig:equivalent_magnetization_on_DWave_additional_plots}
\end{figure*}

\begin{figure*}[ht!]
    \centering
    \includegraphics[width=0.7\textwidth]{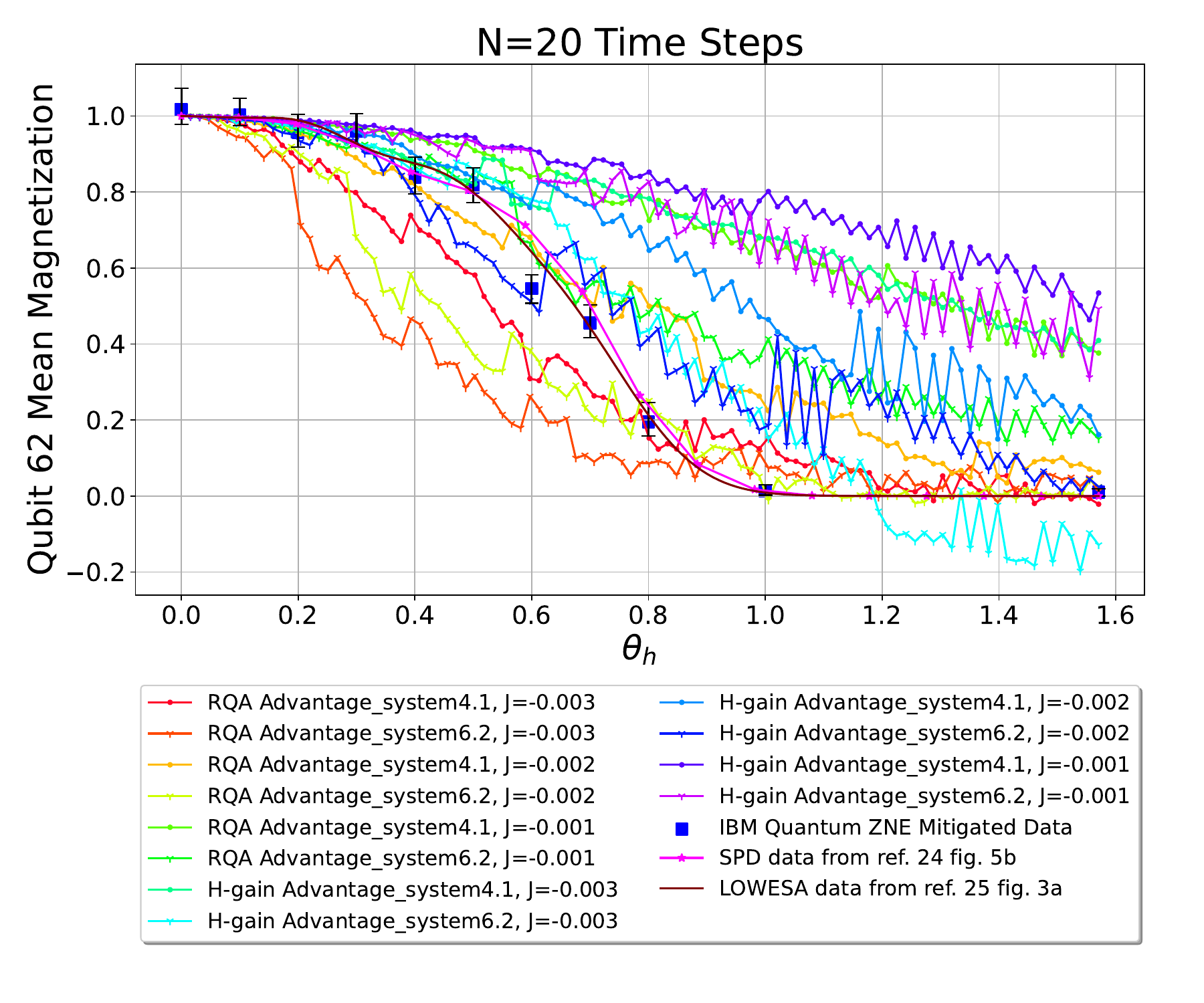}
    \caption{Single site magnetization of qubit $62$ $\braket{Z_{62}}$ as a function of $\theta_h$, for $20$ time steps computed using quantum annealing. This is the same data as Figure~\ref{fig:equivalent_magnetization_on_DWave}, but only plotting the magnetization for qubit $62$ instead of the entire lattice. Reference data of the IBM Quantum ZNE single-site magnetization at $20$ Trotter steps from Ref.~\cite{kim2023evidence} (Figure~4b), the classical simulation from Ref.~\cite{begušić2023fast_2} (Figure~5b, data from \cite{github_data_for_fast_classical_simulation}), and the classical simulation from Ref.~\cite{rudolph2023classical} (Figure~3a) is plotted alongside the quantum annealing magnetization curves. Table \ref{table:20_trotter_step_RMSE} gives the exact RMSE with respect to the LOWESA classical simulation data data, which shows that the lowest error rate simulation data is $0.1273$ from \texttt{DW\_4.1} using reverse annealing and $J=-0.003$. }
    \label{fig:20_Trotter_steps_equivalent_magnetization_on_DWave_qubit_62}
\end{figure*}

\begin{table*}[ht!]
\begin{center}
\begin{tabular}{ |p{1.3cm}|p{1.7cm}|c|p{2.3cm}|c|c| } 
 \hline
 D-Wave device & Simulation technique & Parameters & Total Number of Samples & QPU Time & RMSE \\ 
 \hline
 DW\_4.1 & RQA & $J=-0.003$ & $3,000,000$ & $187.4$ & $0.1273$ \\ 
 \hline
 DW\_6.2 & RQA & $J=-0.003$ & $4,000,000$ & $149.2$ & $0.2555$ \\ 
 \hline
 DW\_4.1 & RQA & $J=-0.002$ & $3,000,000$ & $187.6$ & $0.1513$ \\ 
 \hline
 DW\_6.2 & RQA & $J=-0.002$ & $4,000,000$ & $149.4$ & $0.1772$ \\ 
 \hline
 DW\_4.1 & RQA & $J=-0.001$ & $3,000,000$ & $188.1$ & $0.4141$ \\ 
 \hline
 DW\_6.2 & RQA & $J=-0.001$ & $4,000,000$ & $149.9$ & $0.2068$ \\ 
 \hline
 DW\_4.1 & H-gain & $J=-0.003$ & $3,000,000$ & $1653.3$ & $0.4219$ \\ 
 \hline
 DW\_6.2 & H-gain & $J=-0.003$ & $4,000,000$ & $1631.6$ & $0.1353$ \\ 
 \hline
 DW\_4.1 & H-gain & $J=-0.002$ & $3,000,000$ & $1653.3$ & $0.2904$ \\ 
 \hline
 DW\_6.2 & H-gain & $J=-0.002$ & $4,000,000$ & $1631.6$ & $0.1601$ \\ 
 \hline
 DW\_4.1 & H-gain & $J=-0.001$ & $3,000,000$ & $1653.9$ & $0.5015$ \\ 
 \hline
 DW\_6.2 & H-gain & $J=-0.001$ & $4,000,000$ & $1632.2$ & $0.4226$ \\ 
 \hline
\end{tabular}
\end{center}
\caption{RMSE measures for the single site qubit $62$ magnetization observable from the $127$ qubit quantum annealing simulations, where the reference data is a spline interpolation of the data from Ref.~\cite{rudolph2023classical} (Figure~3a). The Total Number of Samples field is reported as the number of samples used to compute the full magnetization curve for all $\theta_h$ steps - this count includes each independent tiled embedding as a separate sample. QPU time is reported in seconds as the \texttt{QPU access time}, which is the total server side time used to perform the computation, not just annealing time. The H-gain technique QPU times are so large because of the server-side spin reversal transforms. Because the magnetization is for a single site, the number of samples used is also equal to the number of spins over which the magnetization curve is measured. The minimum RMSE across all devices and parameters is $0.1273$.  }
\label{table:20_trotter_step_RMSE}
\end{table*}

\begin{table*}[ht!]
\begin{center}
\begin{tabular}{ |p{1.3cm}|p{1.7cm}|c|p{2.3cm}|c|c| } 
 \hline
 D-Wave device & Simulation technique & Parameters & Total Number of Samples & QPU Time & RMSE \\ 
 \hline
 DW\_4.1 & RQA & $J=-0.0034$ & $30,000,000$ & $6907.0$ & $0.1591$ \\ 
 \hline
 DW\_4.1 & RQA & $J=-0.0033$ & $30,000,000$ & $6907.6$ & $0.1474$ \\ 
 \hline
 DW\_4.1 & RQA & $J=-0.0032$ & $30,000,000$ & $6908.1$ & $0.1409$ \\ 
 \hline
 DW\_4.1 & RQA & $J=-0.0031$ & $30,000,000$ & $6908.6$ & $0.1412$ \\ 
 \hline
 DW\_4.1 & RQA & $J=-0.003$ & $30,000,000$ & $6909.1$ & $0.1303$ \\ 
 \hline
 DW\_4.1 & RQA & $J=-0.0029$ & $30,000,000$ & $6909.8$ & $0.1075$ \\ 
 \hline
 DW\_4.1 & RQA & $J=-0.0028$ & $30,000,000$ & $6910.5$ & $0.0952$ \\ 
 \hline
 DW\_4.1 & RQA & $J=-0.0027$ & $30,000,000$ & $6911.2$ & \textcolor{blue}{$0.0932$} \\ 
 \hline
 DW\_4.1 & RQA & $J=-0.0026$ & $30,000,000$ & $6912.0$ & \textcolor{blue}{$0.0932$} \\ 
 \hline
 DW\_4.1 & RQA & $J=-0.0025$ & $30,000,000$ & $6912.5$ & $0.1102$ \\ 
 \hline
 DW\_4.1 & RQA & $J=-0.0024$ & $30,000,000$ & $6913.8$ & $0.1037$ \\ 
 \hline
 DW\_4.1 & RQA & $J=-0.0023$ & $30,000,000$ & $6914.8$ & $0.0979$ \\ 
 \hline
 DW\_4.1 & RQA & $J=-0.0022$ & $30,000,000$ & $6915.9$ & $0.1057$ \\ 
 \hline
 DW\_4.1 & RQA & $J=-0.0021$ & $30,000,000$ & $6917.1$ & $0.1211$ \\ 
 \hline
 DW\_4.1 & RQA & $J=-0.002$ & $30,000,000$ & $6918.3$ & $0.1532$ \\ 
 \hline
 DW\_4.1 & RQA & $J=-0.0015$ & $30,000,000$ & $6927.6$ & $0.2586$ \\ 
 \hline
 DW\_4.1 & RQA & $J=-0.001$ & $30,000,000$ & $6946.0$ & $0.4038$ \\ 
 \hline
 DW\_4.1 & RQA & $J=-0.0005$ & $30,000,000$ & $7000.3$ & $0.5758$ \\ 
 \hline
 DW\_4.1 & RQA & $J=-0.0001$ & $30,000,000$ & $7411.2$ & $0.7035$ \\ 
 \hline
 DW\_6.2 & RQA & $J=-0.0034$ & $40,000,000$ & $6526.6$ & $0.3176$ \\ 
 \hline
 DW\_6.2 & RQA & $J=-0.0033$ & $40,000,000$ & $6527.2$ & $0.3043$ \\ 
 \hline
 DW\_6.2 & RQA & $J=-0.0032$ & $40,000,000$ & $6527.8$ & $0.2992$ \\ 
 \hline
 DW\_6.2 & RQA & $J=-0.0031$ & $40,000,000$ & $6528.4$ & $0.2835$ \\ 
 \hline
 DW\_6.2 & RQA & $J=-0.003$ & $40,000,000$ & $6529.0$ & $0.272$ \\ 
 \hline
 DW\_6.2 & RQA & $J=-0.0029$ & $40,000,000$ & $6529.7$ & $0.2709$ \\ 
 \hline
 DW\_6.2 & RQA & $J=-0.0028$ & $40,000,000$ & $6530.5$ & $0.2614$ \\ 
 \hline
 DW\_6.2 & RQA & $J=-0.0027$ & $40,000,000$ & $6531.3$ & $0.2445$ \\ 
 \hline
 DW\_6.2 & RQA & $J=-0.0026$ & $40,000,000$ & $6532.2$ & $0.2296$ \\ 
 \hline
 DW\_6.2 & RQA & $J=-0.0025$ & $40,000,000$ & $6533.2$ & $0.2132$ \\ 
 \hline
 DW\_6.2 & RQA & $J=-0.0024$ & $40,000,000$ & $6534.2$ & $0.2165$ \\ 
 \hline
 DW\_6.2 & RQA & $J=-0.0023$ & $40,000,000$ & $6535.3$ & $0.2009$ \\ 
 \hline
 DW\_6.2 & RQA & $J=-0.0022$ & $40,000,000$ & $6536.5$ & $0.1927$ \\ 
 \hline
 DW\_6.2 & RQA & $J=-0.0021$ & $40,000,000$ & $6537.9$ & $0.1805$ \\ 
 \hline
 DW\_6.2 & RQA & $J=-0.002$ & $40,000,000$ & $6539.4$ & $0.1679$ \\ 
 \hline
 DW\_6.2 & RQA & $J=-0.0015$ & $40,000,000$ & $6549.7$ & \textcolor{blue}{$0.1459$} \\ 
 \hline
 DW\_6.2 & RQA & $J=-0.001$ & $40,000,000$ & $6570.2$ & $0.2251$ \\ 
 \hline
 DW\_6.2 & RQA & $J=-0.0005$ & $40,000,000$ & $6630.7$ & $0.4892$ \\ 
 \hline
 DW\_6.2 & RQA & $J=-0.0001$ & $40,000,000$ & $7087.3$ & $0.7037$ \\ 
 \hline
\end{tabular}
\end{center}
\caption{Reduced error rate RMSE measures for the single site qubit $62$ magnetization observable from the $127$ qubit quantum annealing simulations using reverse quantum annealing, with reduced intersample correlation enabled, an increased number of samples, and enumeration over more coupling strengths than what is shown in Table \ref{table:20_trotter_step_RMSE}. The RMSE is again computed from the reference data of a spline interpolation of the simulation data from Ref.~\cite{rudolph2023classical} (Figure~3a). The lowest RMSE measure across all of the devices and parameters is $0.0932$. Blue text denotes the minimum RMSE found across all $J$ coupler strengths found using the two Pegasus graph quantum annealers (the best RMSE found with DW\_4.1 was $0.0932$ for two $J$ values and therefore two of the entries are marked). }
\label{table:20_trotter_step_RMSE_reduced_error}
\end{table*}

\begin{figure*}[ht!]
    \centering
    \includegraphics[width=0.7\textwidth]{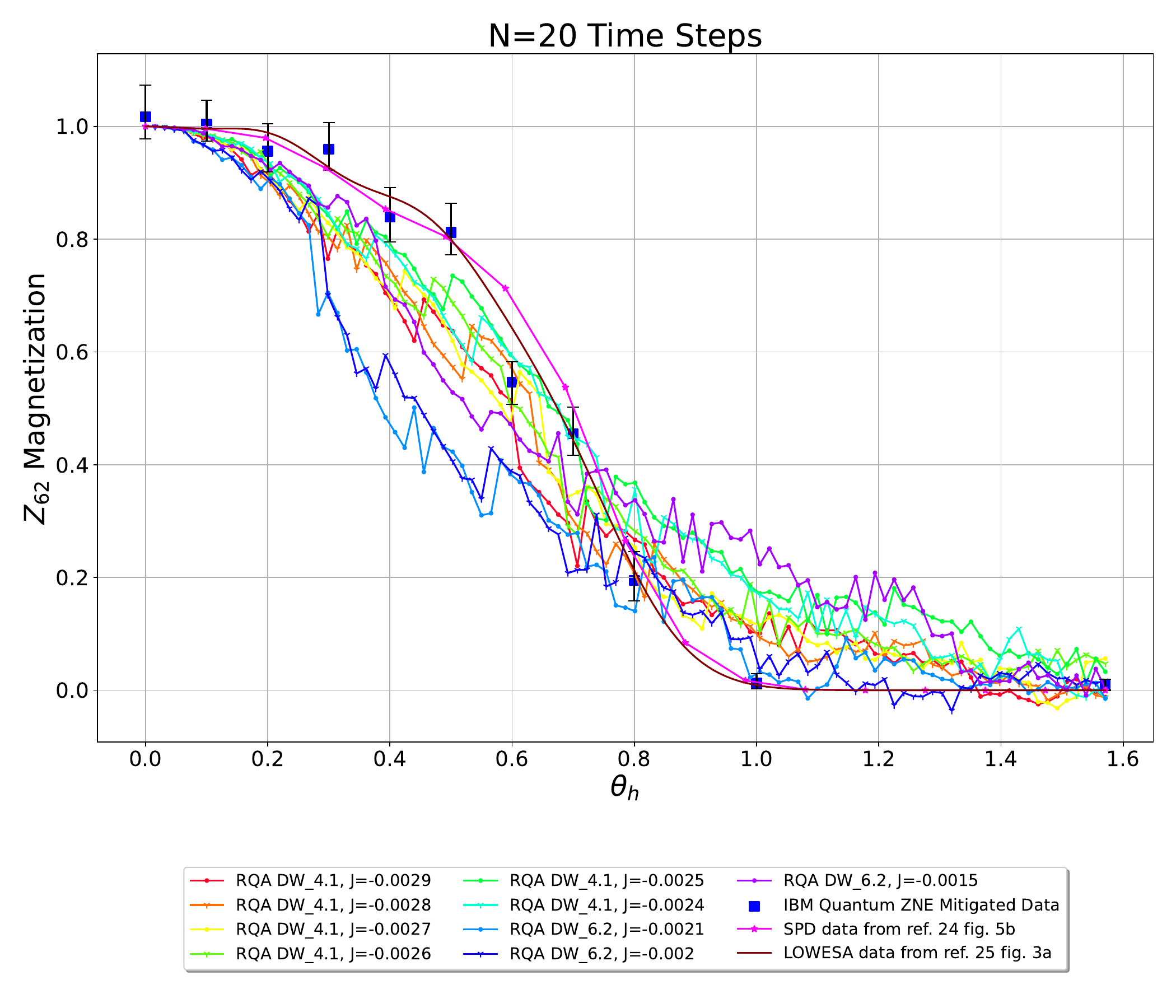}
    \caption{Single site mean magnetization observable $\braket{Z_{62}}$ at $20$ time steps, but with quantum annealing that has reduced error sources -- namely higher shot counts and reduced intersample correlations. This plot shows only a subset of the settings that produced some of the lowest RMSE data. Table \ref{table:20_trotter_step_RMSE_reduced_error} shows the exact error rates these magnetization curves with respect to an existing classical simulation. }
    \label{fig:20_Trotter_steps_equivalent_magnetization_on_DWave_qubit_62_reduced_intersample_correlation}
\end{figure*}

\begin{figure*}[ht!]
    \centering
    \includegraphics[width=0.7\textwidth]{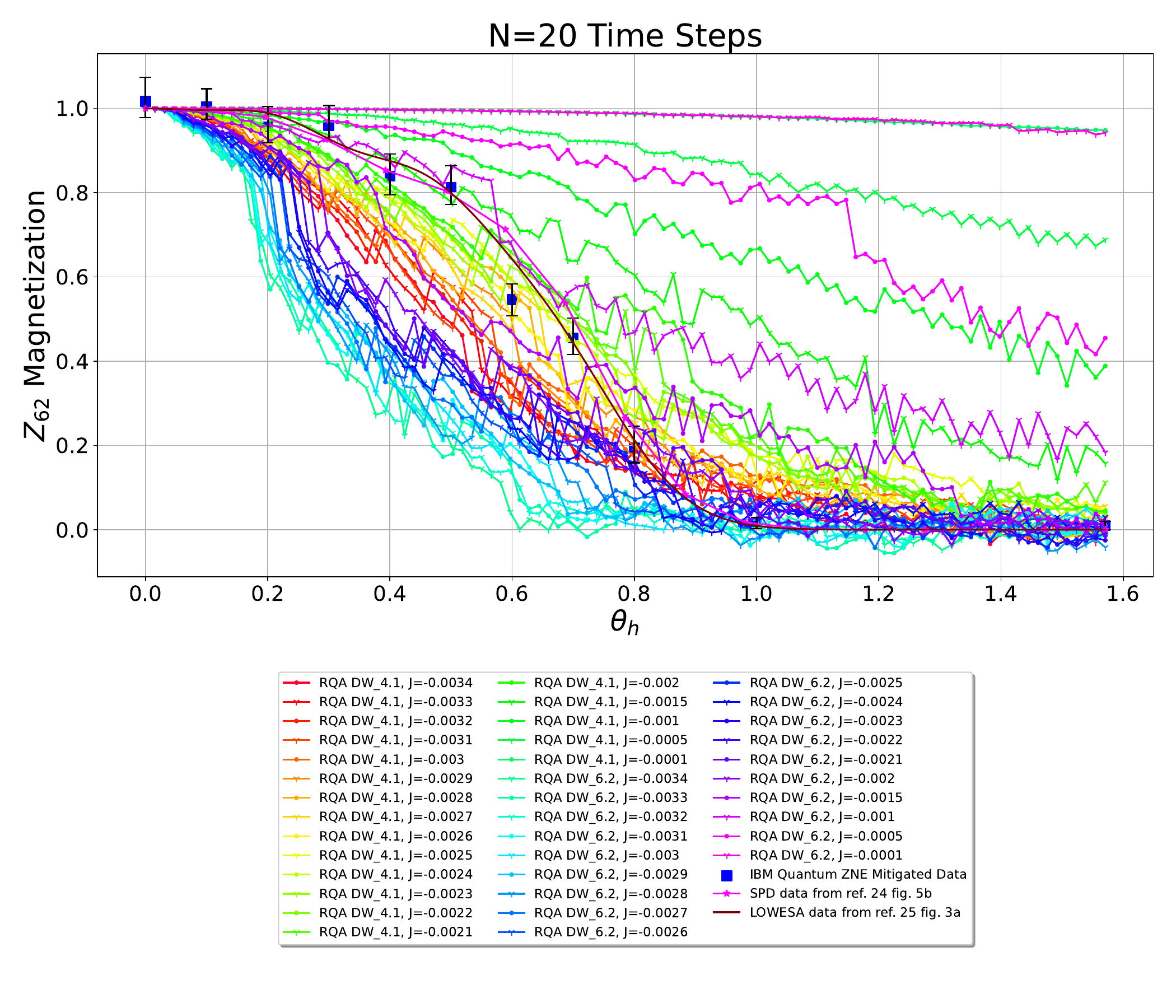}
    \vspace{-0.8cm}
    \caption{Single site mean magnetization observable $\braket{Z_{62}}$ at $20$ time steps computed using reverse quantum annealing and reduced intersample correlations. Table \ref{table:20_trotter_step_RMSE_reduced_error} shows the exact error rates for the complete range of settings. This plot is showing the magnetization curves for a complete parameter sweep of J values - whereas Figure \ref{fig:20_Trotter_steps_equivalent_magnetization_on_DWave_qubit_62_reduced_intersample_correlation} only shows a subset of these settings that were the closest match to the classical LOWESA simulation data.  }
    \label{fig:20_Trotter_steps_equivalent_magnetization_on_DWave_qubit_62_reduced_intersample_correlation_all}
\end{figure*}

%%%%%%%%%%%%%%%%%%%%%%%%%%%%%%%%%%%%%%%%%%%%%%%%%%%%
%%%%%%%%%%%%%%%%%%%%%%%%%%%%%%%%%%%%%%%%%%%%%%%%%%%%
%%%%%%%%%%%%%%%%%%%%%%%%%%%%%%%%%%%%%%%%%%%%%%%%%%%%
\subsection{Greater Than 127 Qubit Heavy-Hex Lattice Quantum Annealing Simulations}
\label{section:methods_QA_larger_lattice}

Because of the size of the current D-Wave Pegasus hardware graphs, it is possible to embed heavy-hex graphs that are larger than $127$ qubits onto the D-Wave devices. However, due to manufacturing defects, and in general because minor-embedding can be quite computationally hard, it is difficult to natively embed (meaning, without minor-embeddings) extremely large heavy-hex lattices. So as to demonstrate the capability of going larger than $127$ qubits, we embed a $384$ heavy-hex lattice onto the hardware graph of \texttt{Advantage\_system4.1}, using the heuristic minor embedding tool \texttt{minorminer} \cite{cai2014practical}. To be consistent with the structured $127$ qubit heavy-hex embedding shown in Figure \ref{fig:Pegasus_embedding}, the goal of this embedding was for it to be exactly hardware compatible, with no chained qubits to form a minor embedding. Specifically, over 1 million randomly initialized embedding attempts were made, where minor-embeddings were discarded, until a hardware native embedding was found. Larger heavy-hex lattices failed to be embedded at a similar level of \texttt{minorminer} attempts. The $384$ node heavy-hex graph is not of a specific IBM quantum processor hardware graph, rather this was custom generated and specifically designed to be roughly square (with $8x8$ hexagon units). 

Figure \ref{fig:384_node_heavy_hex_graph} in Appendix \ref{section:appendix_384_node_heavy_hex} shows the $384$ node heavy-hex graph, along with the embedding onto the hardware Pegasus graph. Unlike the $127$ node heavy-hex embeddings, this embedding is not tiled repeatedly across the D-Wave hardware graph, meaning that only one sample is obtained for each anneal-readout cycle.

%%%%%%%%%%%%%%%%%%%%%%%%%%%%%%%%%%%
%%%%%%%%%%%%%%%%%%%%%%%%%%%%%%%%%%%
%%%%%%%%%%%%%%%%%%%%%%%%%%%%%%%%%%%
\section{Results}
\label{section:results}

Figure~\ref{fig:equivalent_magnetization_on_DWave} plots the equivalent Hamiltonian magnetization $M_z$ (across all qubits) dynamics up to $10,000$ Trotter steps (time steps), for $\theta_h \in (0, \frac{\pi}{2}]$, on two D-Wave quantum annealers using varying programmed $J$ coupling strengths%
\footnote{Note that some of these simulations were performed for Trotter steps that are rounded integer multiples of $\pi$.}.
The quantum annealing simulations have varying magnetization curves for different $N$ and different $J$ values. Different annealing time durations, different ramp durations, and different coupling strengths, lead to different sources of error in the computation, resulting in different magnetization curves, but because we can choose different quantum annealing parameters in order to perform equivalent $N$ Trotter step simulations (see Section \ref{section:methods_Derivation_of_QA_parameters}), we can see what the effects are of changing the programmed quantum annealing parameters, such as the ferromagnetic coupling strength. Currently, classical simulations of the magnetization dynamics of 127 qubit Trotterized circuits for such high number of Trotter steps has not been performed. Full lattice magnetization simulations have also not yet been reported on digital gate model quantum devices (however in Appendix \ref{section:appendix_IBM_Quantum_Hardware_simulations} we perform non-mitigated magnetization dynamics experiments on several IBM Quantum processors in order to assess what the mean magnetization observable would be). Therefore, we do not have a good basis for comparison for $N$ Trotter steps greater than $20$. However, we can compare against the experimental results and classical simulation results for single site magnetization at $20$ Trotter steps, which is shown in Figure~\ref{fig:20_Trotter_steps_equivalent_magnetization_on_DWave_qubit_62}. The quantum annealing results for $N=20$ are somewhat consistent with these single site magnetization plots but are quite noisy. In an ideal quantum annealing simulation of this system, the different choices of the $J$ coupling strength would result in the same magnetization observables being computed for each $N$ and each $\theta_h$; the differences in the D-Wave results are due to various sources of error affecting the computation in different ways. This error is primarily due to the extremely short annealing times, and fast ramps, being used in these simulations (on the order of $1$ microsecond, see Figure~\ref{fig:equivalent_anneal_parameters_to_IBMQ_Hamiltonian}). The other primary source of error is the limited precision of the coefficient encoding on the D-Wave device; $J$ values close to zero are heavily encountering this effect, which results in the state not changing very much from the initial spin up state, which we see occurring especially for $J=-0.001$ in Figure~\ref{fig:20_Trotter_steps_equivalent_magnetization_on_DWave_qubit_62}. The minimum root mean square error (RMSE) for the single site magnetization, with reference to the LOWESA \cite{rudolph2023classical} classical simulation data with a spline interpolation, is $0.1273$, which was computed on \texttt{DW\_4.1} with $J=-0.003$ using reverse quantum annealing (the range of possible RMSE measure for magnetization in the range of $[0, 1]$ is $[0, 1]$). Table~\ref{table:20_trotter_step_RMSE} enumerates the RMSE measures for all of the $127$ qubit $N=20$ simulations. 

In order to examine the single site magnetization properties for the entire lattice, in Figure~\ref{fig:single_site_magnetization_heatmaps} in Appendix \ref{section:appendix_single_site_magnetization_heatmaps} we plot single site magnetization for the entire heavy-hex lattice for a subset of the $N=126$ time step quantum annealing results. Additionally, we can compare these $127$ qubit results against $27$ qubit circuit simulations, which can be brute-force computed very easily. Figure~\ref{fig:27_qubit_circuit_magnetization} shows the mean lattice magnetization for a $27$ qubit heavy-hex lattice, computed using classical circuit simulation. The $27$ qubit circuit simulations of Figure~\ref{fig:27_qubit_circuit_magnetization} show good agreement with the $127$ qubit quantum annealing magnetization curves of Figures~\ref{fig:equivalent_magnetization_on_DWave}, \ref{fig:equivalent_magnetization_on_DWave_additional_plots}, especially for a high number of Trotter steps.

In terms of computation time, the total QPU time (measured as \emph{QPU Access Time} to generate all of the magnetization data in Figure~\ref{fig:equivalent_magnetization_on_DWave} for $N=10,000$ time steps was $6594.35$ seconds (note that this includes the server-side spin reversal transforms for the h-gain state encoding data, which costs significantly more computation time compared to having no server-side spin reversal transforms). This is significantly faster than the digital gate model Trotterized circuit simulations performed using ZNE quantum error mitigation \cite{kim2023evidence}, and performed for a larger number of time steps than on all of the IBM Quantum processors. 

Figure~\ref{fig:20_Trotter_steps_equivalent_magnetization_on_DWave_qubit_62_reduced_intersample_correlation} reports single site magnetization results executed on D-Wave quantum annealing hardware, but now using the \texttt{reduce\_intersample\_correlation} flag, which is intended to reduce the amount of correlations in each readout batch (which can be caused by spin bath polarization), along with an order of magnitude more shots per setting resulting in diminished shot noise. Figure~\ref{fig:20_Trotter_steps_equivalent_magnetization_on_DWave_qubit_62_reduced_intersample_correlation} plots a subset of the results for settings that gave lowest RMSE with respect to existing classical simulation data. Figure~\ref{fig:20_Trotter_steps_equivalent_magnetization_on_DWave_qubit_62_reduced_intersample_correlation_all} shows a full parameter sweep of varying J values across the two D-Wave quantum processors - and Table~\ref{table:20_trotter_step_RMSE_reduced_error} show the exact RMSE quantities for all of these settings. Figure~\ref{fig:20_Trotter_steps_equivalent_magnetization_on_DWave_qubit_62_reduced_intersample_correlation_all}, like Figure~\ref{fig:20_Trotter_steps_equivalent_magnetization_on_DWave_qubit_62} shows that as J becomes too small we approach the hardware precision limit and no longer get a meaningful signal as $\theta_h$ increases, and instead the magnetization converges to nearly $1$. These values, (for example, $J=-0.0001$) shows very clearly where the precision limit is for these problems.

Figure~\ref{fig:IBM_Quantum_hardware_circuit_results} in Appendix \ref{section:appendix_IBM_Quantum_Hardware_simulations} plot IBM Quantum processor mean magnetization for Trotter steps of $4$, $5$, $20$, $50$, $100$, and $200$, that were computed using no error mitigation and no error suppression, as a direct comparison against the (whole lattice) mean magnetization data that was computed using quantum annealing in Figure~\ref{fig:equivalent_magnetization_on_DWave} and Figure~\ref{fig:equivalent_magnetization_on_DWave_additional_plots}. Figure~\ref{fig:IBM_Quantum_hardware_circuit_results} shows that the at the high Trotter steps, the computation degrades significantly due to noise, whereas the equivalent simulations at high $N$ using quantum annealing in Figure~\ref{fig:equivalent_magnetization_on_DWave} do not. 

Although in ref. \cite{kim2023evidence}, ZNE was able to extend the computational reach of the gate model quantum processor, ZNE and quantum error mitigation in general can not arbitrarily extend the computational capability of a noisy quantum processor \cite{pelofske2023increasing, quek2023exponentially, Takagi_2022} -- the primary limiting factor is the error rate of the hardware, followed by the in-general exponential scaling cost of quantum error mitigation, as a function of circuit size and depth, for obtaining measurements of observables for a fixed accuracy. In particular, for sufficiently high noise in the computation, as shown in Figure~\ref{fig:IBM_Quantum_hardware_circuit_results} (for example at $N=200$ Trotter steps), ZNE would not be able to recover meaningful signal.

\begin{figure*}[ht!]
    \centering
    \includegraphics[width=0.49\textwidth]{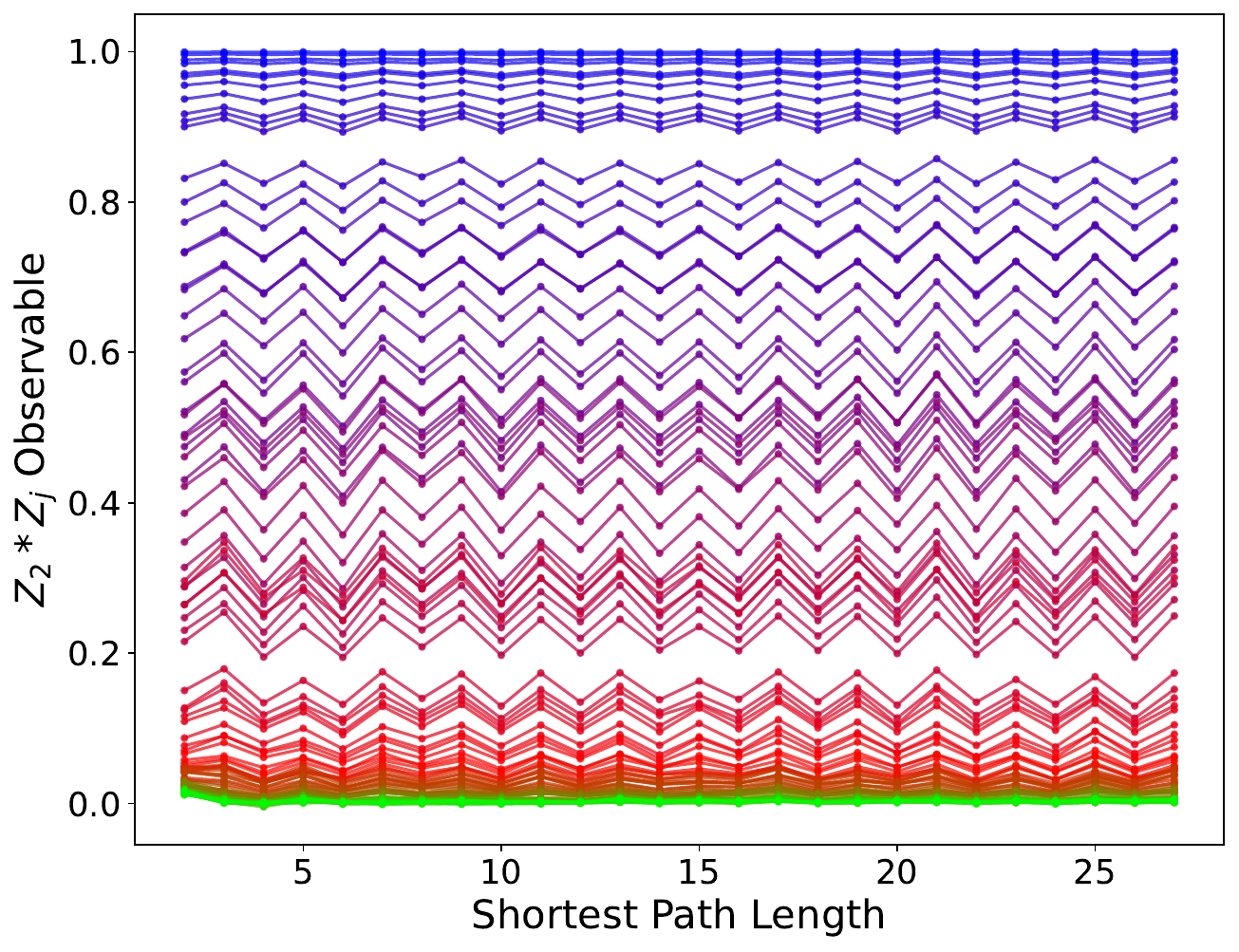}
    \includegraphics[width=0.49\textwidth]{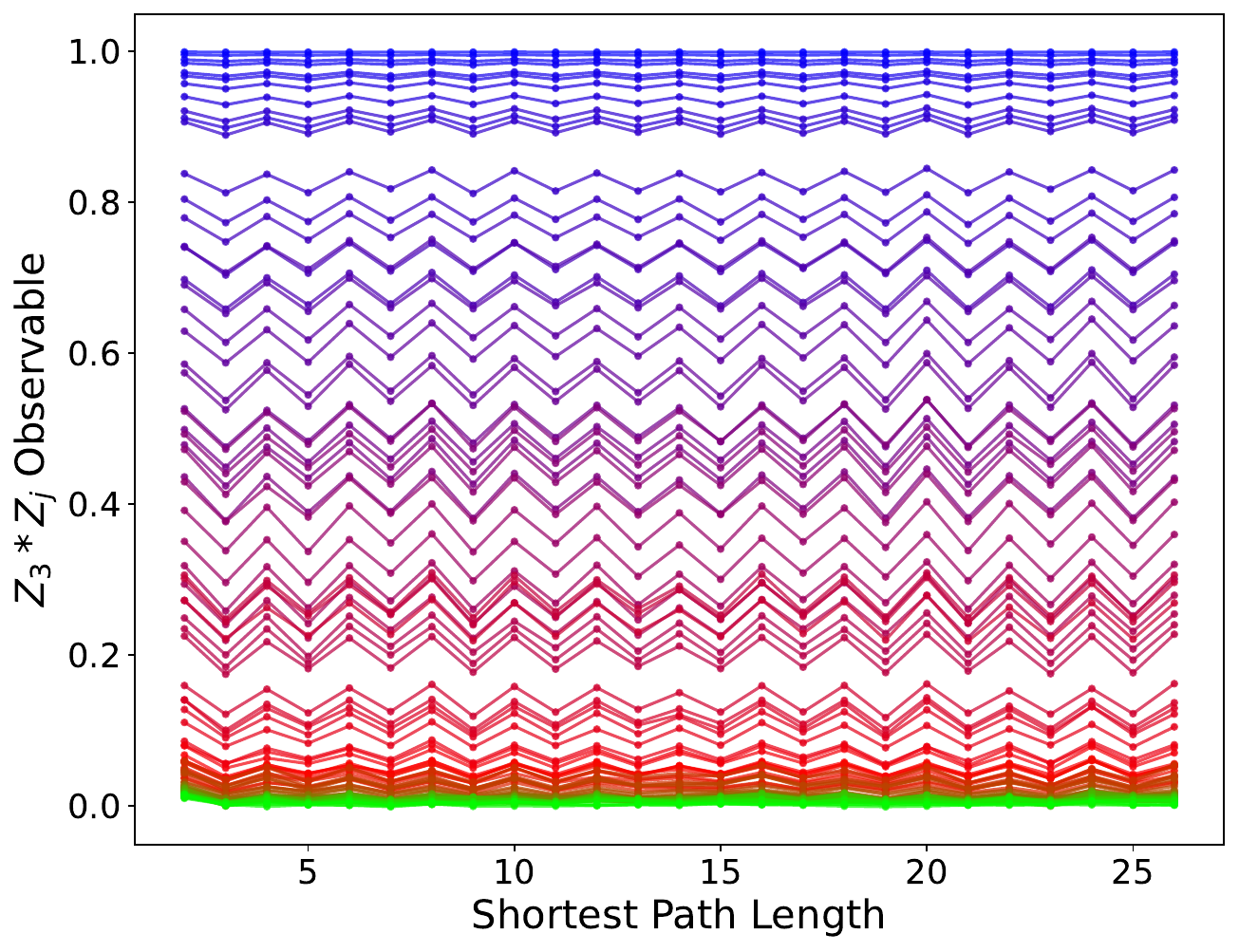}
    \includegraphics[width=0.49\textwidth]{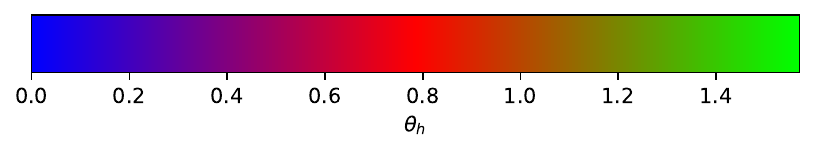}
    \vspace{-0.4cm}
    \caption{Spin-spin higher order observable results from the reverse quantum annealing runs for $N=20$, $J=-0.0027$ on \texttt{Advantage\_system4.1} with reduced intersample correlation (computed from a subset of the mean lattice magnetization data shown in Figure \ref{fig:20_Trotter_steps_equivalent_magnetization_on_DWave_qubit_62_reduced_intersample_correlation}). The two sub-plots show the $Z_i Z_j$ (averaged) observable on the y-axis, where $i$ is fixed to qubit $2$ (left-plot), and qubit $3$ (right-plot). Each plot contains $100$ horizontal lines, corresponding to the $100$ different $\theta_h$ intervals which are color coded by the colormap shown below the plots. The x-axis encodes the (shortest) path length from the qubit $i$ to the qubit $j$ - each spin-spin observable datapoint on the y-axis is averaged over all spin-spin products for the all qubits $j$ that are that distance away from $i$. The aim of these plots is to show what higher-order observable correlations, if any, exist between spins that are spatially disconnected in the lattice. }
    \label{fig:spin_spin_correlation_function_observable}
\end{figure*}

Figure~\ref{fig:spin_spin_correlation_function_observable} plots some example higher order observables from the $N=20$ stepsize, $J=-0.0027$ simulations on \texttt{Advantage\_system4.1}. These plots are showing spin-spin observables $Z_i \cdot Z_j$. The goal is to identify what correlations, if any, exist between spins that are spatially disconnected in the hardware lattice. This is accomplished by first fixing one of the spins, $i$ to be a specific qubit. Then, finding the shortest paths from that qubit $i$ to all other $126$ qubits in the lattice, using the Dijkstra shortest path algorithm. Then, we compute the product between the spin $i$ and the spin $j$ ($j$ may be a neighbor of $i$, or somewhere far away in the lattice) for each measured sample. We take the sum over these weight-2 $Z$ observables across all samples and embeddings, and plot the measures as a function of the shortest path length between $i$ and $j$. Figure \ref{fig:spin_spin_correlation_function_observable} shows two figures; one for $i=2$ and one for $i=3$. When $\theta_h$ is near zero, similar to the bulk magnetization measure, $Z_i Z_j$ is near to $1$. And, when $\theta_h$ is near to $\frac{\pi}{2}$, $Z_i Z_j$ converges to $0$, also similar to the mean magnetization measure. In between $\theta_h$ being $0$ and $\frac{\pi}{2}$, there are clearly defined periodic oscillations. These oscillations correspond to whether the $j$ qubit has degree 2 or 3 in the heavy-hex graph; and whether $i$ is degree 2 or degree 3 determines the orientation of these oscillations, as seen by the two sub-plots of Figure \ref{fig:spin_spin_correlation_function_observable}. This type of dependence on the degree of variables in the Ising model has been seen before, for different systems, in the context of quantum annealing \cite{PhysRevResearch.5.013224}. Lastly, Figure~\ref{fig:spin_spin_correlation_function_observable} also shows that this spin-spin observable, although it fluctuates, on average stays approximately the same as the path length increases - there is not a clear slope of increasing or decreasing as a function of the path length. 

\begin{figure*}[ht!]
    \centering
    \includegraphics[width=0.24\textwidth]{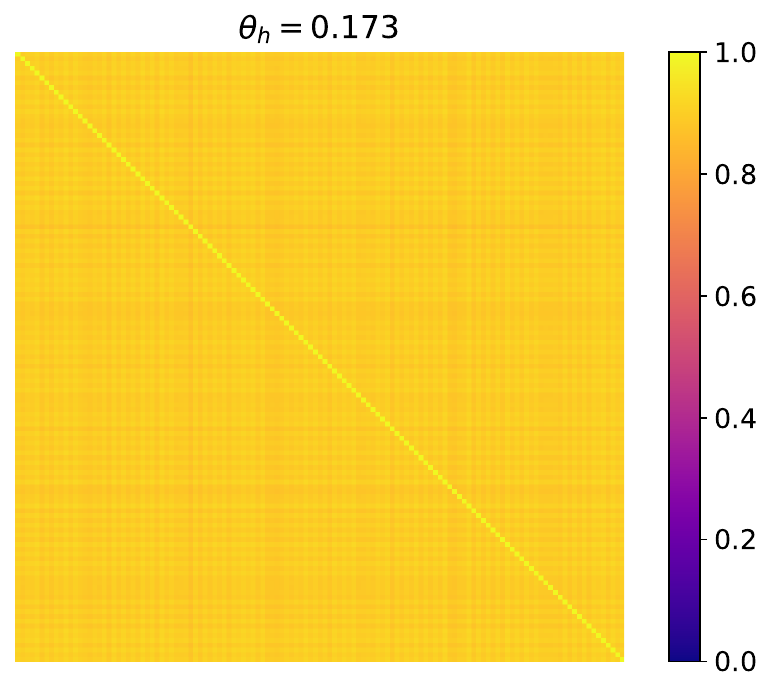}
    \includegraphics[width=0.24\textwidth]{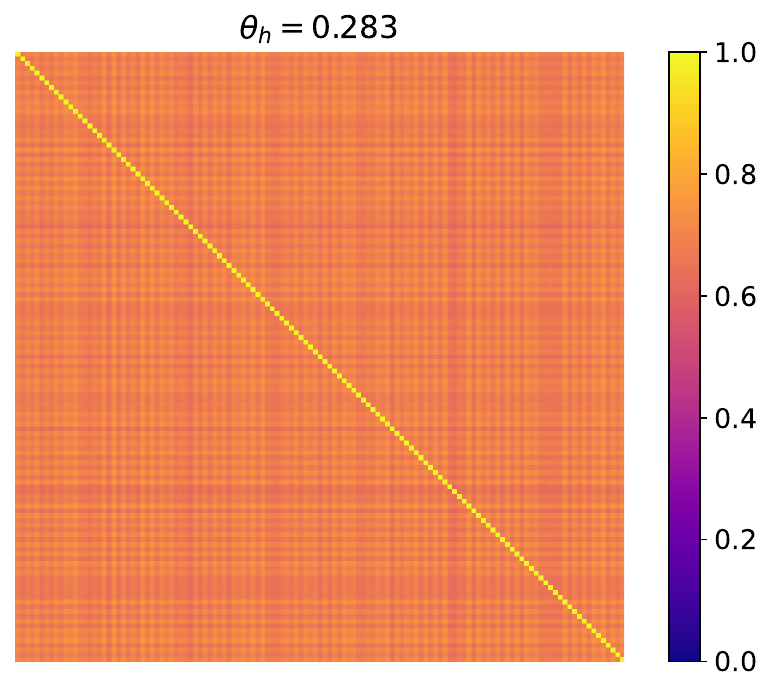}
    \includegraphics[width=0.24\textwidth]{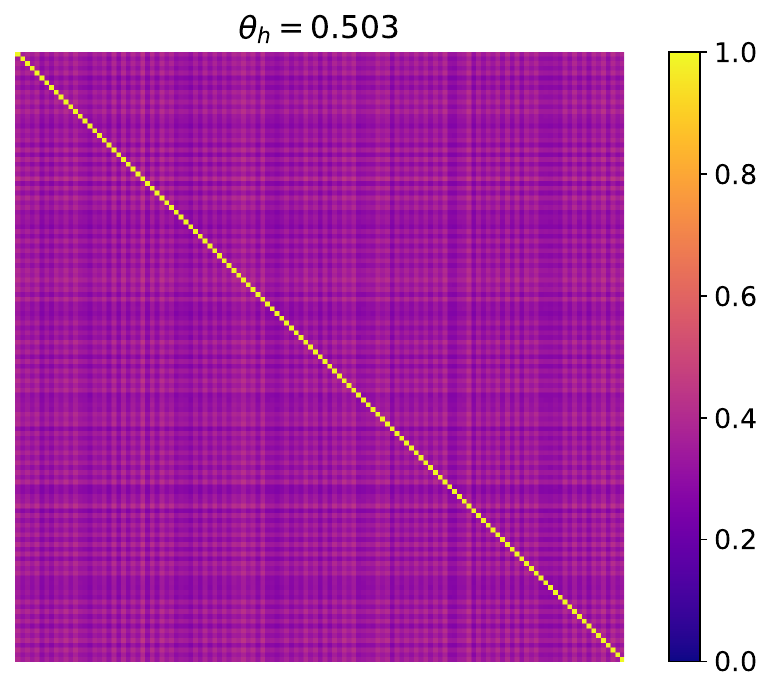}
    \includegraphics[width=0.24\textwidth]{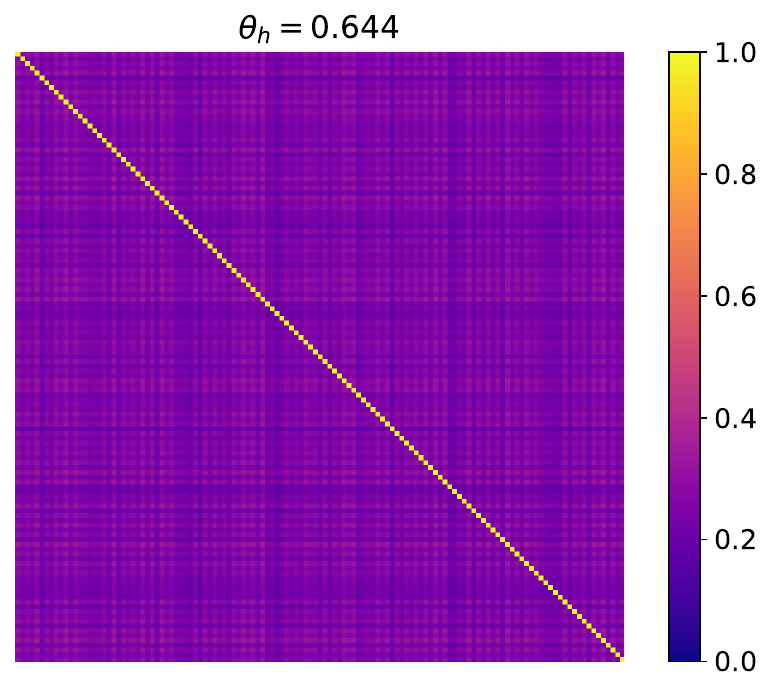}
    \caption{$Z_i Z_j$ higher order observable correlation matrix results from the reverse quantum annealing runs for $N=20$, $J=-0.0027$ on \texttt{Advantage\_system4.1} with reduced intersample correlation (computed from a subset of the mean lattice magnetization data shown in Figure \ref{fig:20_Trotter_steps_equivalent_magnetization_on_DWave_qubit_62_reduced_intersample_correlation}) for four different $\theta_h$ values. These matrices are $127$ rows by $127$ columns and are symmetric along the diagonal and the diagonal values necessarily always have a value of $1$ - the off diagonal values show spin-spin correlations for spins that are spatially separated in the lattice. }
    \label{fig:spin_spin_correlation_function_observable_matrix}
\end{figure*}

Figure~\ref{fig:spin_spin_correlation_function_observable_matrix} shows the full spin-spin observables $Z_i \cdot Z_j$ matrix for all $127$ spins, for four specific $\theta_h$ values, for the same simulation that Figure~\ref{fig:spin_spin_correlation_function_observable} shows. In these matrices there are trends of rows and columns of spin agreement and disagreement (denoted by lighter or darker shaded coloring). At low $\theta_h$ (Figure~\ref{fig:spin_spin_correlation_function_observable_matrix} left), there is high spin agreement, and then as $\theta_h$ increases the spins disagree more as the magnetization $M_Z$ of the lattice tends toward $0$. 

Both Figure~\ref{fig:spin_spin_correlation_function_observable} and Figure~\ref{fig:spin_spin_correlation_function_observable_matrix}  are showing higher order whole-lattice $Z$-basis observables from the data used to compute the single-site magnetization observable in Figure \ref{fig:20_Trotter_steps_equivalent_magnetization_on_DWave_qubit_62_reduced_intersample_correlation}, since in all of these simulations the states of all of the qubits in the lattice were measured the end of the simulation. 

\begin{figure*}[ht!]
    \centering
    \includegraphics[width=0.32\textwidth]{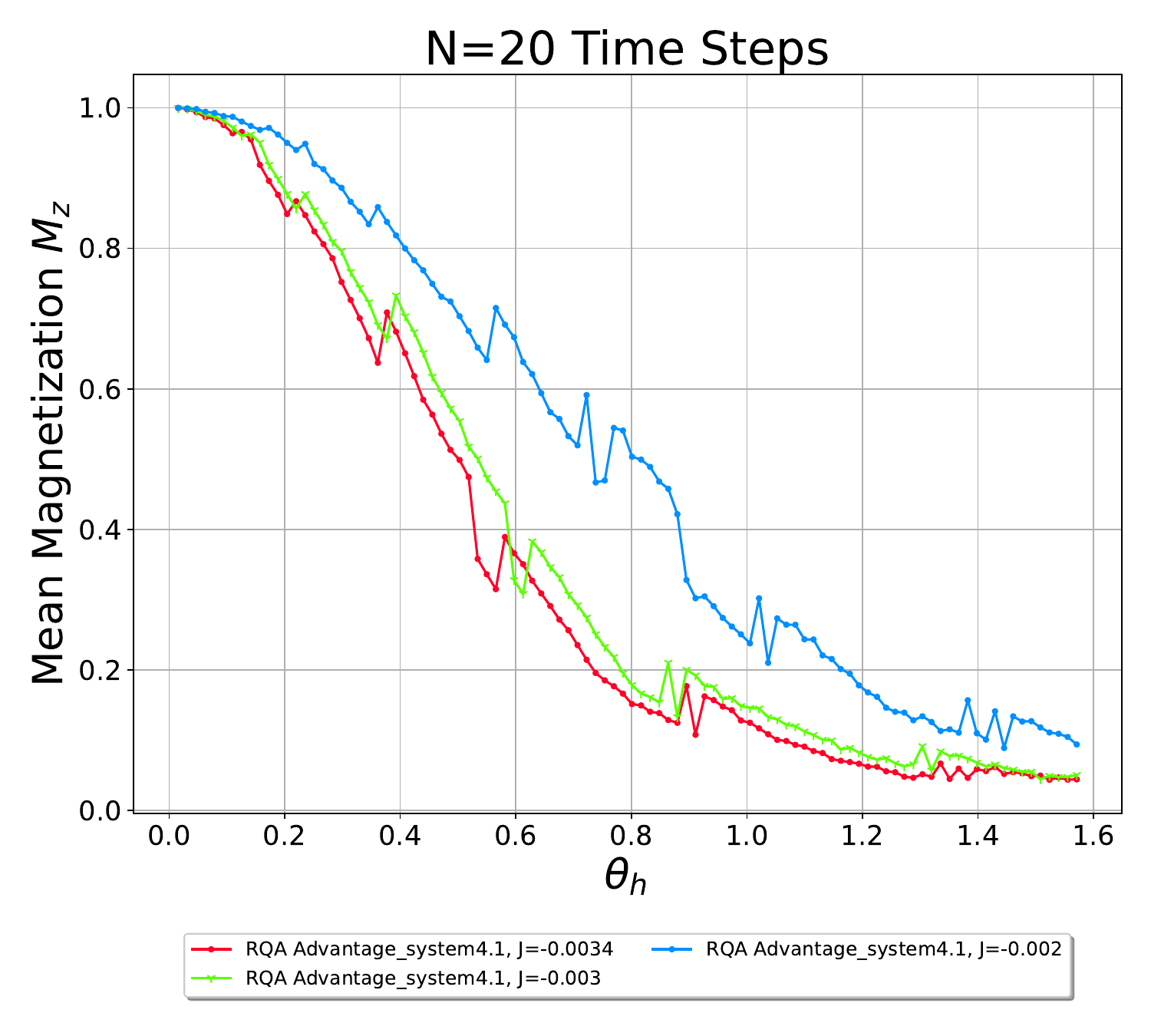}
    \includegraphics[width=0.32\textwidth]{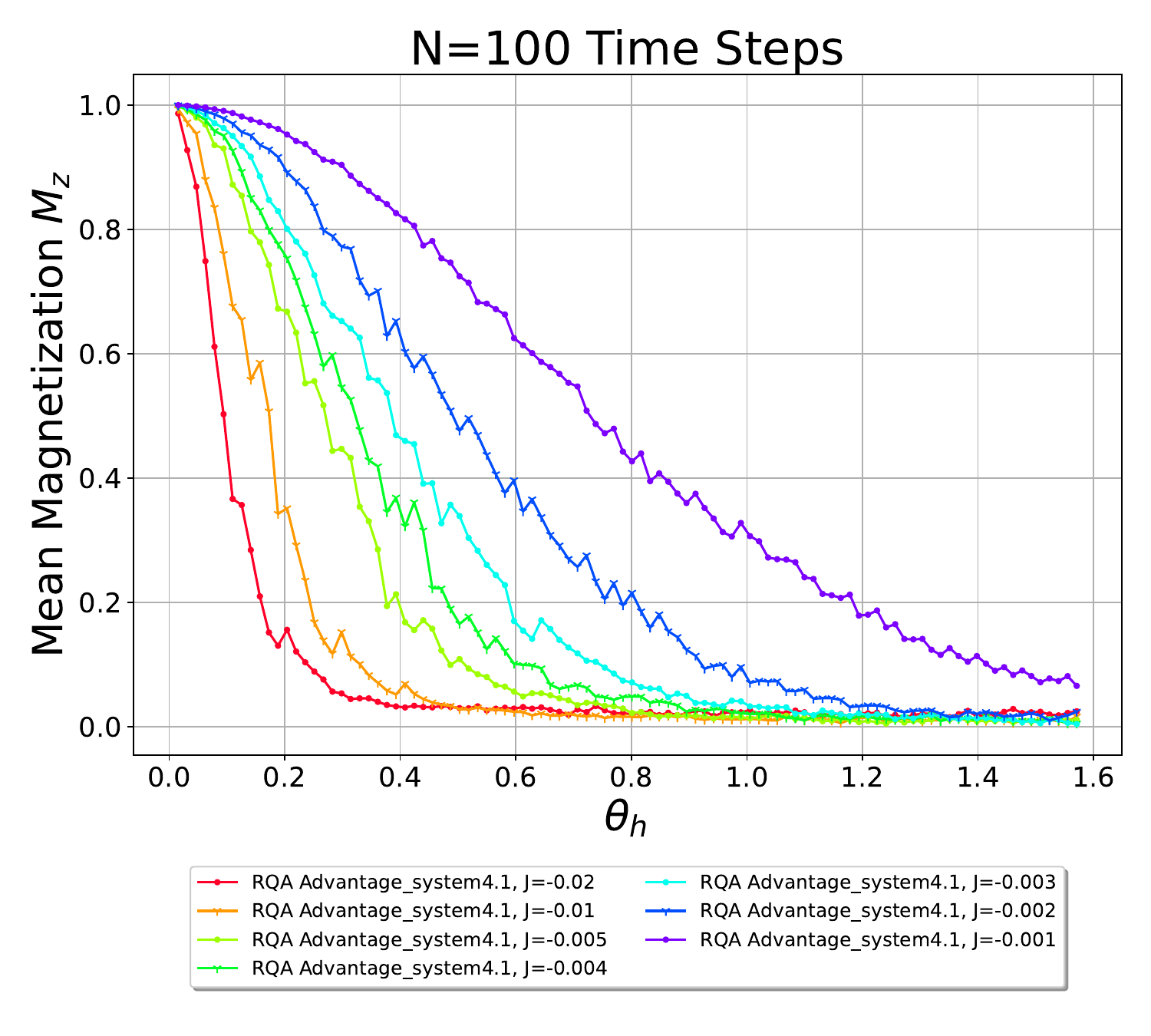}
    \includegraphics[width=0.32\textwidth]{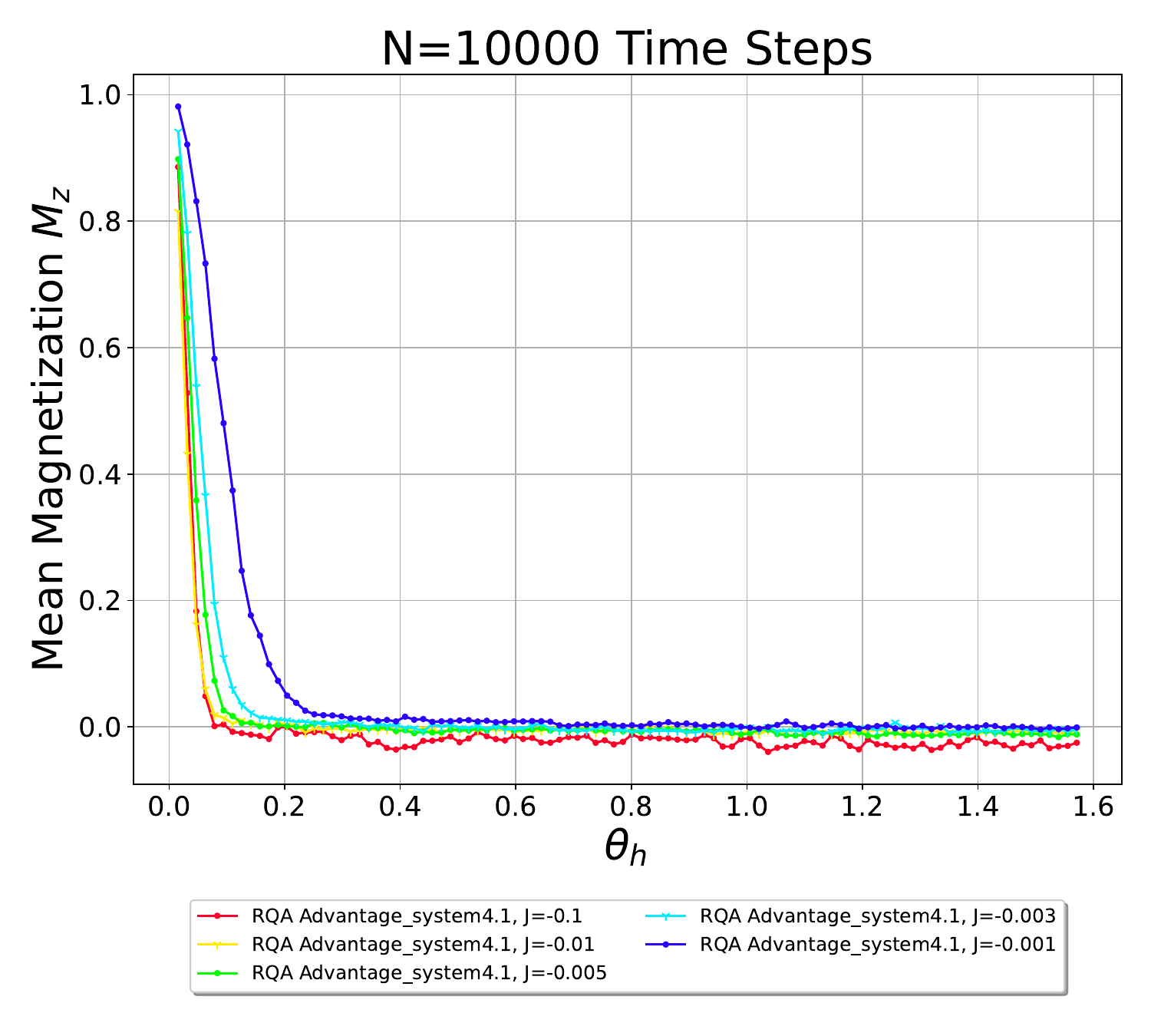}
    \caption{Mean lattice magnetization results as a function of $\theta_h$ for the $384$ node heavy-hex graph, on \texttt{Advantage\_system4.1}, for three different time steps ($N=20, 100, 10000$) with varying ferromagnetic coupling strength. }
    \label{fig:mean_magnetization_384_nodes}
\end{figure*}

Figure~\ref{fig:mean_magnetization_384_nodes} shows mean magnetization results for the $384$ node heavy-hex graph. These results demonstrate that larger heavy-hex systems than $127$ variables can be simulated on the D-Wave hardware. Interestingly, these mean curves results are very similar to the magnetization found from the $127$ variable results (see Figures \ref{fig:equivalent_magnetization_on_DWave} and \ref{fig:equivalent_magnetization_on_DWave_additional_plots}). This suggests that the effect of increasing system size has minimal impact on the mean magnetization dynamics, for this particular transverse field Ising model simulation. This observation of only minimal changes of the observables as the system size increases has been seen in the numerical simulations of refs. \cite{patra2023efficient, tindall2023efficient}.

Appendix~\ref{section:appendix_fixed_AT_s_magnetization} contains D-Wave quantum annealing simulations which show magnetization dynamics results on the same two D-Wave quantum annealers as a function of the anneal fraction $s$, using fixed anneal times. These magnetization result curves are similar to the results presented in this section, and moreover shows very consistent magnetization phase-change curves when the coupling strengths are weak.

%%%%%%%%%%%%%%%%%%%%%%%%%%%%%%%%%%%
%%%%%%%%%%%%%%%%%%%%%%%%%%%%%%%%%%%
%%%%%%%%%%%%%%%%%%%%%%%%%%%%%%%%%%%
\section{Discussion and Conclusion}
\label{section:discussion}

We have demonstrated that current analog quantum computers, specifically programmable D-Wave quantum annealers, can efficiently simulate the Ising model magnetization observable that was measured on the $127$ qubit Eagle IBM Quantum processor \texttt{ibm\_kyiv} with the help of ZNE post processing on Trotterized circuits in Ref.~\cite{kim2023evidence}. We have demonstrated this capability on two D-Wave quantum annealers, using two different Hamiltonian simulation techniques, for equivalent Trotter steps (time steps) of $20, 50, 100$, up to $10,000$. These computations were performed using significantly less QPU time compared to the digital gate model ZNE experimental results \cite{kim2023evidence}. This simulation was only possible because there exists an equivalence between the Trotterized Transverse Field Ising Model circuit simulation and the D-Wave quantum annealer Hamiltonian, and because the heavy-hex graph can be directly embedded onto the Pegasus graph topology of the D-Wave chips. These simulations push the capabilities of the current D-Wave quantum annealers, namely in the available annealing times and the programmable coefficient precision. Importantly, these quantum annealing simulations do contain a variety of sources of error, and there are many ways that the simulation quality could be improved \cite{chern2023tutorial}. We have shown that equivalent Trotterized magnetization dynamics can be efficiently performed using current D-Wave quantum annealers, in particular for simulations that are equivalent to hundreds of Trotter steps, which is outside of the computational capability of current heavy-hex IBM Quantum superconducting qubit quantum computers \cite{pelofske2023short, pelofske2023qavsqaoa, kim2023evidence}, as shown by Figure~\ref{fig:IBM_Quantum_hardware_circuit_results} in Appendix~\ref{section:appendix_IBM_Quantum_Hardware_simulations}. Figure~\ref{fig:IBM_Quantum_hardware_circuit_results} shows magnetization dynamics results, computed on $27$, $127$, and $133$ qubit IBM Quantum processors, using no error mitigation strategies.

Because of the error rate of the Trotter decomposition used in Ref.~\cite{kim2023evidence} compared to ideal Hamiltonian simulation, for a large number of Trotter steps we expect that there is a divergence from the ideal Hamiltonian simulation dynamics. Programmable quantum annealers could be used for these types of Hamiltonian simulation dynamics (at least for a restricted set of observables that do not require change of computational basis of the qubits). However, the aim of our study was to show how the equivalent Trotterized Hamiltonian dynamics could be performed on D-Wave quantum annealers. We leave more extensive Hamiltonian dynamics simulation on these types of problems open to future research (both using quantum annealing, and improved Trotter decompositions executed on digital gate model quantum computers). 

For potential future research on simulations with larger system sizes, magnetization dynamics on significantly larger heavy-hex (or, non-heavy-hex) graphs could be implemented on the current Pegasus hardware graphs of current D-Wave quantum annealers (see Figure~\ref{fig:Pegasus_embedding}). 

The existing efficient classical simulations of these Hamiltonian dynamics circuits \cite{kechedzhi2023effective, tindall2023efficient, begušić2023fast, liao2023simulation, begušić2023fast_2, rudolph2023classical, shao2023simulating, patra2023efficient} have not computed mean lattice magnetization, or single site magnetization, for simulations with a large number of Trotter steps (e.g. greater than several hundred, or thousands of time steps). It seems likely that, given the efficiency of the existing methods, high quality classical simulations of a high number of Trotter steps (e.g. $200$) could be performed. We encourage additional study of simulations of these Trotterized circuits for a large number of Trotter steps so as to better understand the boundary of classical computation for this problem type.

%%%%%%%%%%%%%%%%%%%%%%%%%%%%%%%%%%%
%%%%%%%%%%%%%%%%%%%%%%%%%%%%%%%%%%%
%%%%%%%%%%%%%%%%%%%%%%%%%%%%%%%%%%%
\section{Acknowledgments}
\label{section:acknowledgments}
This work was supported by the U.S. Department of Energy through the Los Alamos National Laboratory. Los Alamos National Laboratory is operated by Triad National Security, LLC, for the National Nuclear Security Administration of U.S. Department of Energy (Contract No. 89233218CNA000001). This research used resources provided by the Los Alamos National Laboratory Institutional Computing Program, which is supported by the U.S. Department of Energy National Nuclear Security Administration under Contract No.~89233218CNA000001. Research presented in this article was supported by the NNSA's Advanced Simulation and Computing Beyond Moore's Law Program at Los Alamos National Laboratory. The research presented in this article was supported by the Laboratory Directed Research and Development program of Los Alamos National Laboratory under project numbers 20210114ER and 20230049DR. LANL report number LA-UR-23-27359. 

We acknowledge the use of IBM Quantum services for this work. The views expressed are those of the authors, and do not reflect the official policy or position of IBM or the IBM Quantum team. 

Thank you to Tameem Albash, Marc Vuffray, and Carleton Coffrin for several helpful discussions on this project. Thanks to the folks at AQC 2023 for interesting discussions on quantum annealing. Thanks to D-Wave Technical Support for help with the quantum annealer system specifications. Thanks to the authors of Refs.~\cite{begušić2023fast_2, rudolph2023classical} for providing the classical simulation data of the single site magnetization. 

The figures in this article were generated using matplotlib \cite{thomas_a_caswell_2021_5194481, Hunter:2007}, networkx \cite{hagberg2008exploring}, and Qiskit \cite{Qiskit} in Python 3.

%%%% REFERENCES
%\clearpage
\setlength\bibitemsep{0pt}
\printbibliography
\clearpage

%%%%%%%%%%%%%%%%%%%%%%%%%%%%%%%%%%%%%%%%%%%%%%%%%%%%%%%%%%%%%%%%%%%%%%
%%%%%%%%%%%%%%%%%%%%%%%%%%%%%%%%%%%%%%%%%%%%%%%%%%%%%%%%%%%%%%%%%%%%%%
%%%%%%%%%%%%%%%%%%%%%%%%%%%%%%%%%%%%%%%%%%%%%%%%%%%%%%%%%%%%%%%%%%%%%%
\appendix
\section{Classical $27$ Qubit Heavy-hex Circuit Magnetization Simulation}
\label{section:appendix_27_qubit_simulations}

\begin{figure*}[ht!]
    \centering
    \includegraphics[width=0.7\textwidth]{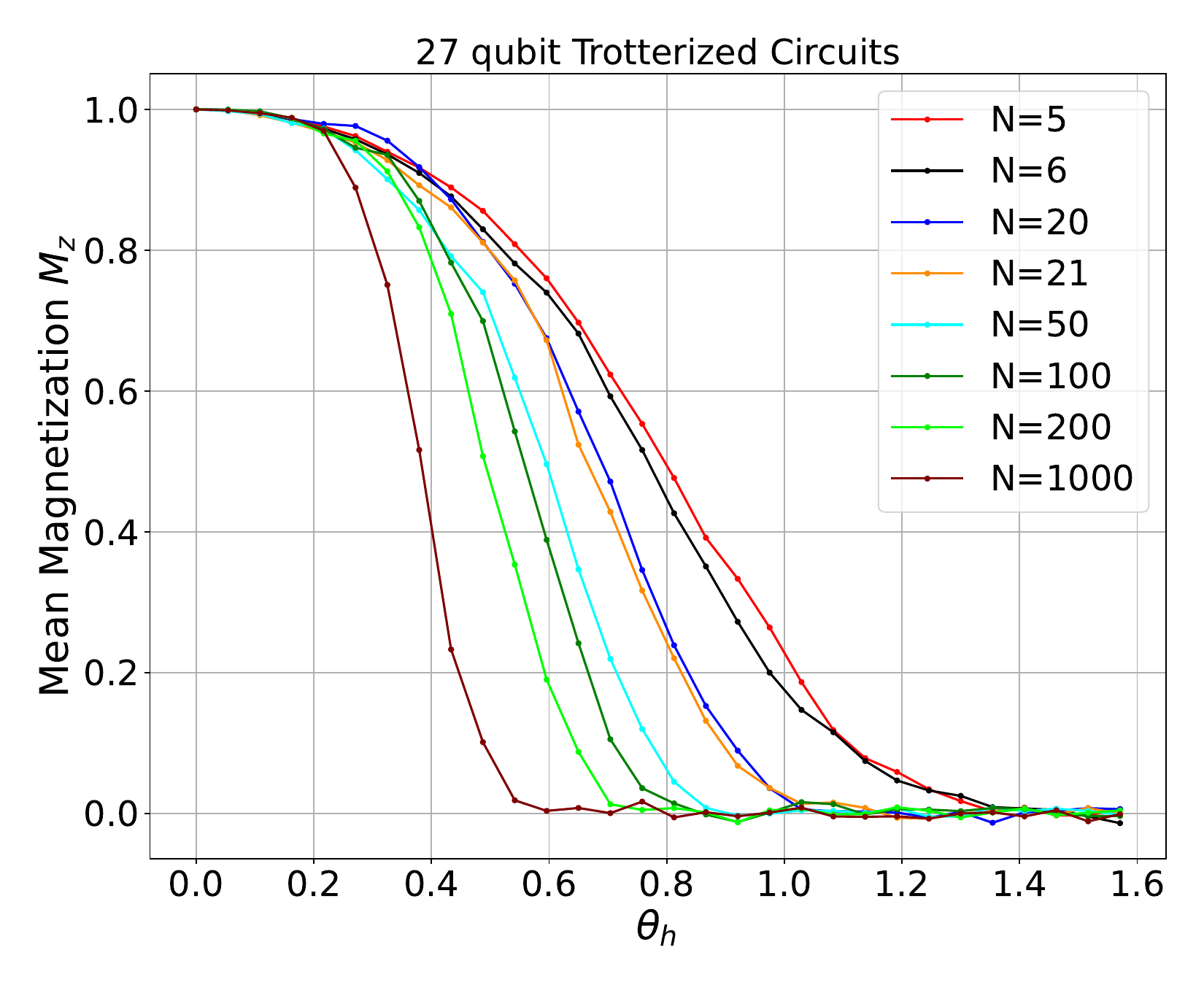}
    \caption{Exact classical circuit simulation mean magnetization for $27$ qubit heavy-hex Trotterized circuits, of the same form as Ref.~\cite{kim2023evidence}, for an increasing number of Trotter steps $N$. $30$ linearly spaced Rx rotation angles of $\theta_h \in [0, \frac{\pi}{2}]$ are simulated for each $N$. $1000$ shots are measured for each parameter combination. }
    \label{fig:27_qubit_circuit_magnetization}
\end{figure*}

\begin{figure*}[ht!]
    \centering
    \includegraphics[width=0.7\textwidth]{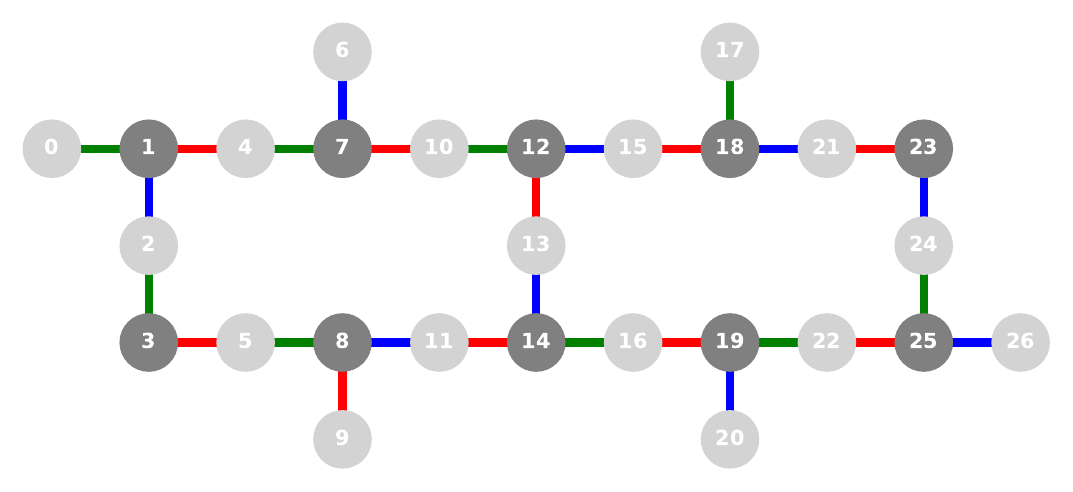}
    \caption{$27$ qubit heavy-hex hardware connectivity. Bi-partition is denoted by dark and light grey qubit coloring. 3-edge-coloring shown by red, blue, and green edges. }
    \label{fig:27_qubit_heavy_hex_edge_coloring}
\end{figure*}

Figure~\ref{fig:27_qubit_circuit_magnetization} plots mean Magnetization for a number of different Trotter steps $N$ as a function of $\theta_h$. Circuits constructed using the simple 3-edge coloring that is possible on heavy-hex graphs \cite{kim2023evidence, pelofske2023qavsqaoa, pelofske2023short}. The heavy-hex hardware connectivity used is shown in Figure~\ref{fig:27_qubit_heavy_hex_edge_coloring}, including the heavy-hex bi-partition and 3-edge-coloring. Each $\theta_h$ steps is simulated using classical simulation in Qiskit \cite{Qiskit}, with $1000$ shots. The $\{0, 1\}$ qubit state measurements are mapped to spins via $1 \rightarrow -1$ and $0 \rightarrow +1$.

%%%%%%%%%%%%%%%%%%%%%%%%%%%%%%%%%%%%%%%%%%%%%%%%%%%%%%%%%%%%%%%%%%%%%%
%%%%%%%%%%%%%%%%%%%%%%%%%%%%%%%%%%%%%%%%%%%%%%%%%%%%%%%%%%%%%%%%%%%%%%
%%%%%%%%%%%%%%%%%%%%%%%%%%%%%%%%%%%%%%%%%%%%%%%%%%%%%%%%%%%%%%%%%%%%%%
\section{Trotterized Circuit Results on IBM Quantum Processors}
\label{section:appendix_IBM_Quantum_Hardware_simulations}

\begin{figure*}[ht!]
    \centering
    \includegraphics[width=0.49\textwidth]{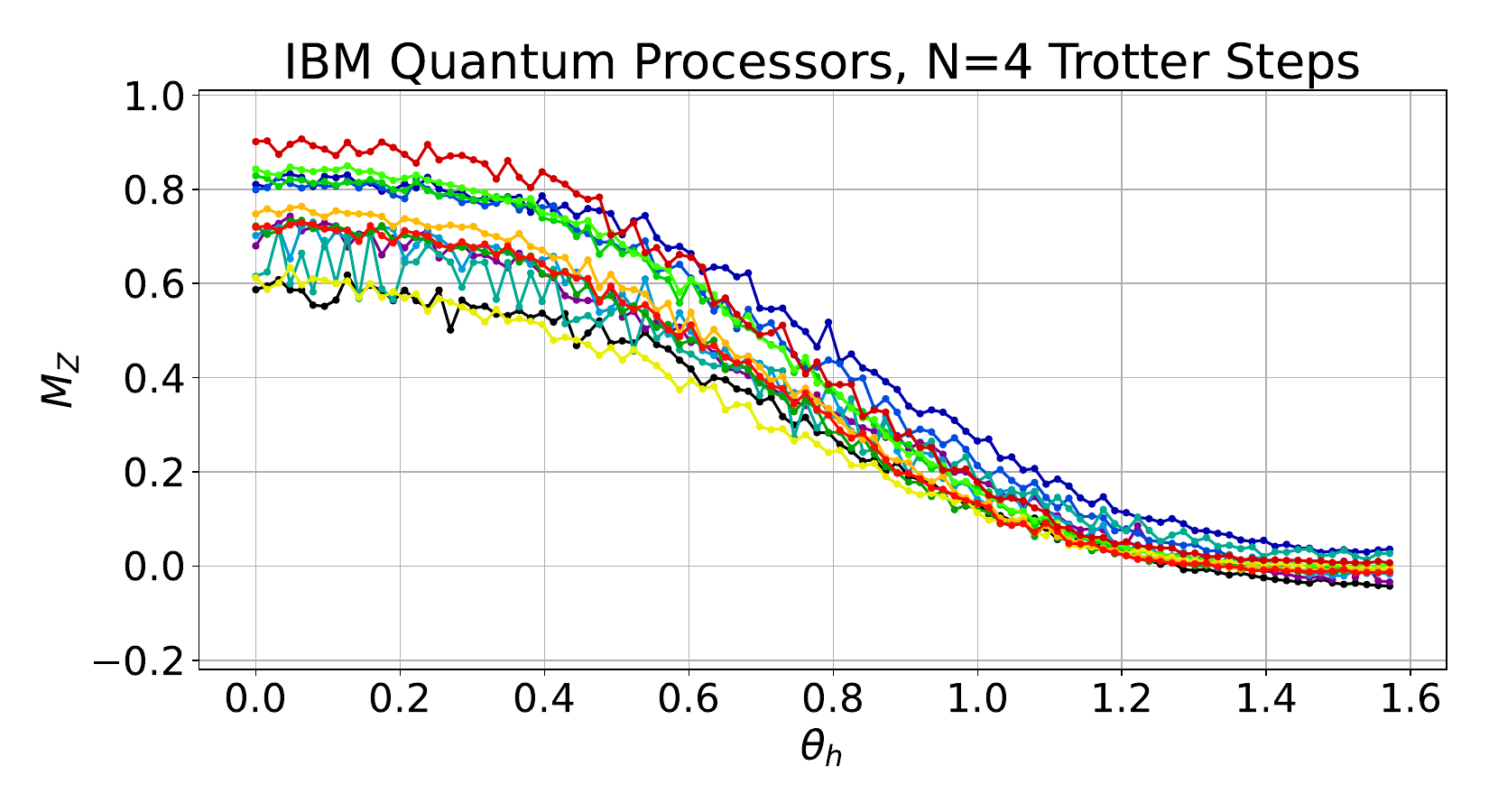}
    \includegraphics[width=0.49\textwidth]{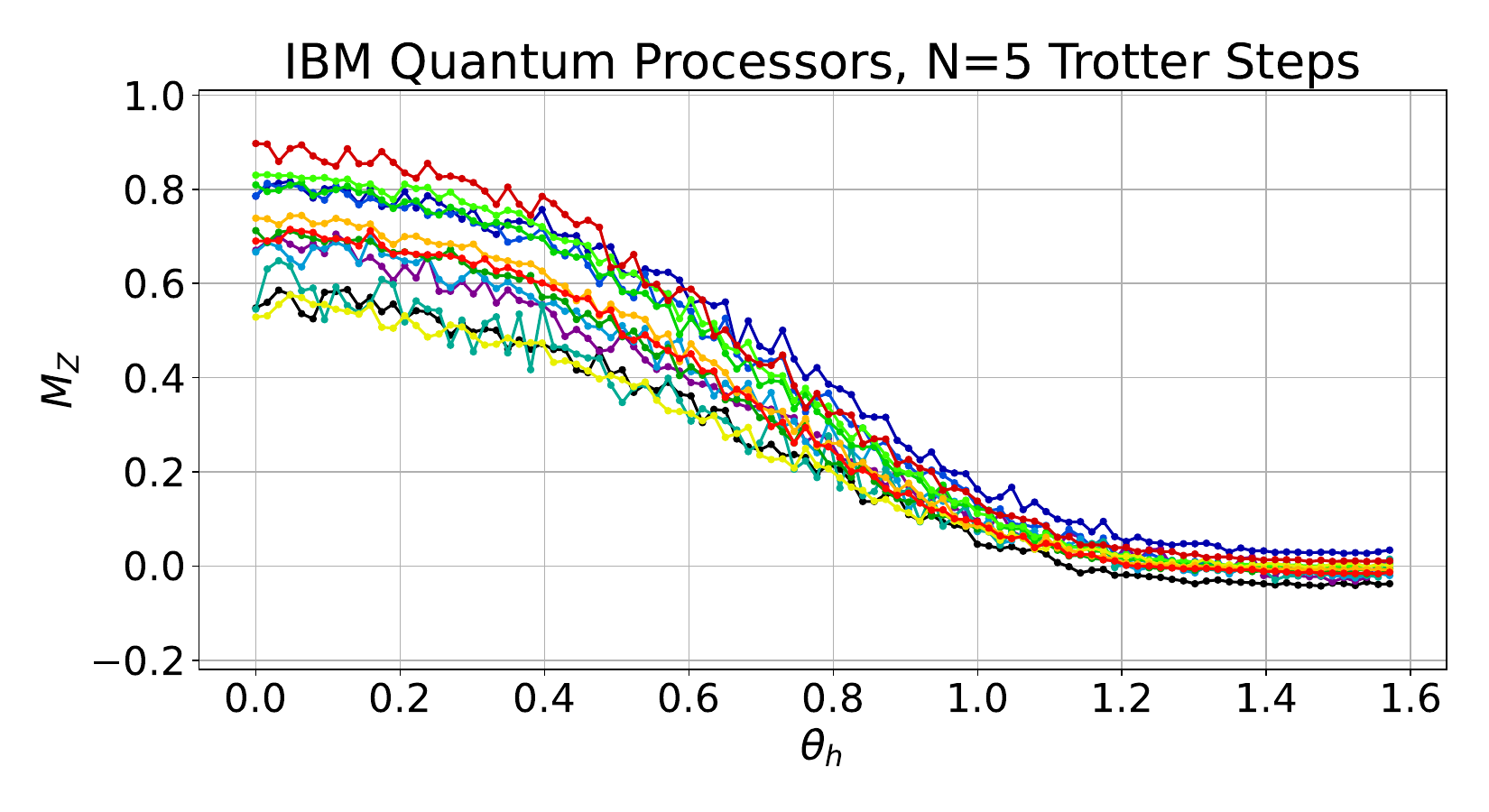}
    \includegraphics[width=0.49\textwidth]{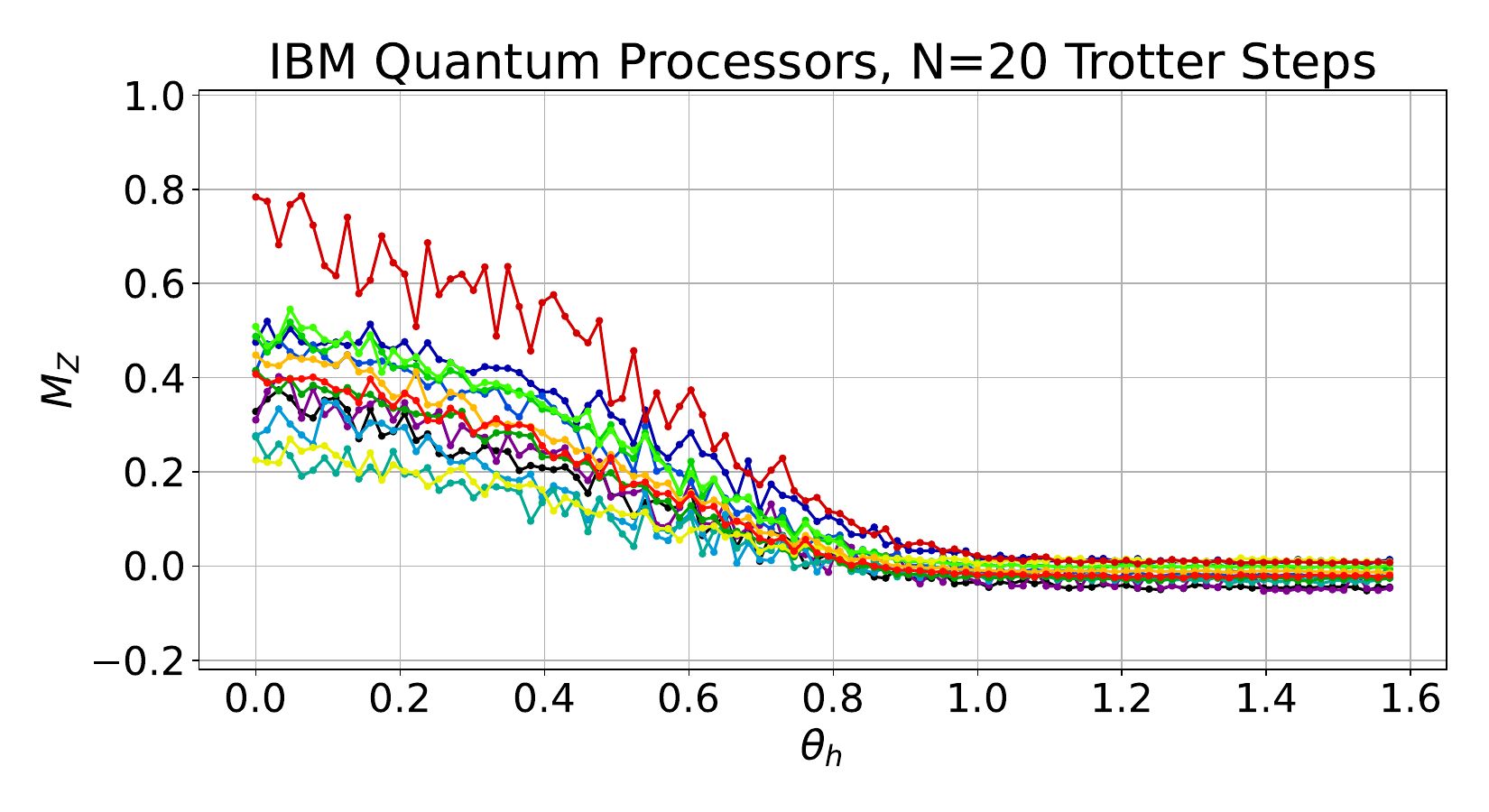}
    \includegraphics[width=0.49\textwidth]{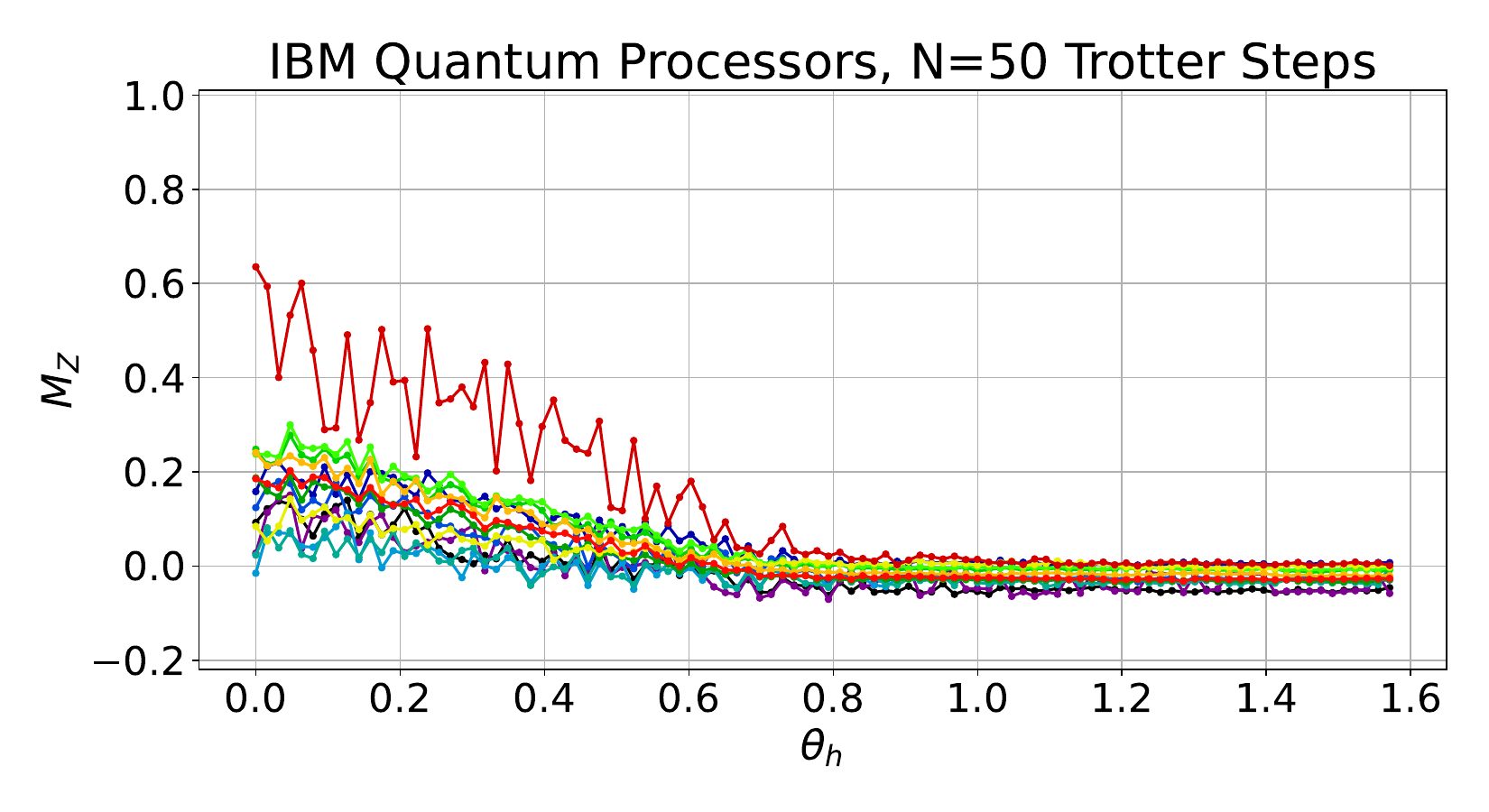}
    \includegraphics[width=0.49\textwidth]{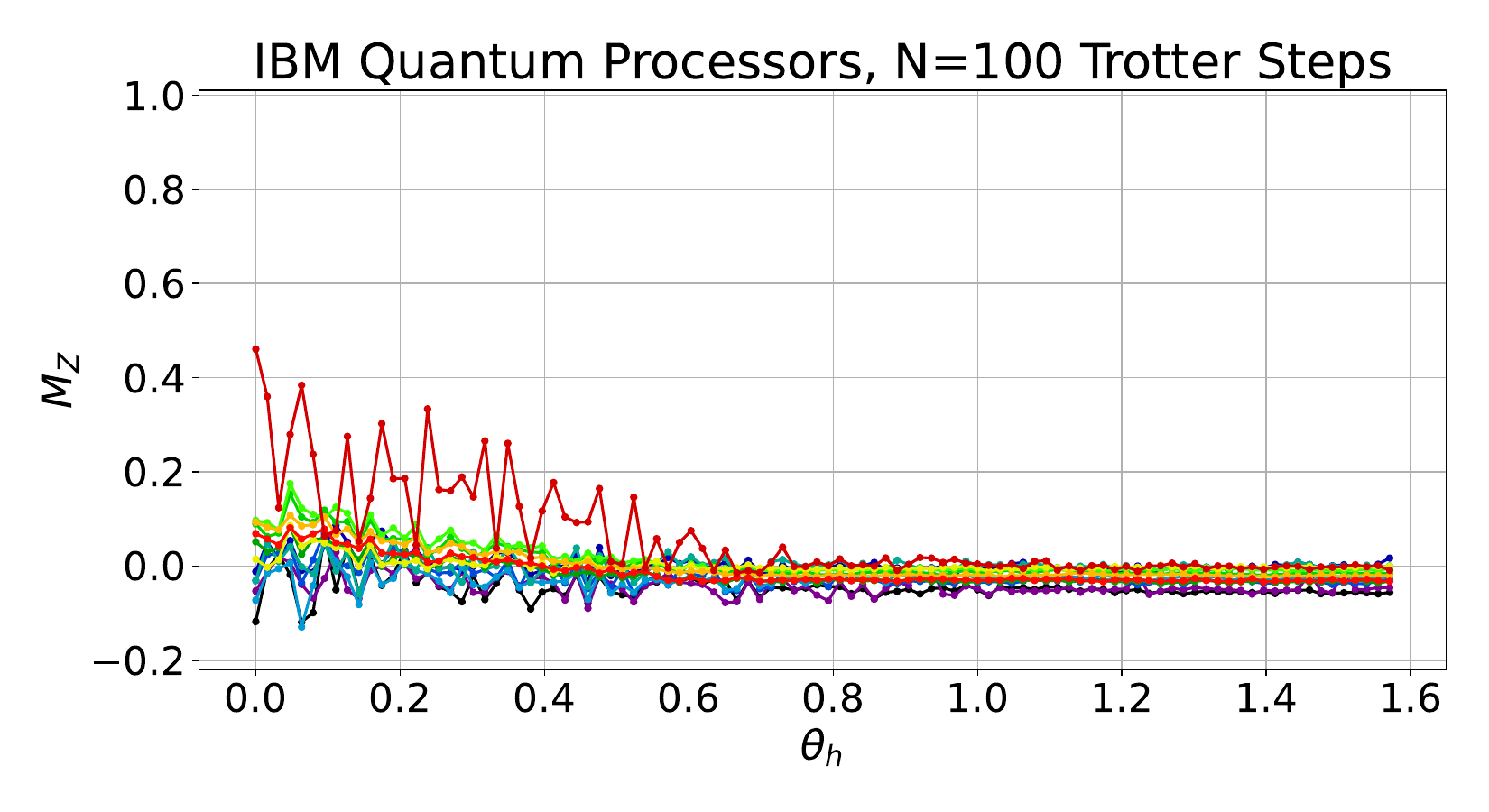}
    \includegraphics[width=0.49\textwidth]{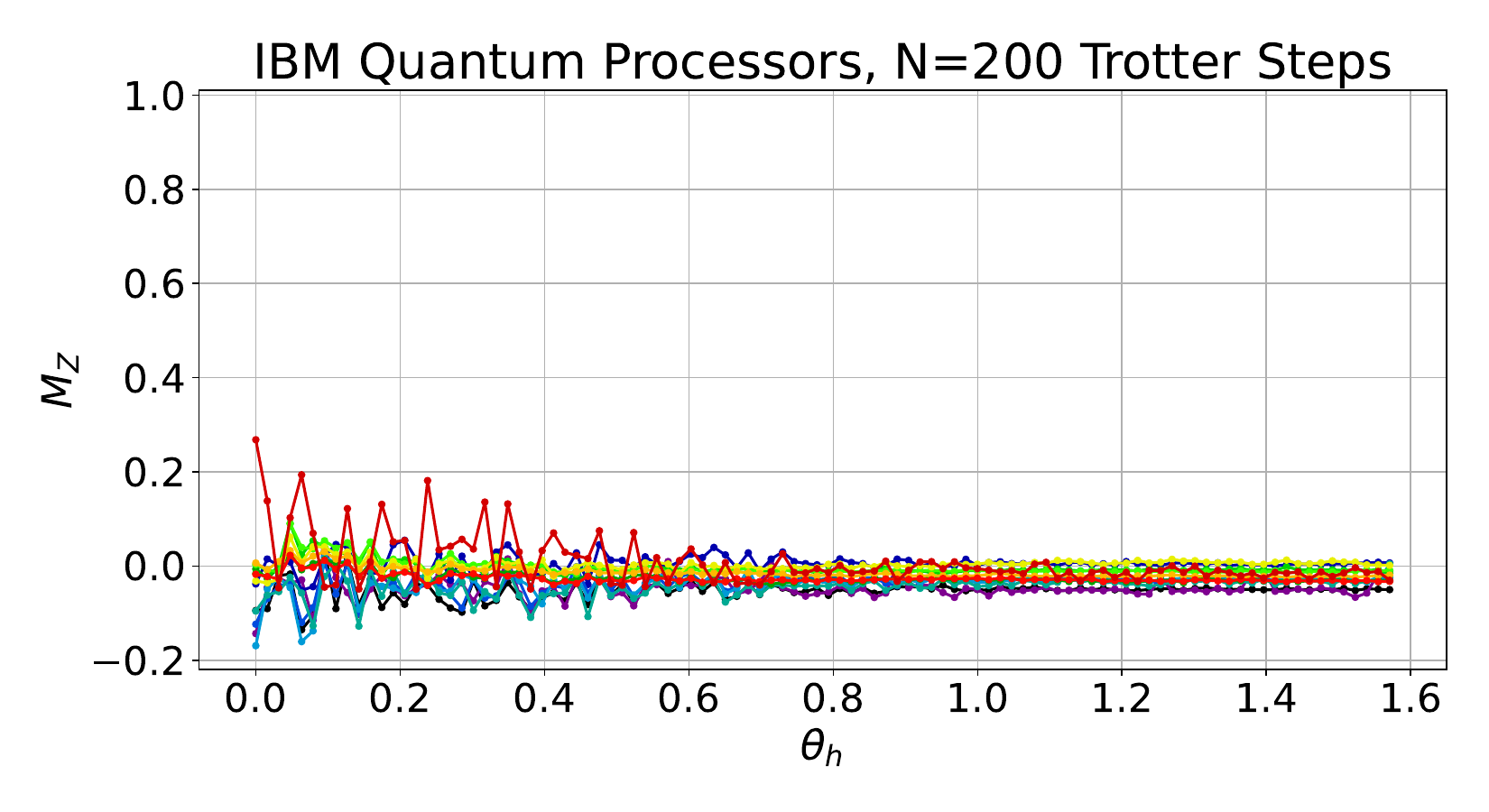}
    \includegraphics[width=0.40\textwidth]{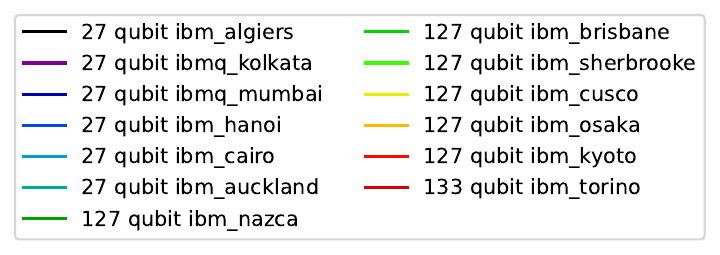}
    \caption{Whole-chip $27$, $127$, $133$ qubit heavy-hex IBM Quantum processor Trotterized circuit magnetization results for $4, 5, 20, 50, 100, 200$ Trotter steps reported as average lattice magnetization $M_Z$ as a function of $\theta_h$. No error mitigation or error suppression techniques were used. Each $\theta_h$ point is the mean lattice magnetization averaged over $10,000$ shots.  }
    \label{fig:IBM_Quantum_hardware_circuit_results}
\end{figure*}

\begin{figure*}[ht!]
    \centering
    \includegraphics[width=0.95\textwidth]{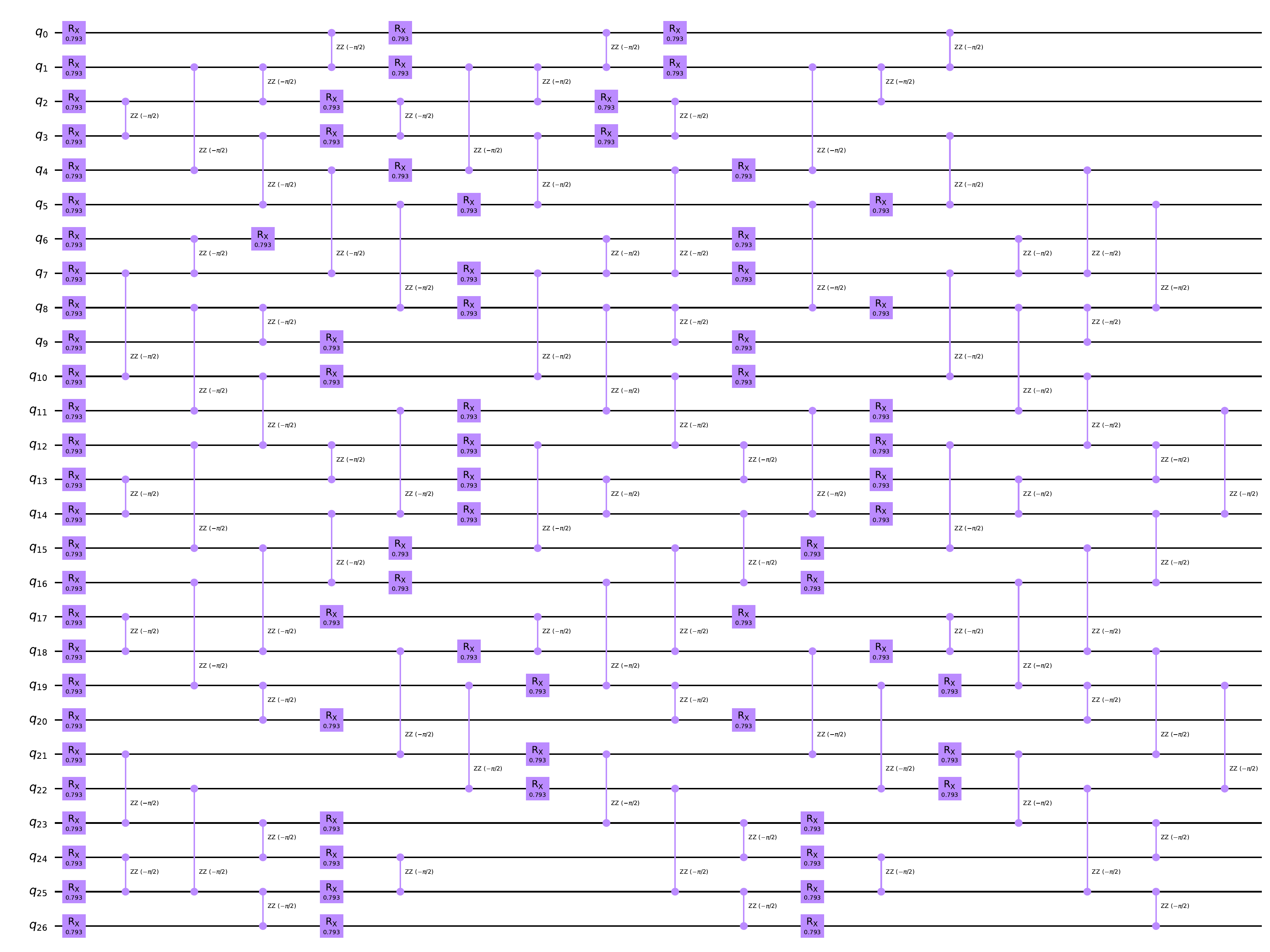}
    \caption{Compiled circuit onto a $27$ qubit heavy-hex lattice for $3$ Trotter steps with a fixed RZZ angle and a fixed RX angle. }
    \label{fig:27_qubit_trotterized_circuit}
\end{figure*}

Figure~\ref{fig:IBM_Quantum_hardware_circuit_results} show mean lattice magnetization Trotterized circuit simulations derived from the methods in ref.~\cite{kim2023evidence}, but with no error mitigation or error suppression techniques, on several $27$, $127$, and $133$ qubit IBM Quantum superconducting processors, for varying $\theta_h$ angles, for Trotter steps of $4$ up to $200$. The circuits were executed on the IBM quantum processors using Qiskit \cite{Qiskit} to adapt the circuits to the required hardware quantum gateset. All of the superconducting qubit processors shown in these plots have heavy-hex hardware graphs, more processor details are summarized in Table~\ref{table:IBM_processor_summary_table}. Figure~\ref{fig:27_qubit_trotterized_circuit} shows a circuit rendering of one of these Trotterized circuits, operating on a $27$ qubit heavy-hex grid for exactly $3$ Trotter steps (measurements on the qubits are not shown, but the states of all $27$ qubits are measured at the end of the circuit).

\begin{table*}[h!]
\centering
\begin{tabular}{|c | c | c | c| c |} 
 \hline
 QPU Name & Processor Type & Hardware Gateset & Number of Qubits & Number of two qubit gates \\ 
 \hline
 \hline
 \texttt{ibm\_algiers} & Falcon r5.11 & CX, ID, RZ, SX, X & 27 & 28 \\ 
 \hline
 \texttt{ibmq\_kolkata} & Falcon r5.11 & CX, ID, RZ, SX, X & 27 & 28 \\ 
 \hline
 \texttt{ibmq\_mumbai} & Falcon r5.10 & CX, ID, RZ, SX, X & 27 & 28 \\ 
 \hline
 \texttt{ibm\_hanoi} & Falcon r5.11 & CX, ID, RZ, SX, X & 27 & 28 \\ 
 \hline
 \texttt{ibm\_cairo} & Falcon r5.11 & CX, ID, RZ, SX, X & 27 & 28 \\ 
 \hline
 \texttt{ibm\_auckland} & Falcon r5.11 & CX, ID, RZ, SX, X & 27 & 28 \\ 
 \hline
 \texttt{ibm\_nazca} & Eagle r3 & ECR, ID, RZ, SX, X & 127 & 144 \\ 
 \hline
 \texttt{ibm\_sherbrooke} & Eagle r3 & ECR, ID, RZ, SX, X & 127 & 144 \\ 
 \hline
 \texttt{ibm\_brisbane} & Eagle r3 & ECR, ID, RZ, SX, X & 127 & 144 \\ 
 \hline
 \texttt{ibm\_cusco} & Eagle r3 & ECR, ID, RZ, SX, X & 127 & 144 \\ 
 \hline
 \texttt{ibm\_osaka} & Eagle r3 & ECR, ID, RZ, SX, X & 127 & 144 \\ 
 \hline
 \texttt{ibm\_kyoto} & Eagle r3 & ECR, ID, RZ, SX, X & 127 & 144 \\ 
 \hline
 \texttt{ibm\_torino} & Heron r1 & CZ, ID, RZ, SX, X & 133 & 150 \\ 
 \hline
\end{tabular}
\caption{IBM superconducting qubit processor summary for the Trotterized circuit simulations used in this study for $4$ up to $200$ Trotter steps in order to compare against the D-Wave quantum annealer simulations (the mean lattice magnetization results are shown in Figure~\ref{fig:IBM_Quantum_hardware_circuit_results}). The number of two qubit gates is equivalent to the number of edges in the hardware graph. CX, ECR~\cite{sheldon2016procedure}, and CZ are the three distinct two qubit (entangling) quantum gate operations supported on these QPUs; RZ, SX, X, and ID are the single qubit gate operations. }
\label{table:IBM_processor_summary_table}
\end{table*}

%%%%%%%%%%%%%%%%%%%%%%%%%%%%%%%%%%%%%%%%%%%%%%%%%%%%%%%%%%%%%%%%%%%%%%
%%%%%%%%%%%%%%%%%%%%%%%%%%%%%%%%%%%%%%%%%%%%%%%%%%%%%%%%%%%%%%%%%%%%%%
%%%%%%%%%%%%%%%%%%%%%%%%%%%%%%%%%%%%%%%%%%%%%%%%%%%%%%%%%%%%%%%%%%%%%%
\section{384 Node Heavy-hex Graph}
\label{section:appendix_384_node_heavy_hex}

\begin{figure*}[h!]
    \centering
    \includegraphics[width=0.49\textwidth]{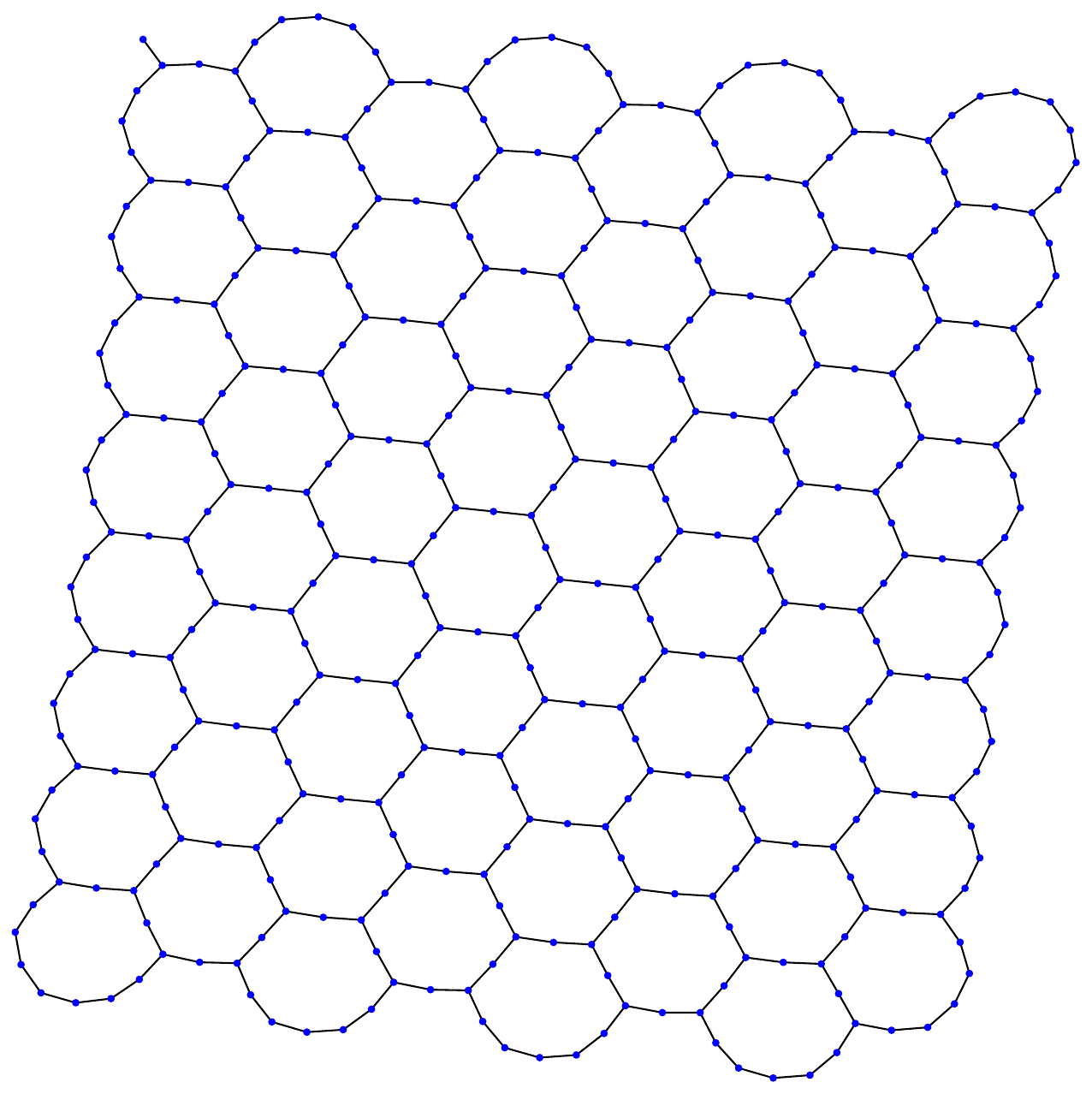}
    \includegraphics[width=0.49\textwidth]{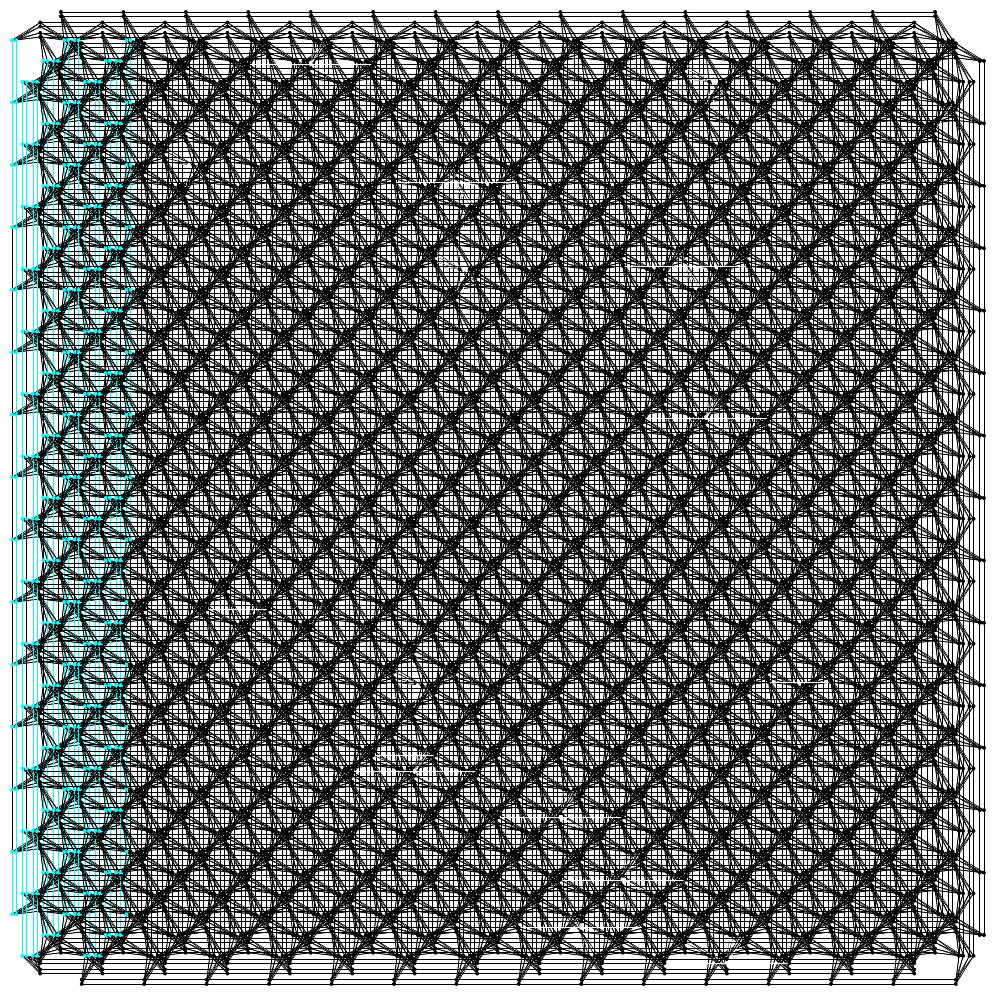}
    \caption{$384$ variable heavy-hex graph (left), and the native embedding of that heavy-hex graph onto the hardware Pegasus graph of \texttt{Advantage\_system4.1} shown by cyan nodes and edges (right). }
    \label{fig:384_node_heavy_hex_graph}
\end{figure*}

Figure \ref{fig:384_node_heavy_hex_graph} shows the $384$ node heavy-hex graph, along with the embedding onto the D-Wave hardware Pegasus graph.

%%%%%%%%%%%%%%%%%%%%%%%%%%%%%%%%%%%%%%%%%%%%%%%%%%%%%%%%%%%%%%%%%%%%%%
%%%%%%%%%%%%%%%%%%%%%%%%%%%%%%%%%%%%%%%%%%%%%%%%%%%%%%%%%%%%%%%%%%%%%%
%%%%%%%%%%%%%%%%%%%%%%%%%%%%%%%%%%%%%%%%%%%%%%%%%%%%%%%%%%%%%%%%%%%%%%
\section{D-Wave QPU Schedule Characteristics}
\label{section:appendix_DWave_calibration_data}

\begin{figure*}[p!]
    \centering
    \includegraphics[width=0.49\textwidth]{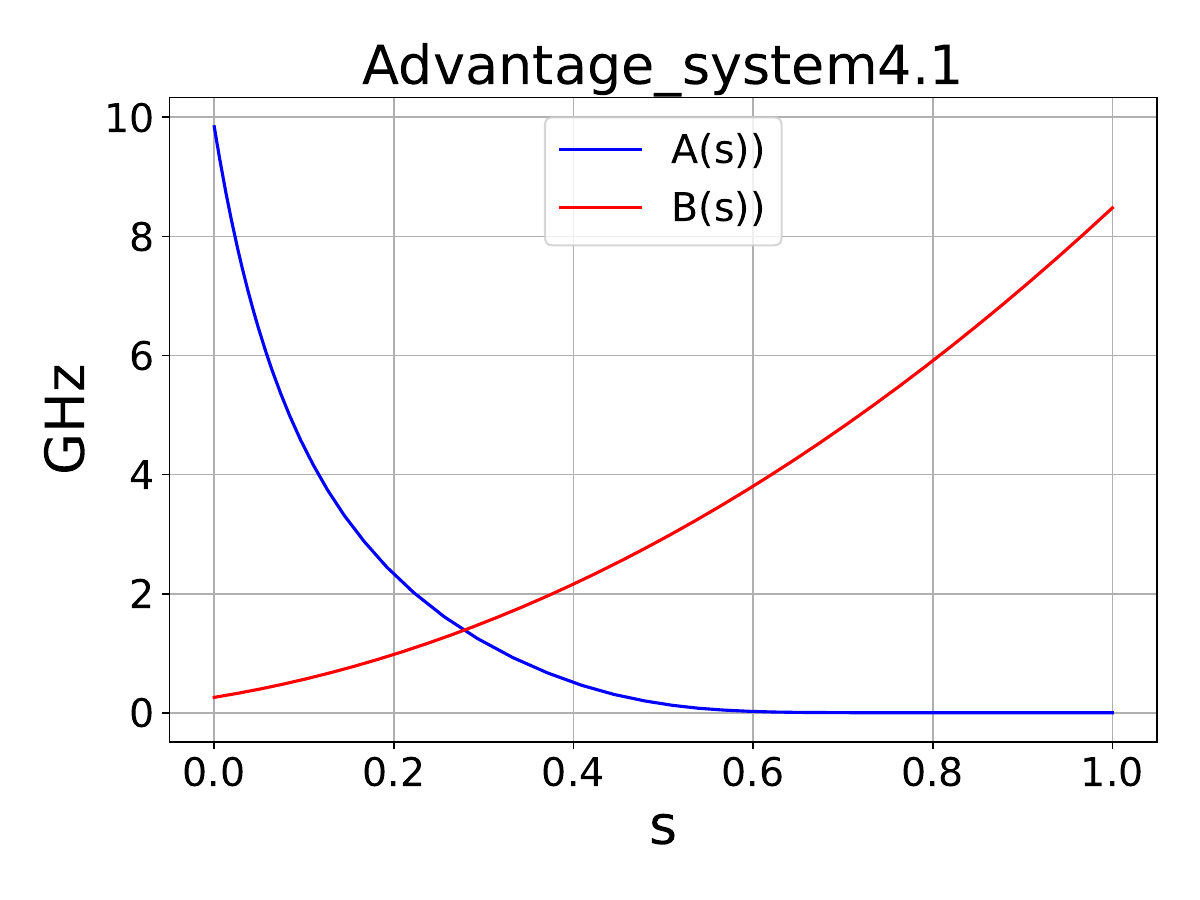}
    \includegraphics[width=0.49\textwidth]{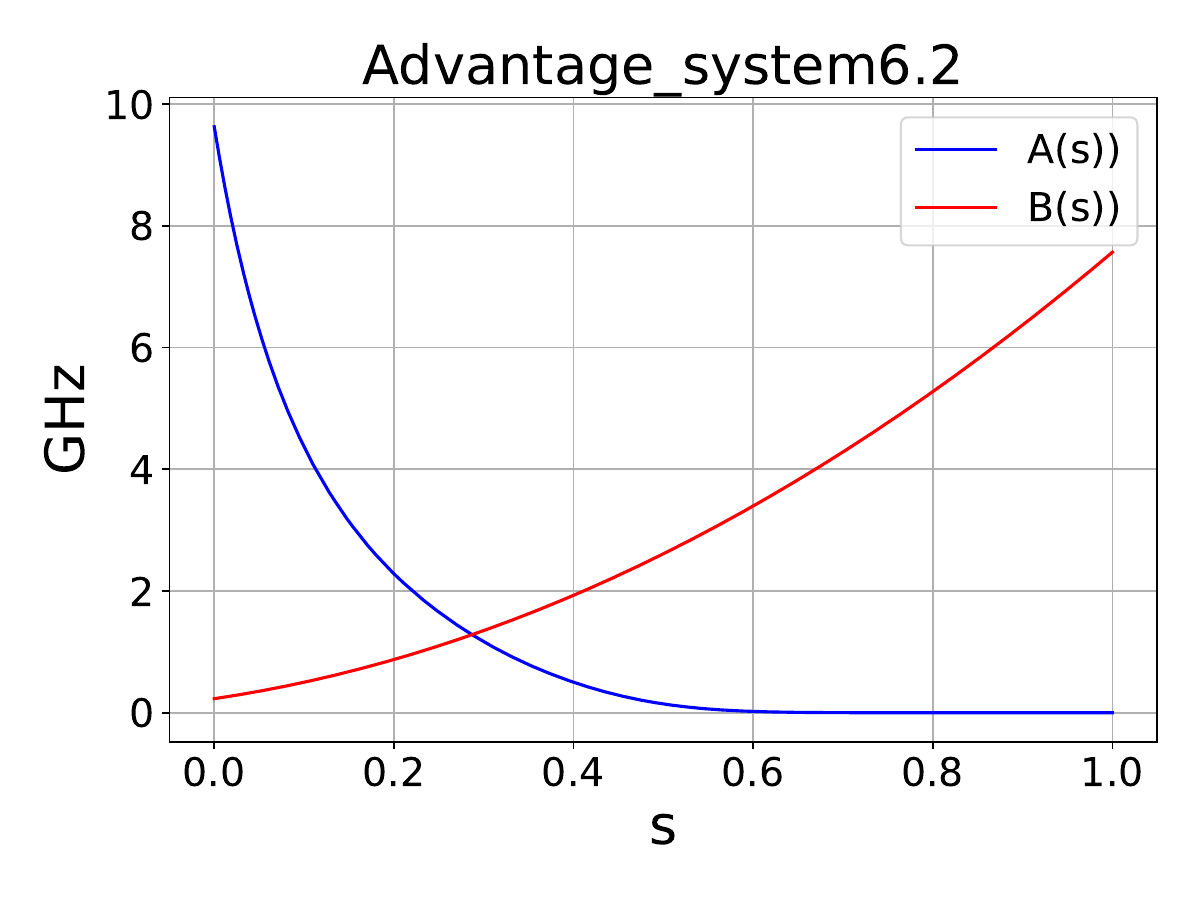}
    \vspace{-5ex}
    \caption{D-Wave QPU anneal schedule calibration plots.}
    \label{fig:DWave_calibration_data_plots}
\end{figure*}

Figure~\ref{fig:DWave_calibration_data_plots} plots the vendor-provided anneal schedule characteristics defining the $A(s)$ and $B(s)$ energy scales as a function of the anneal fractions $s$. This is the calibration data used when computing the equivalent anneal schedules to the TFIM Trotterized circuits (Section \ref{section:methods_Derivation_of_QA_parameters}). 

\begin{figure*}[p!]
    \centering
    \includegraphics[width=0.50\textwidth]{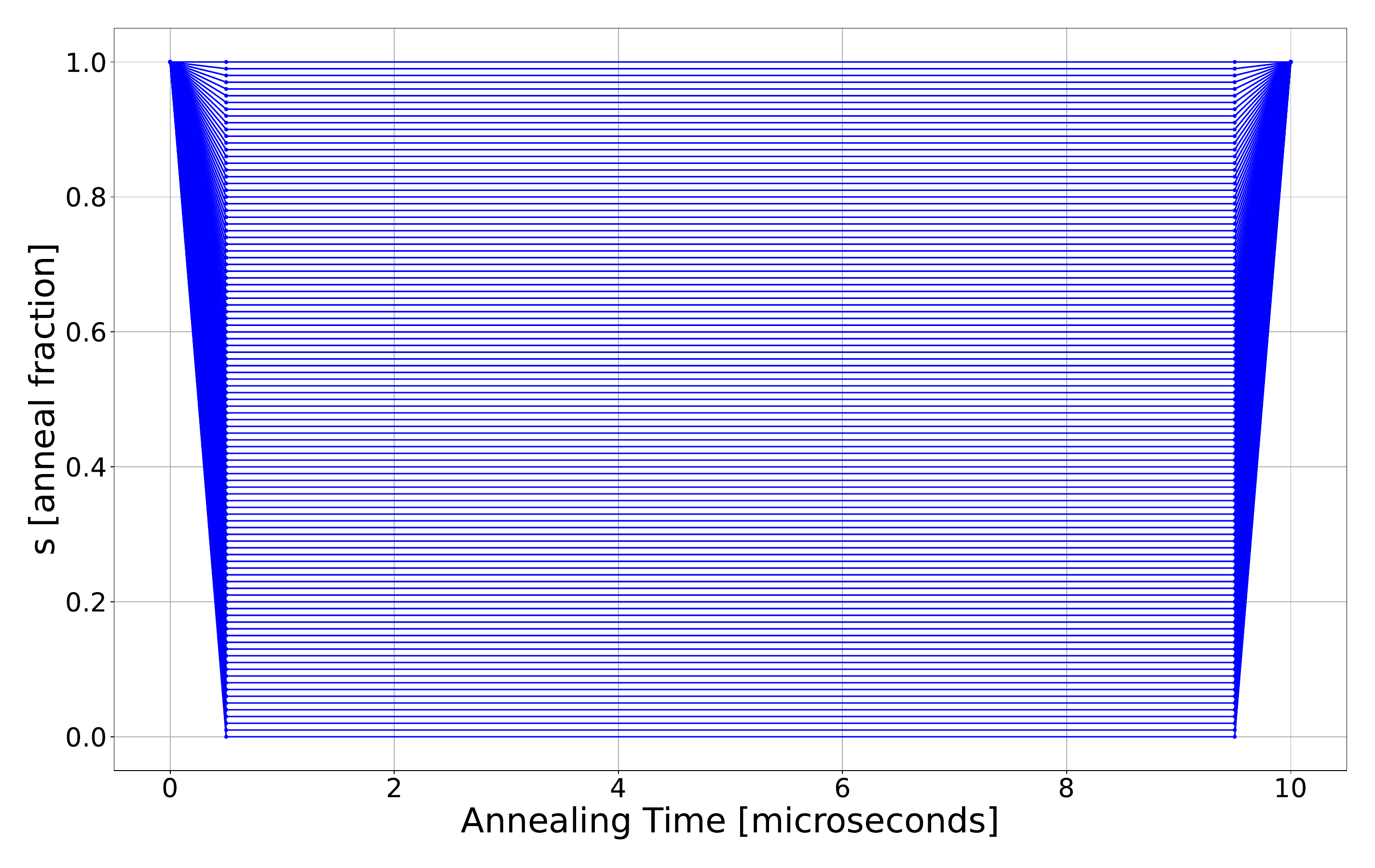}
    \vspace{-1ex}
    \caption{Full range of reverse annealing schedules used for simulating magnetization dynamics, using a fixed ramp duration of $0.5$ microseconds and varying the anneal fraction at which the pause occurs (specifically $100$ different $s$ values in linear increments $\in [0, 1]$). This plot uses the example total annealing time of $10$ microseconds, but for annealing times other than $10$ microseconds, the same schedules are used but the ramps down to the pause and up the measurement are always fixed to have a duration of $500$ nanoseconds.}
    \label{fig:reverse_annealing_schedules}
\end{figure*}

\begin{figure*}[p!]
    \centering
    \includegraphics[width=0.49\textwidth]{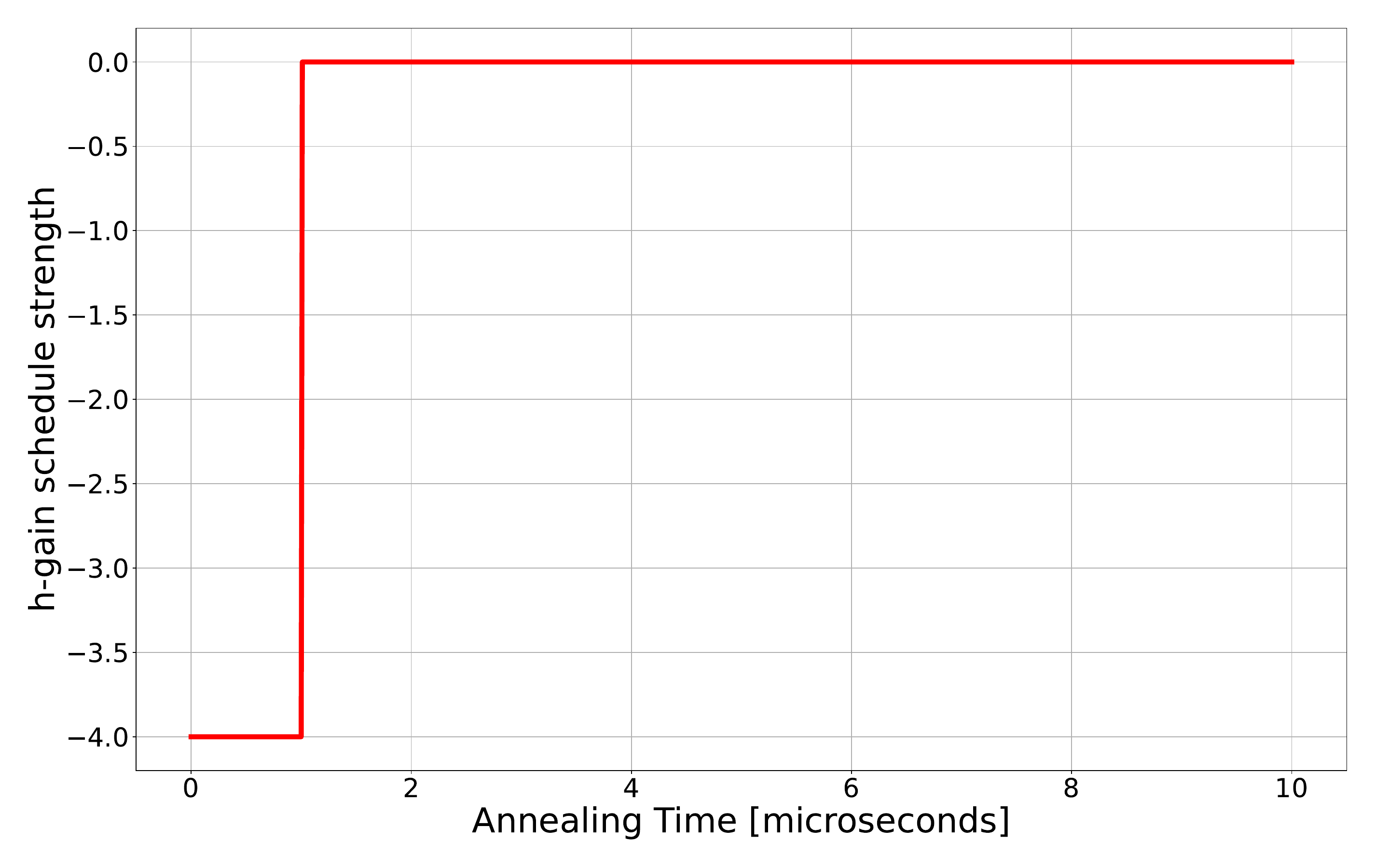}
    \includegraphics[width=0.49\textwidth]{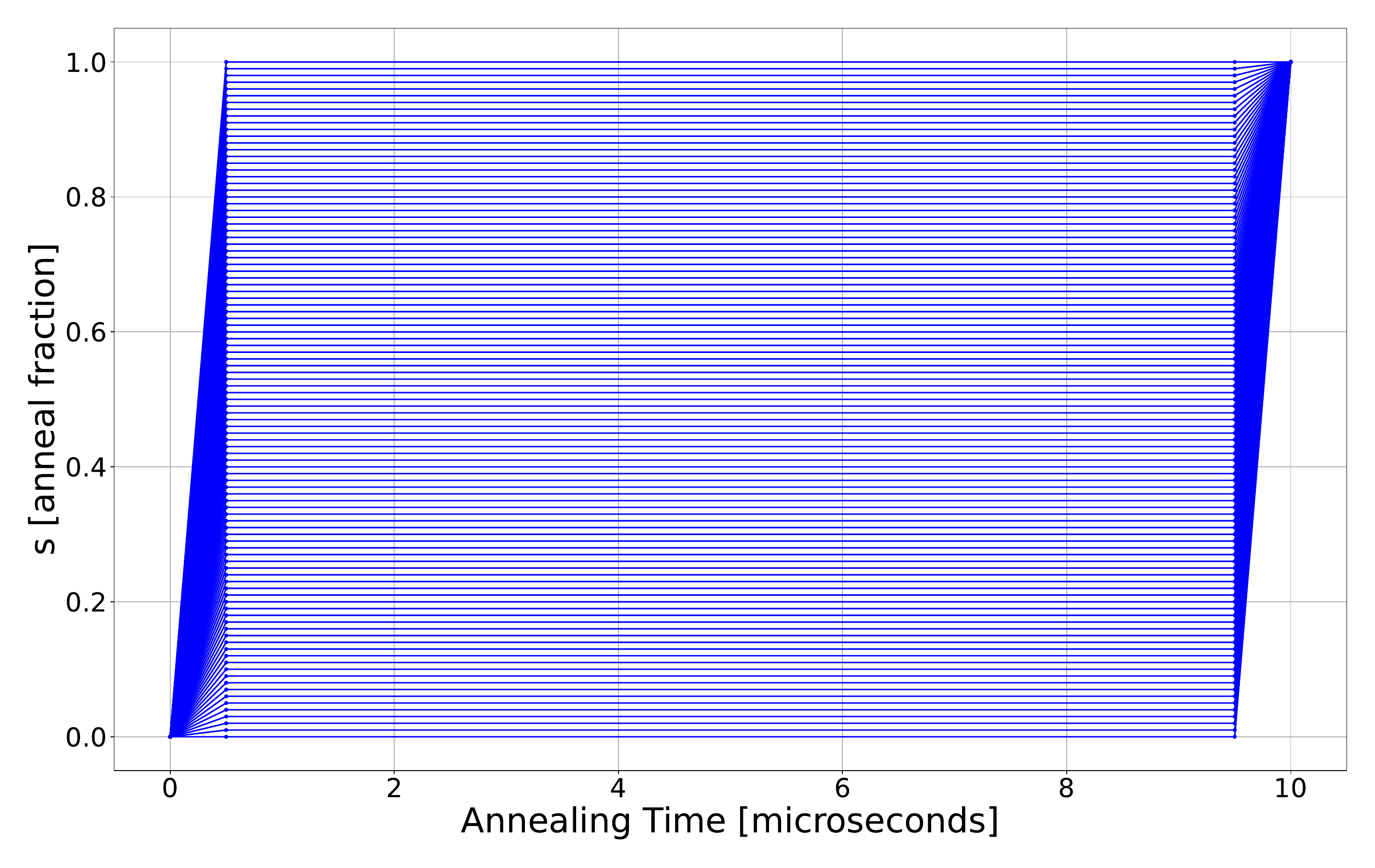}
    \vspace{-1ex}
    \caption{Full range of forward annealing schedules (right) used for the magnetization dynamics with fixed schedules, in conjunction with a fixed h-gain schedule (left). Note that the h-gain schedule pulse at the beginning of the anneal is fixed in its duration to always be exactly $1$ microsecond, with a ramp-down to $0$ time of $10$ nanoseconds, and is also set the maximum programmable h-gain schedule strength possible on the D-Wave device; for \texttt{Advantage\_system6.2} this is $-4$, but for \texttt{Advantage\_system4.1} this is $-3$. The forward annealing schedules are applied to other annealing times besides $10$ microseconds, but the ramp durations are always set to $500$ nanoseconds. }
    \label{fig:h_gain_state_encoding_schedules}
\end{figure*}

%%%%%%%%%%%%%%%%%%%%%%%%%%%%%%%%%%%%%%%%%%%%%%%%%%%%
%%%%%%%%%%%%%%%%%%%%%%%%%%%%%%%%%%%%%%%%%%%%%%%%%%%%
%%%%%%%%%%%%%%%%%%%%%%%%%%%%%%%%%%%%%%%%%%%%%%%%%%%%
\section{Fixed Annealing Time Magnetization Dynamics}
\label{section:appendix_fixed_AT_s_magnetization}

The magnetization dynamics reported in Section \ref{section:results} are specifically replicating the same experimental parameters used for the Trotterized IBM Quantum experiment \cite{kim2023evidence}, using D-Wave quantum annealers. However, there are many other ways that quantum annealers could be used to examine magnetization dynamics on the ferromagnetic TFIM. In this section, we report mean lattice magnetization measures using the same two simulation methods as before (reverse quantum annealing and h-gain state encoding), but here we fix the total anneal time, the ramp durations, and the anneal fractions. Figure~\ref{fig:reverse_annealing_schedules} shows what these fixed reverse quantum annealing schedules, and Figure~\ref{fig:h_gain_state_encoding_schedules} shows the fixed h-gain state encoding schedules. The experimental procedure in this case is to vary the anneal fraction $s$ and measure the resulting magnetization. Therefore, this is not directly equivalent to the Trotterized experiments (as outlined in Section \ref{section:methods_Derivation_of_QA_parameters}), however this is performing Hamiltonian dynamics simulation of same ferromagnetic model in a transverse field. Instead of specifying the anneal schedules to have a slope that is steepest allowed on the machine, in this section we set the ramp durations to be fixed at $0.5$ microseconds which is the maximum ramp slope when reverse annealing at $s=0$, since the sweep from $s=0$ to $s=1$ is a larger range compared to the more limited anneal fraction results of Section \ref{section:results}.

Figure~\ref{fig:reverse_annealing_schedules} shows the full range of reverse annealing schedules (for all values of the anneal fraction $s$) that are used in these fixed annealing time simulations. 

The h-gain field used for the fixed annealing time h-gain state encoding method is shown in Figure~\ref{fig:h_gain_state_encoding_schedules} (left). The h-gain state encoding handles the spin up state, and the for the Hamiltonian dynamics we can again simply pause the schedule for a long duration at an anneal fraction $s$, then we must quench and measure the qubit states. Thus, this protocol does not require reverse annealing -- instead we can initialize the anneal as in a standard forward anneal. The corresponding anneal schedules are shown in Figure~\ref{fig:h_gain_state_encoding_schedules} (right). The time of the applied h-gain field was chosen to be $1$ microsecond for all annealing times in order to ensure that the system is fully saturated with the spin up state, but this means that the total paused annealing time is $AT-1.51$ microseconds (without the strong applied h-gain field), as opposed to the fixed time reverse annealing protocol where the total paused annealing time is $AT-1$ microseconds. As before, $100$ random server-side spin reversal transforms are applied per $1,000$ anneal batch for all h-gain state encoding simulations.

\begin{figure*}[ht!]
    \centering
    \includegraphics[width=0.47\textwidth]{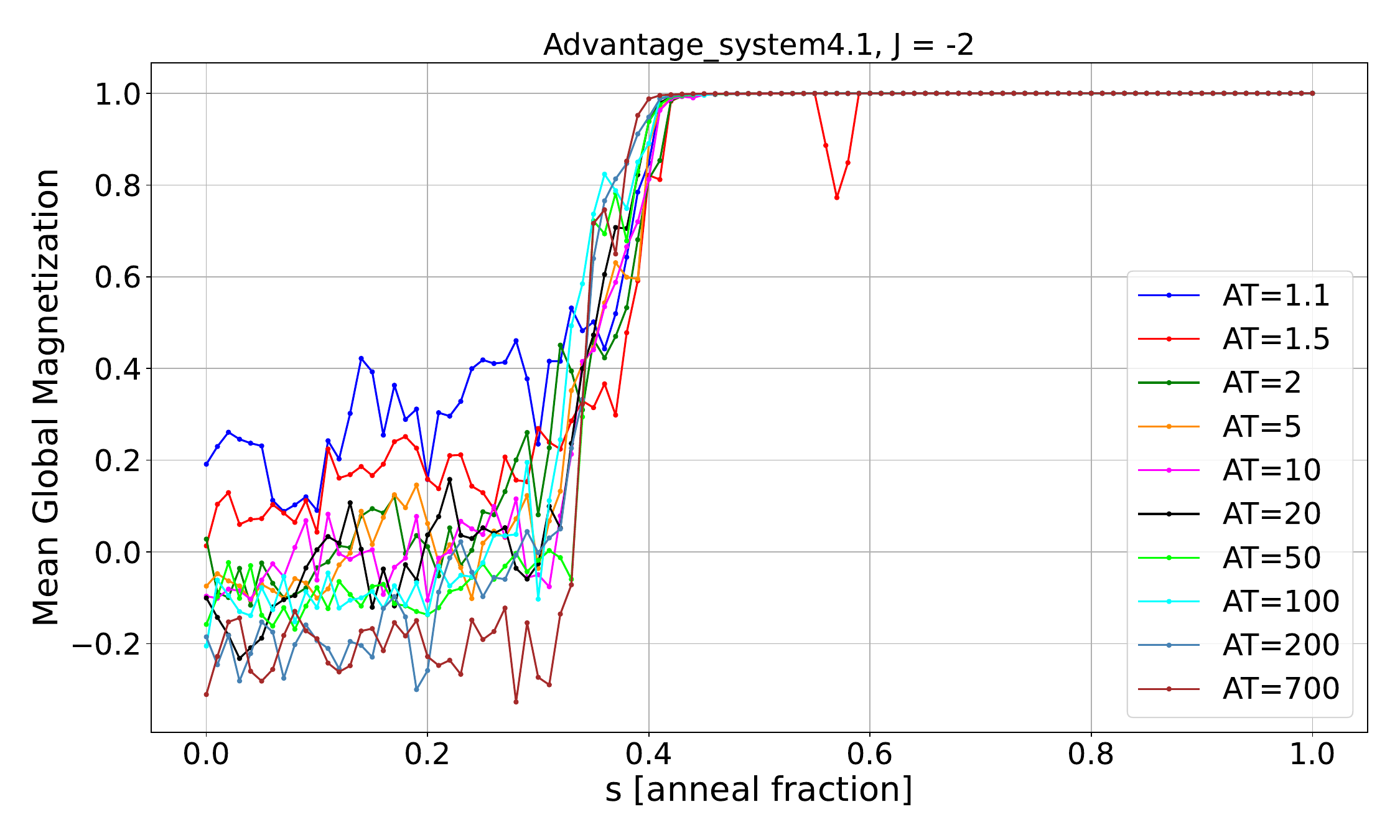}
    \includegraphics[width=0.47\textwidth]{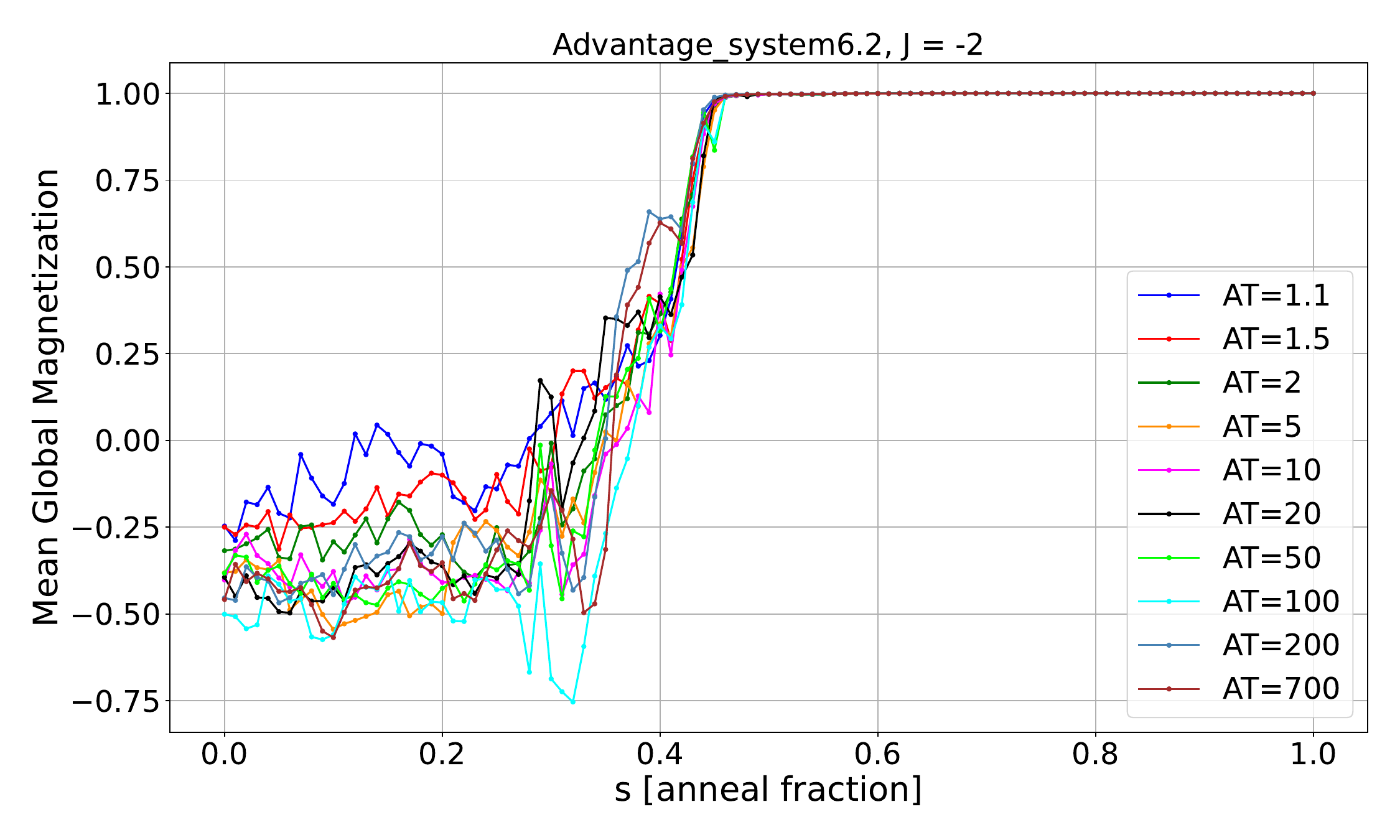}\\
    \includegraphics[width=0.47\textwidth]{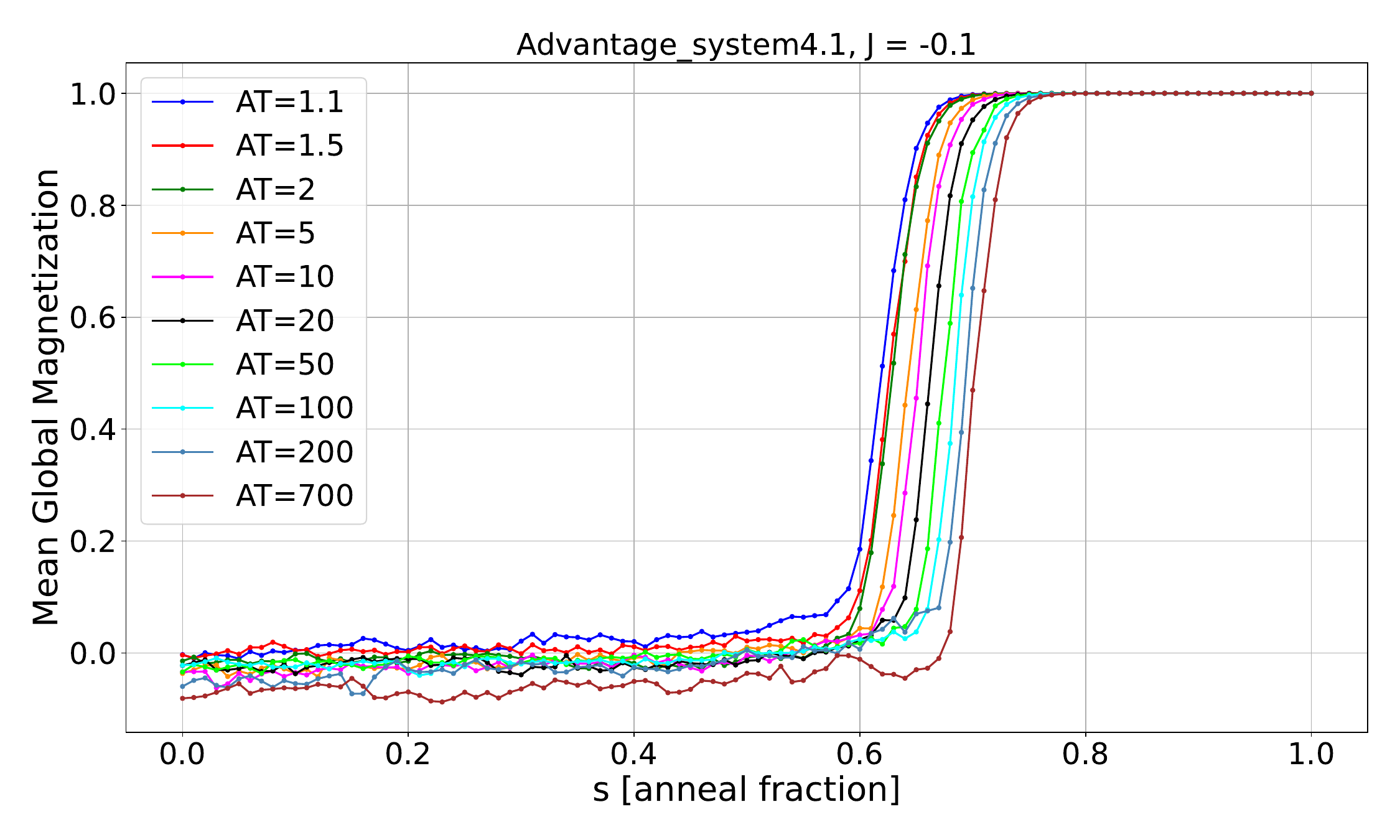}
    \includegraphics[width=0.47\textwidth]{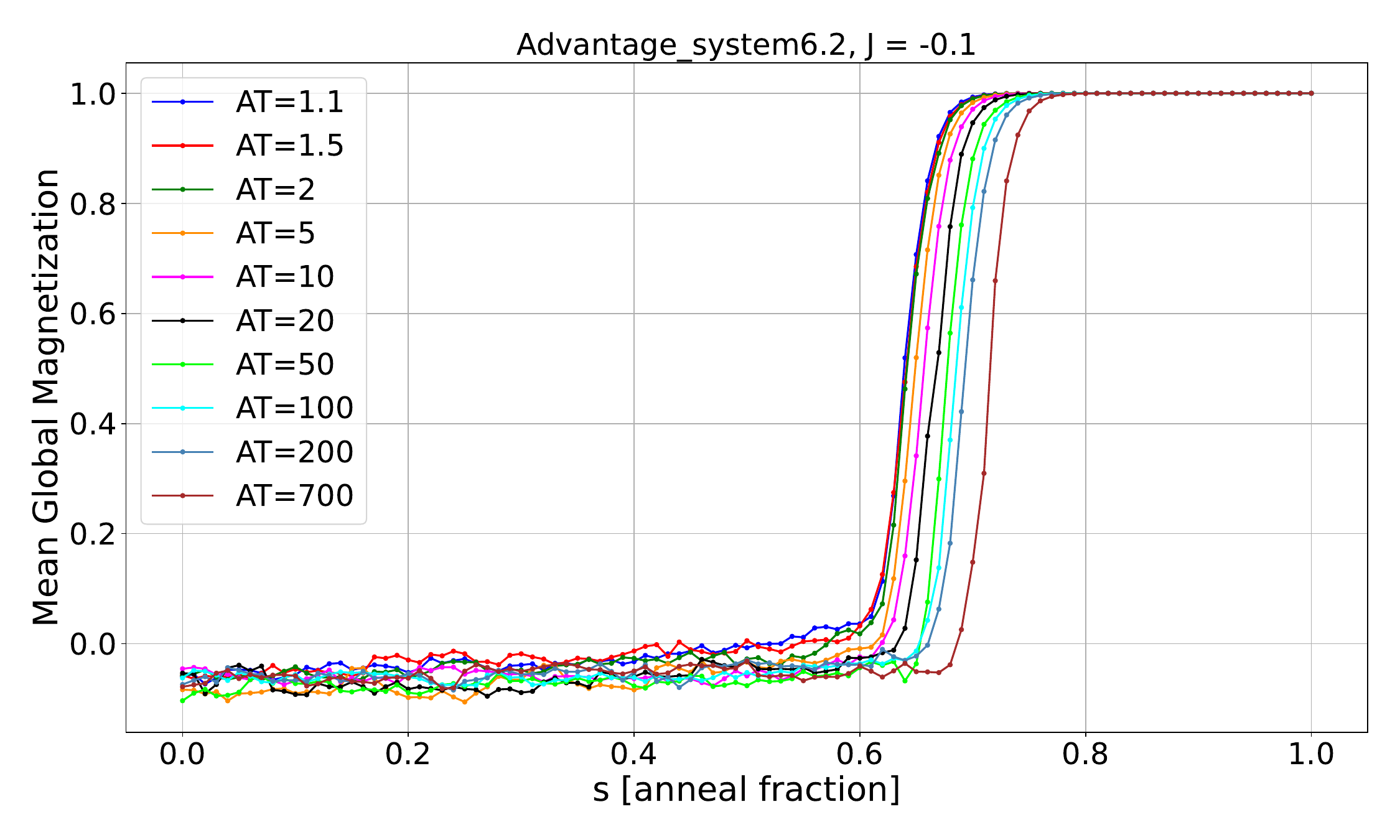}\\
    \includegraphics[width=0.47\textwidth]{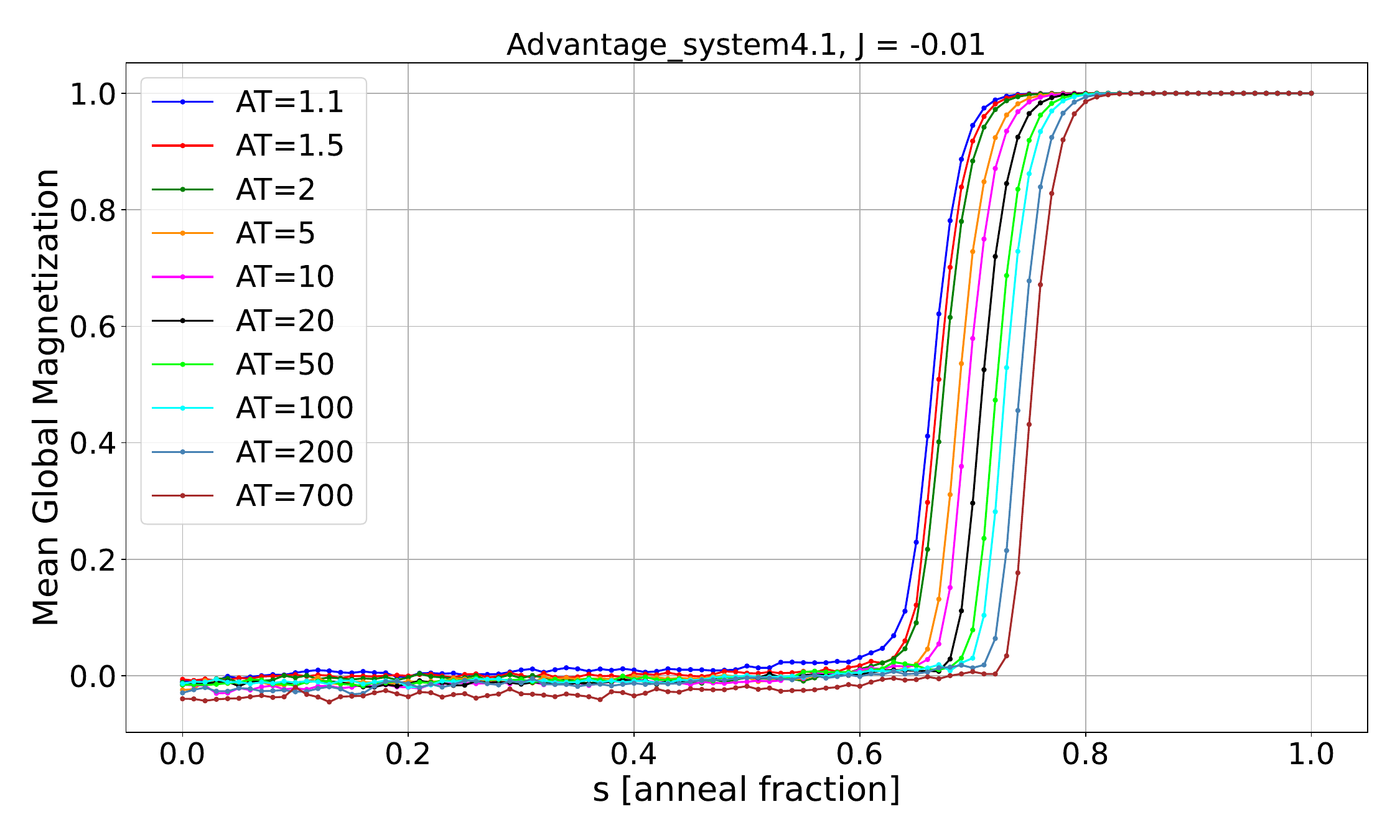}
    \includegraphics[width=0.47\textwidth]{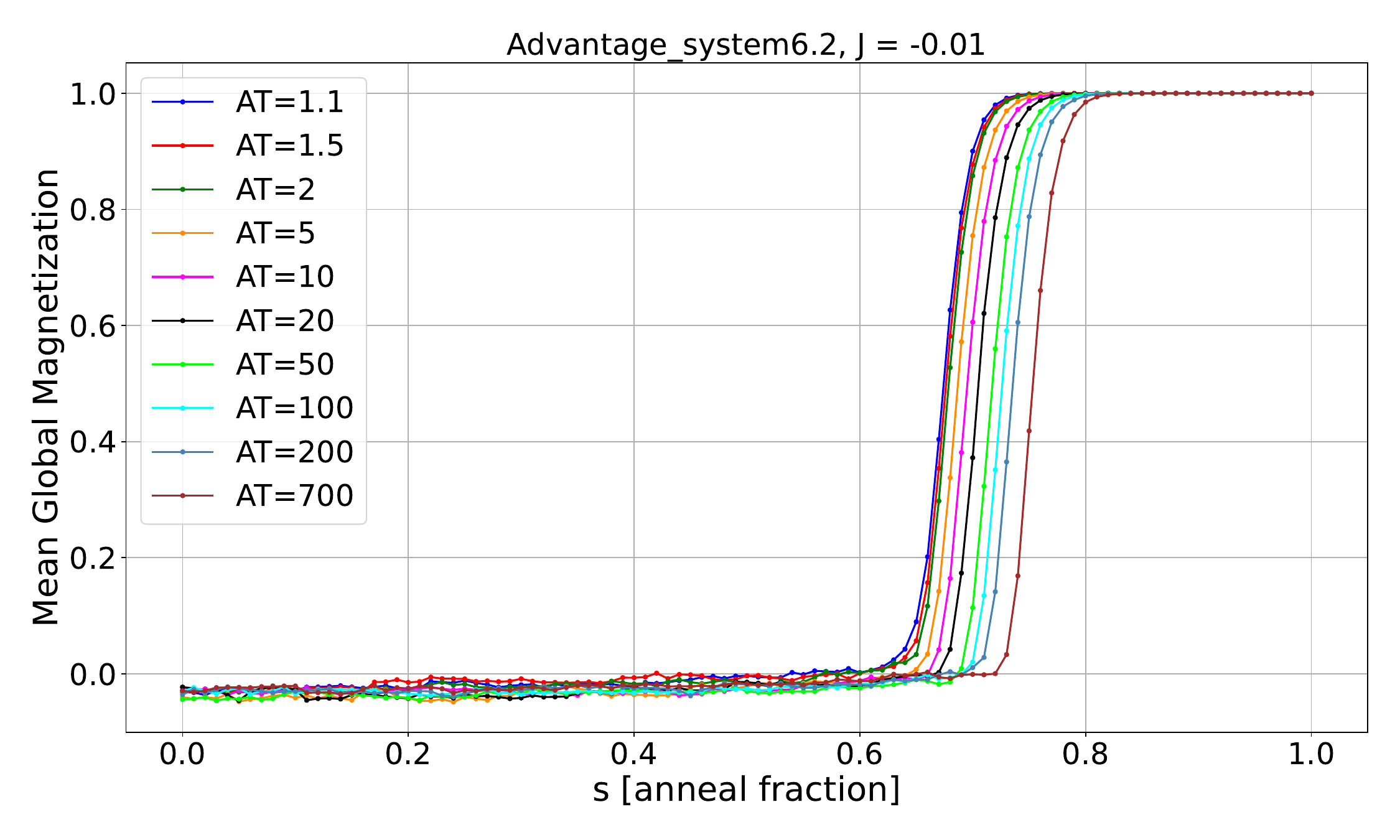}
    \includegraphics[width=0.47\textwidth]{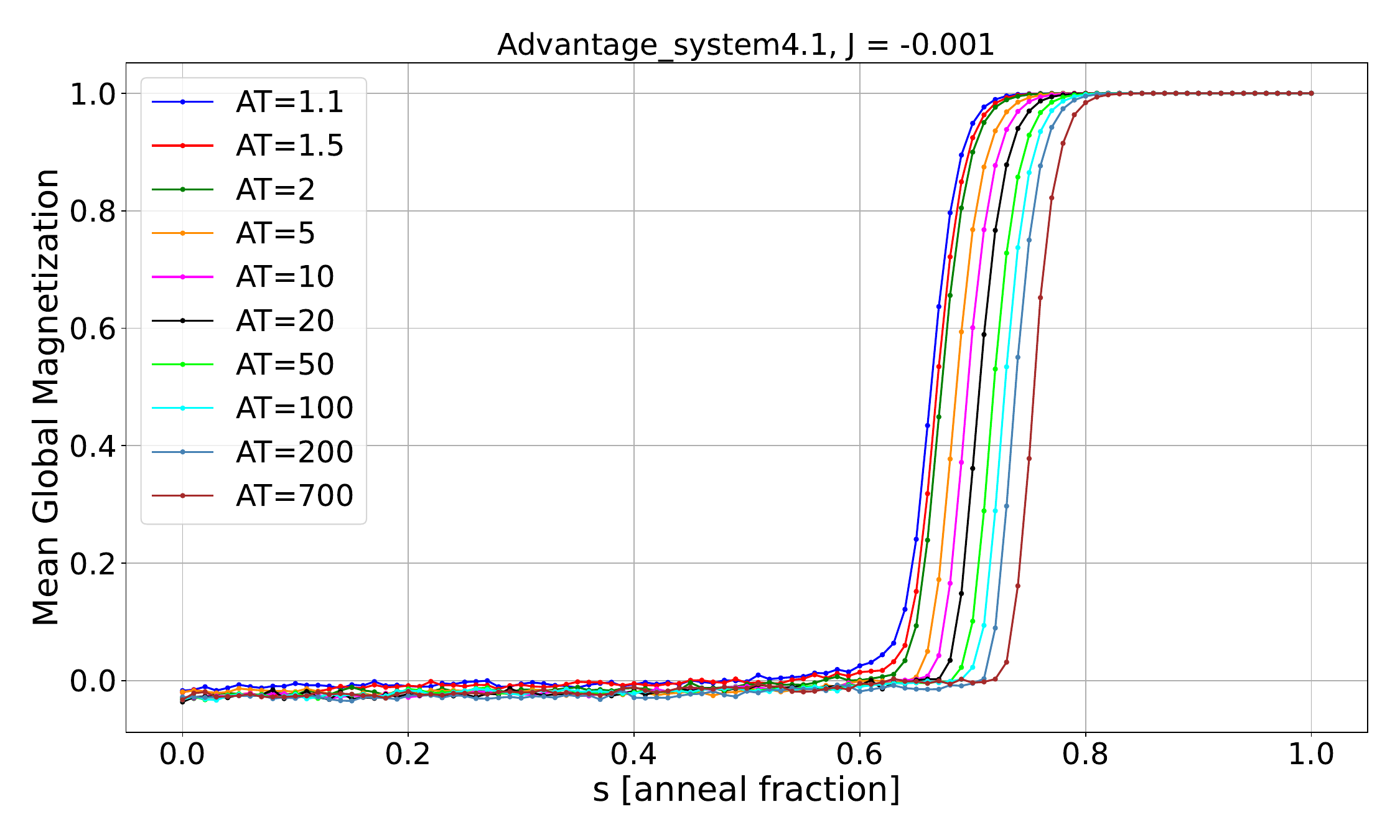}
    \includegraphics[width=0.47\textwidth]{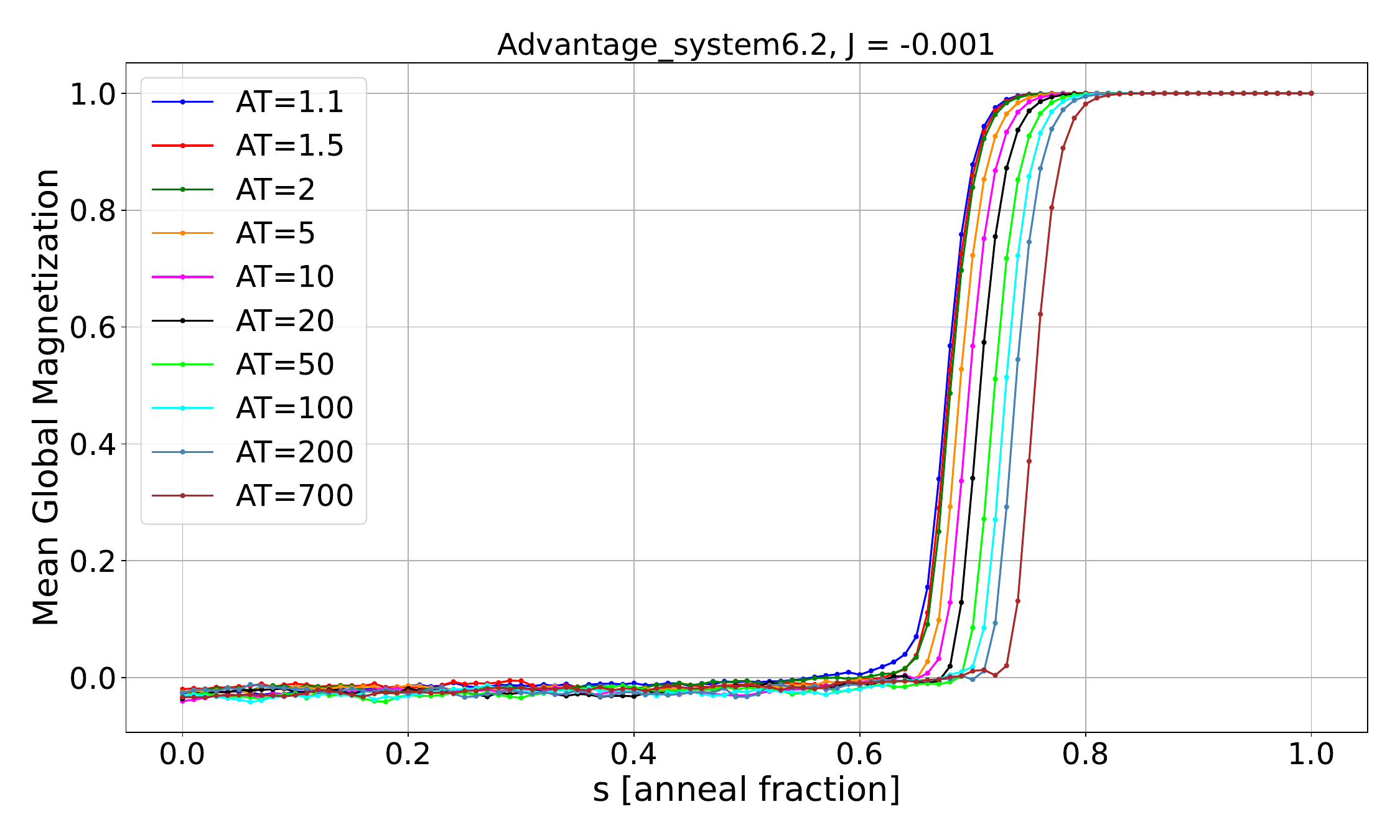}
    \caption{Mean magnetization plots using the reverse annealing dynamics simulation method on the two D-Wave quantum annealers; (y-axis) is the mean spin across all variables and samples, (x-axis) is anneal fraction $s$. Annealing times of $1.1, 1.5, 2, 5, 10, 20, 50, 100, 200, 700$ microseconds are tested; each line corresponds to a different annealing time. }
    \label{fig:global_magnetization_plots_reverse_annealing}
\end{figure*}

\begin{figure*}[h!]
    \centering
    \includegraphics[width=0.49\textwidth]{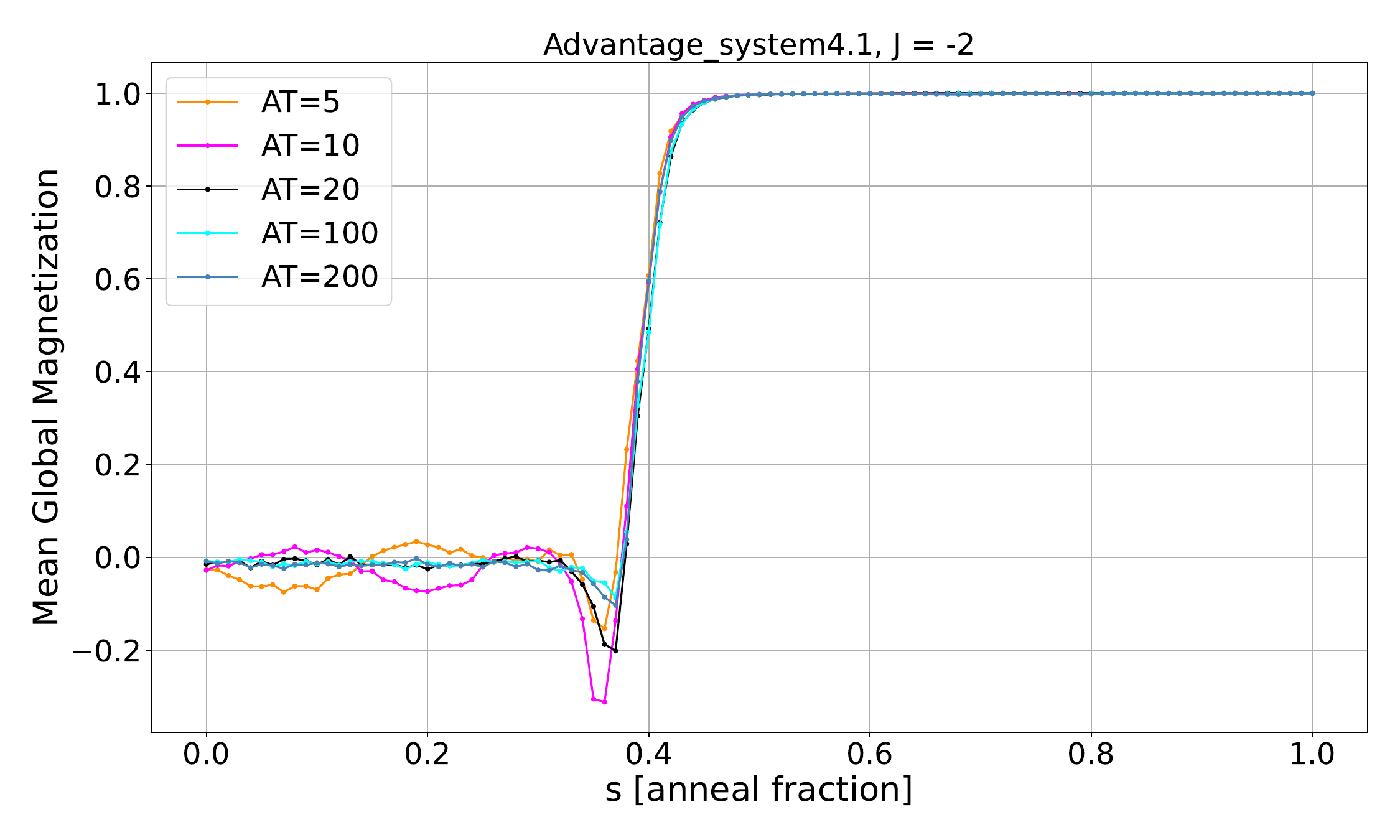}
    \includegraphics[width=0.49\textwidth]{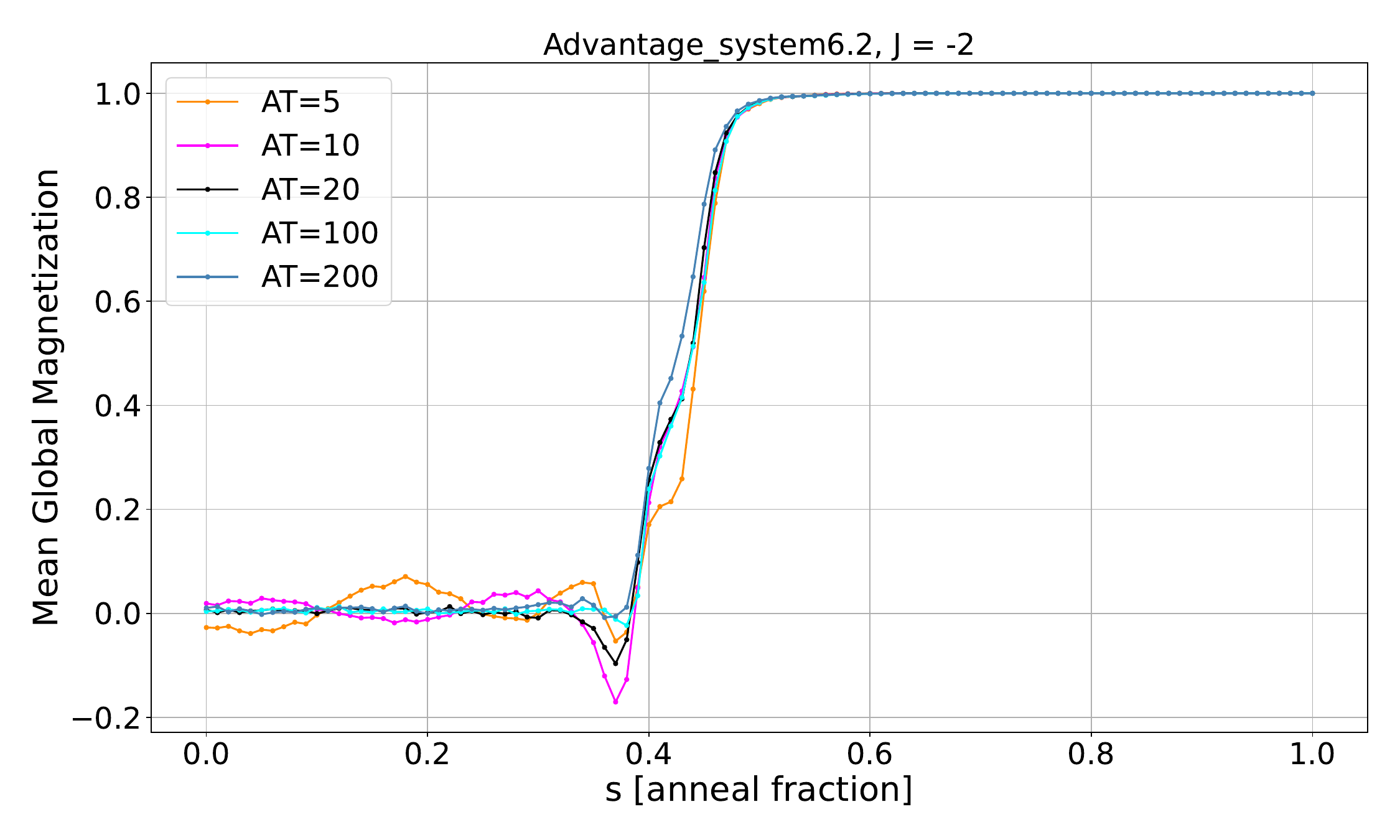}\\
    \includegraphics[width=0.49\textwidth]{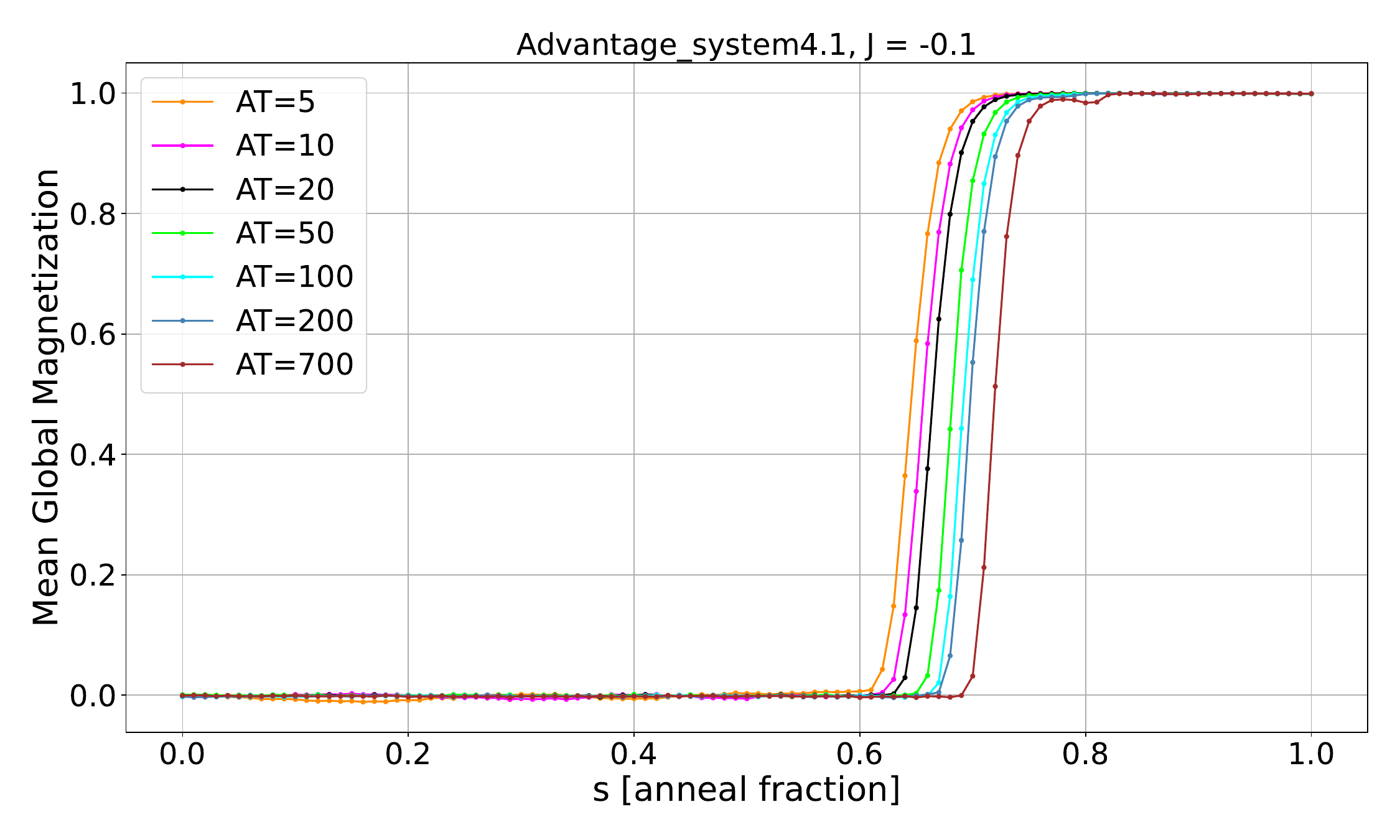}
    \includegraphics[width=0.49\textwidth]{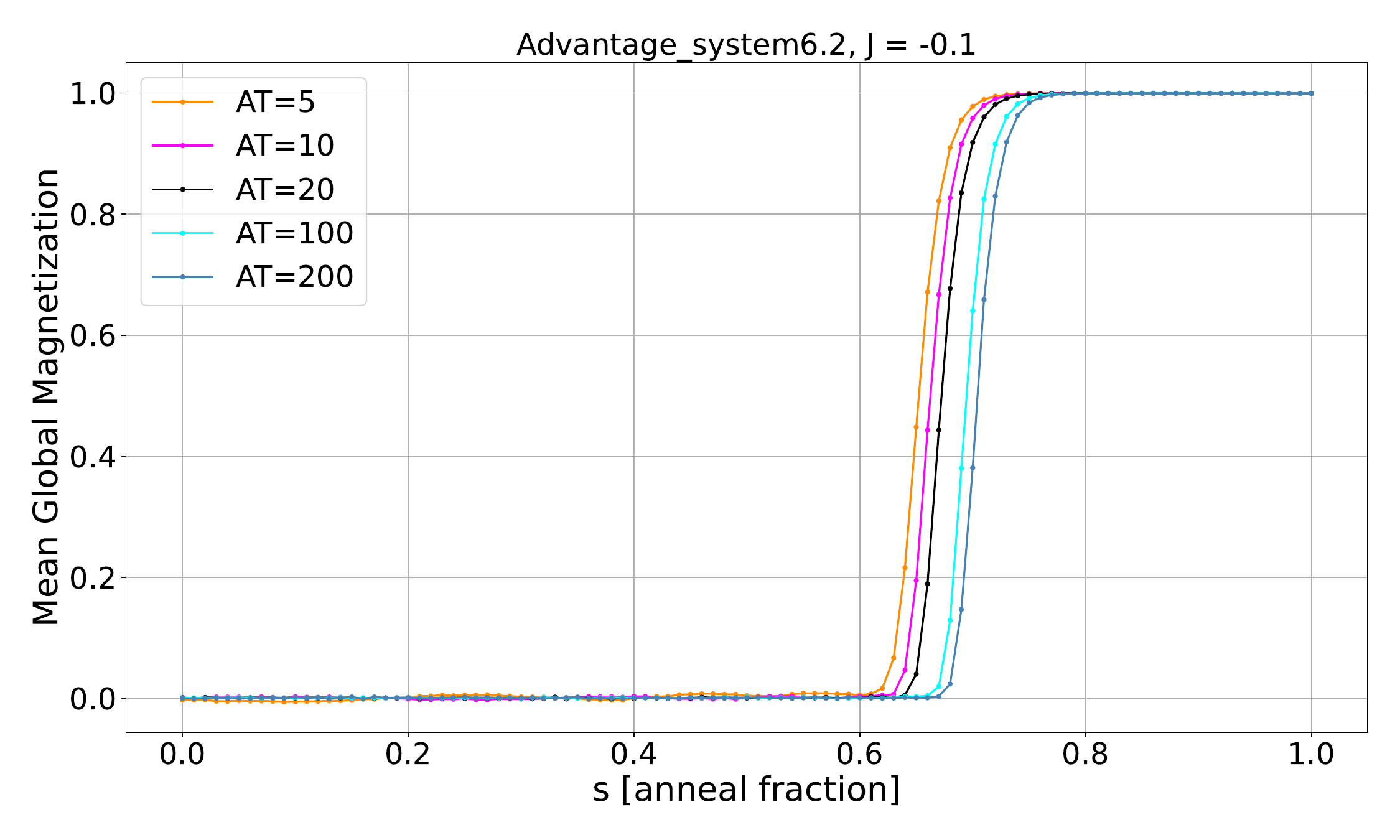}\\
    \includegraphics[width=0.49\textwidth]{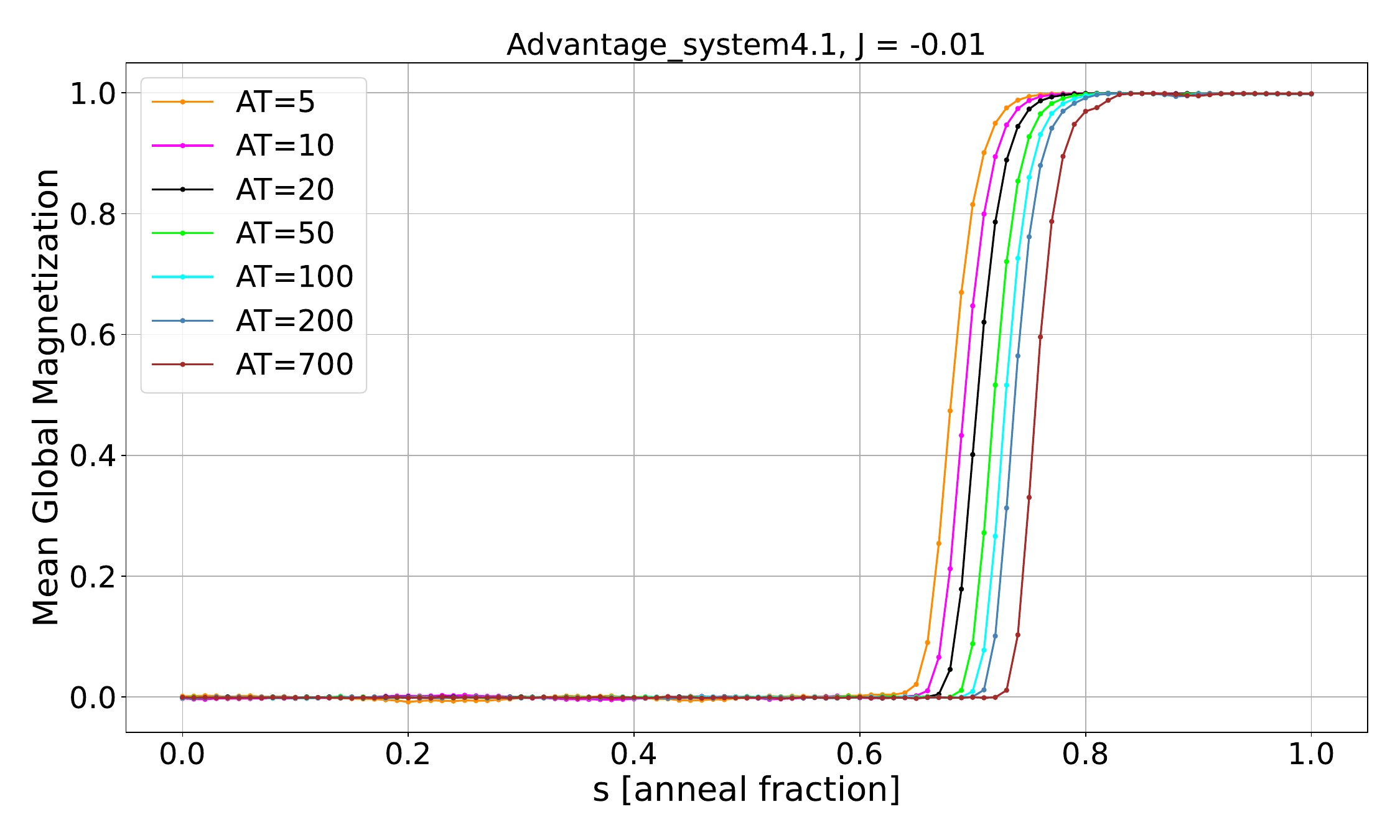}
    \includegraphics[width=0.49\textwidth]{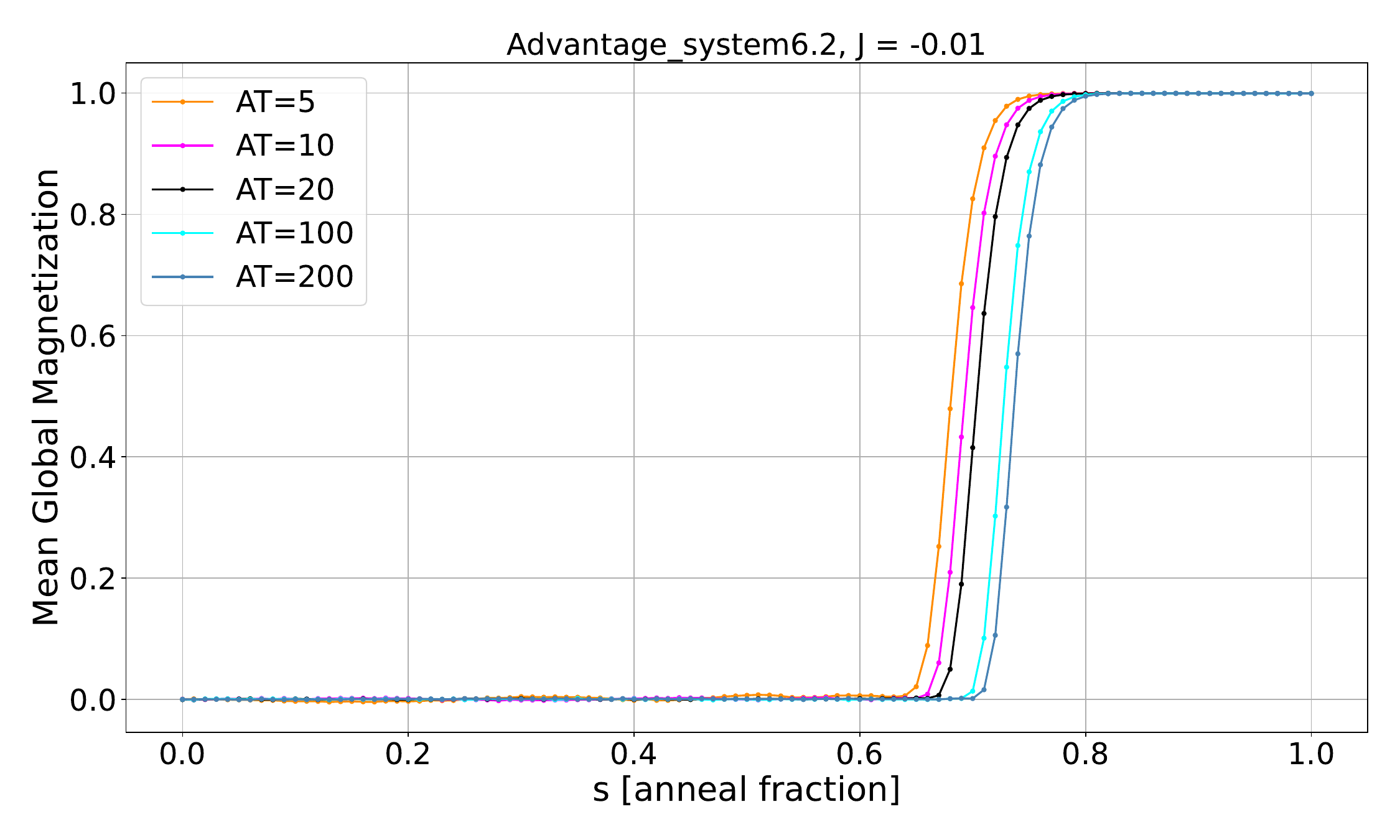}
    \includegraphics[width=0.49\textwidth]{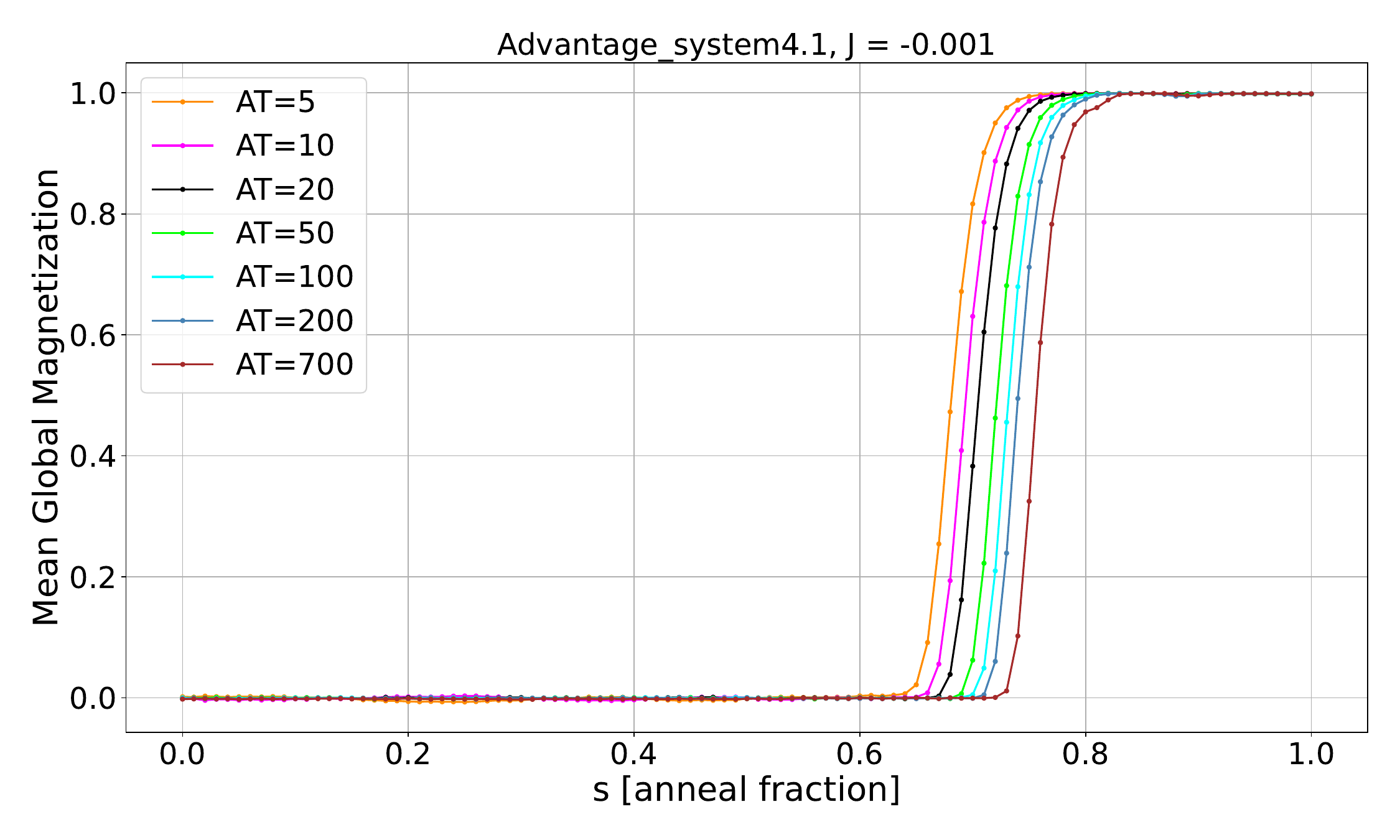}
    \caption{Mean magnetization plots using the h-gain state encoding method on the two D-Wave quantum annealers; (y-axis) is the mean spin across all variables and sampled, (x-axis) is anneal fraction $s$.  }
    \label{fig:global_magnetization_plots_reverse_annealing_h_gain}
\end{figure*}

Figure~\ref{fig:global_magnetization_plots_reverse_annealing} shows mean lattice magnetization (across all samples) as a function of $s \in [0, 1]$ in steps of $0.01$ for varying annealing times and varying programmed coupler strengths ($J= -0.001, -0.01, -0.1, -2$) on two D-Wave quantum annealers, using the reverse annealing Hamiltonian dynamics simulation method. The annealing times used in Figure~\ref{fig:global_magnetization_plots_reverse_annealing} have a total paused anneal time (e.g. time paused at the specified anneal fraction $s$) for $AT-1$ microseconds where $AT$ is the total annealing time. Figure~\ref{fig:global_magnetization_plots_reverse_annealing_h_gain} shows the same simulations, but using the h-gain state encoding technique to perform the Hamiltonian dynamics simulation (with server side spin reversal transforms). In both Figure~\ref{fig:global_magnetization_plots_reverse_annealing} and Figure~\ref{fig:global_magnetization_plots_reverse_annealing_h_gain}, as the programmed energy scale is decreased, the magnetization at lower $s$ becomes closer to $0$. In the magnetization dynamics shown in Figures \ref{fig:global_magnetization_plots_reverse_annealing} and \ref{fig:global_magnetization_plots_reverse_annealing_h_gain}, a total of $10,000$ samples are obtained (using $10$ independent device calls) for each point plotted in the sub-figures (the magnetization observable computed for each point is the mean spin of all measured qubits, across all $10,000$ samples). Note that for Figure~\ref{fig:global_magnetization_plots_reverse_annealing_h_gain}, when $J=-2$ this is accomplished by turning on the device coefficient autoscaling, which also scales up the programmed qubit weights. 

The range of programmed energy scales Figure~\ref{fig:global_magnetization_plots_reverse_annealing} show that the anneal fraction $s$ at which the global magnetization changes is determined by the programmed energy scales. The energy scales at $J=-2$ have the phase change happening at approximately $s=0.4$, whereas the $J=-0.01$ energy scale has the magnetization change beginning at $s=0.8$.

The high coupling strength plots in Figure~\ref{fig:global_magnetization_plots_reverse_annealing} show high magnetization variability at small anneal fractions. A possible explanation could be that the spin bath polarization effect \cite{lanting2020probing} is causing self correlations within each anneal cycle, which causes instability in the resulting measurements. 

The observed fixed $s$ and fixed annealing time curves follow a similar pattern, but not identical, to the Trotterized circuit magnetization results \cite{kim2023evidence}, and a similar pattern to the quantum annealing simulation of those same systems that we show in Section \ref{section:results}. We note that Ref.~\cite{Izquierdo_2021} observed similar magnetization for a 1-d Ising chain in the transverse field Ising model using reverse quantum annealing and fast quenching on a D-Wave quantum annealer.

\section{Single Site Heavy-Hex Lattice Magnetization Heatmaps}
\label{section:appendix_single_site_magnetization_heatmaps}

\begin{figure*}[ht!]
    \centering
    \includegraphics[width=0.45\textwidth]{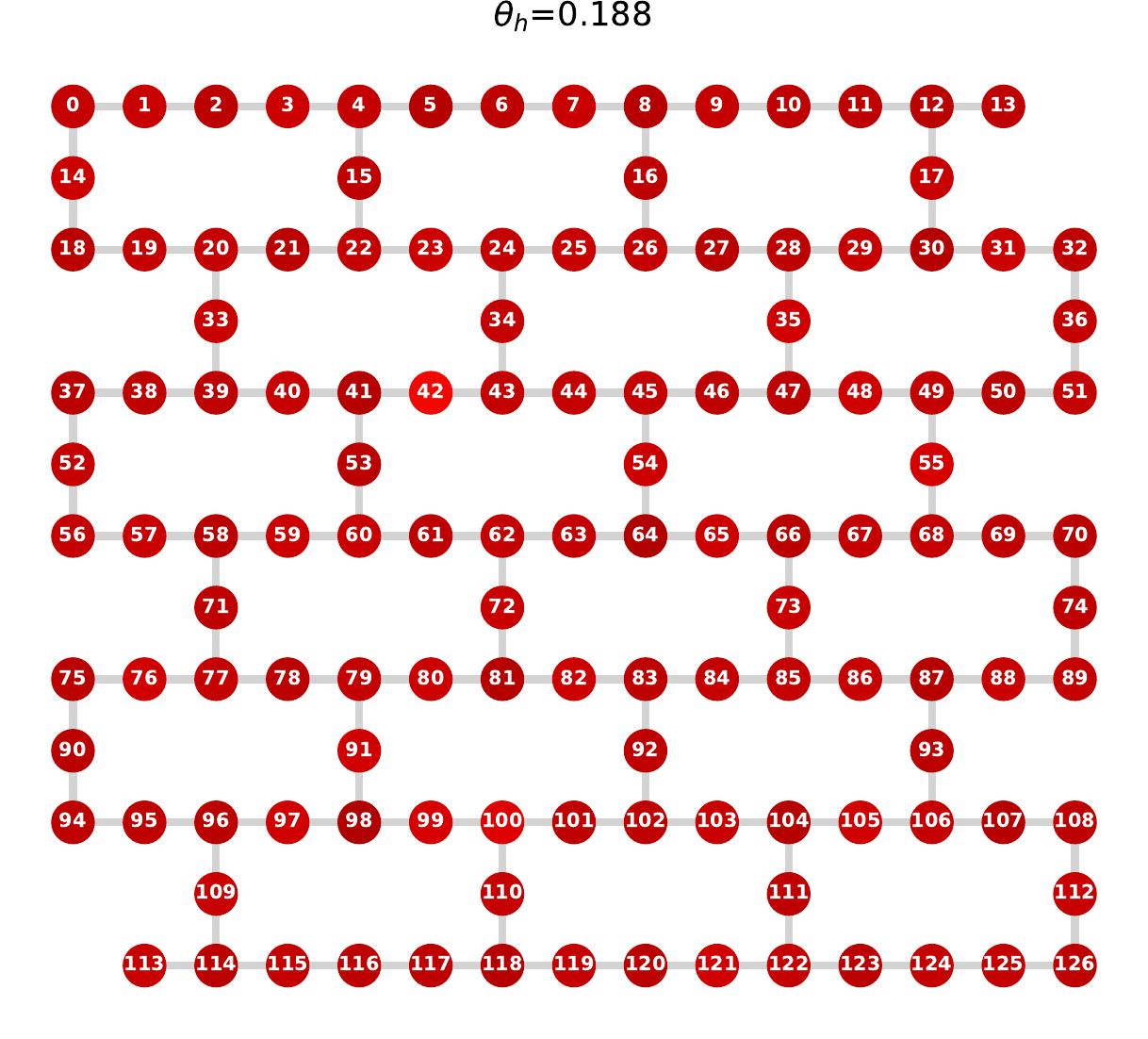}
    \includegraphics[width=0.45\textwidth]{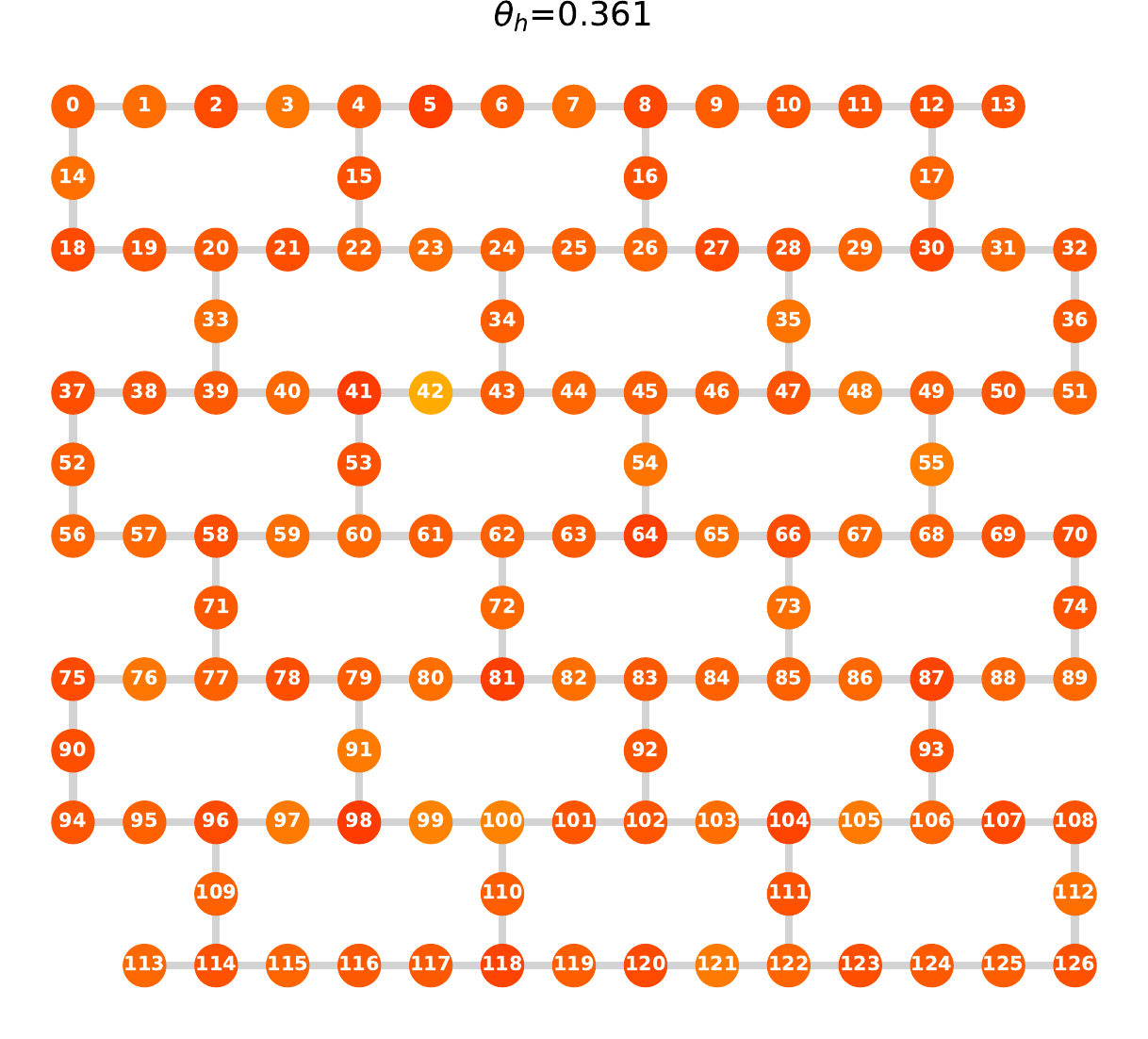}
    \includegraphics[width=0.45\textwidth]
    {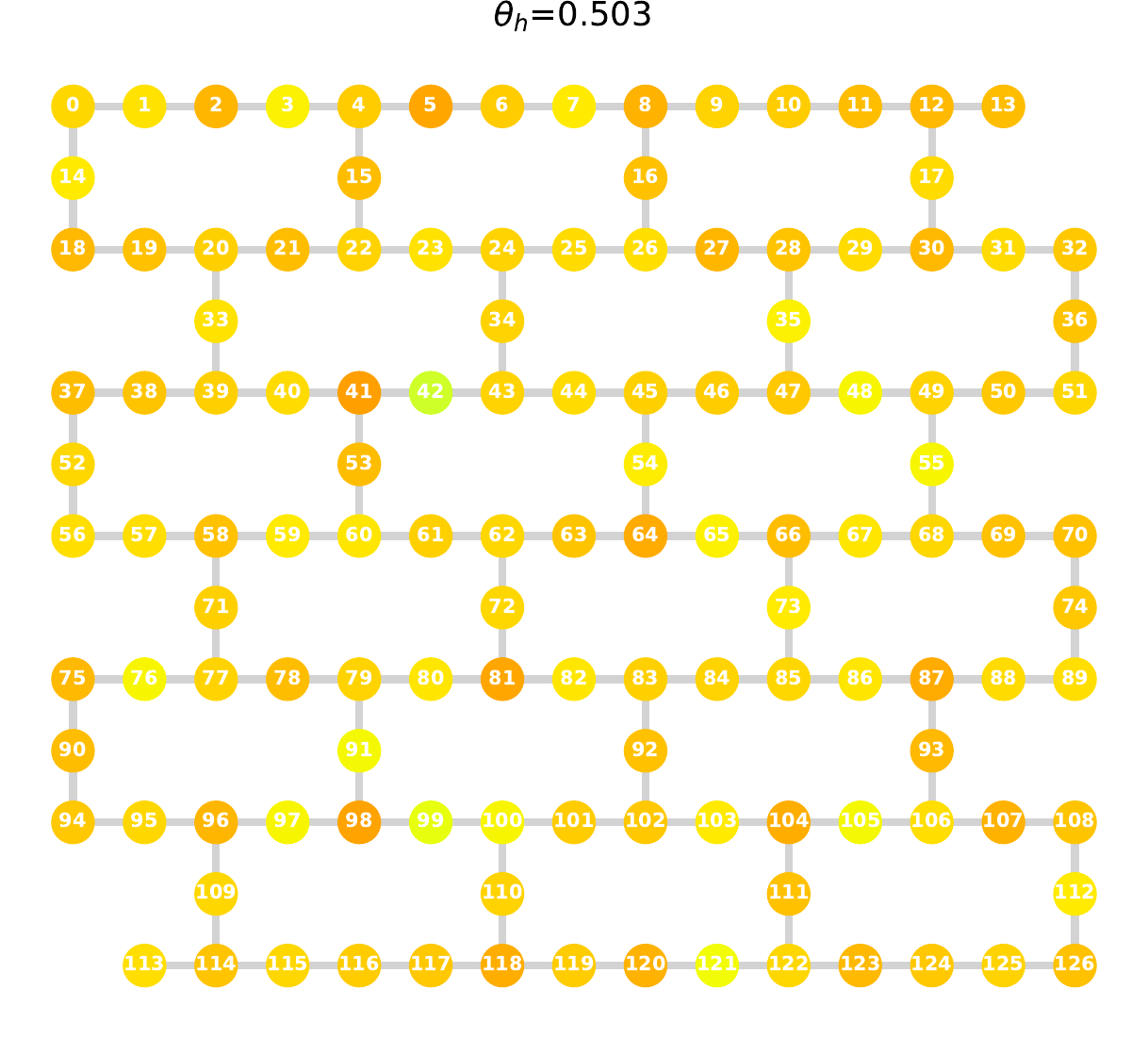}
    \includegraphics[width=0.45\textwidth]{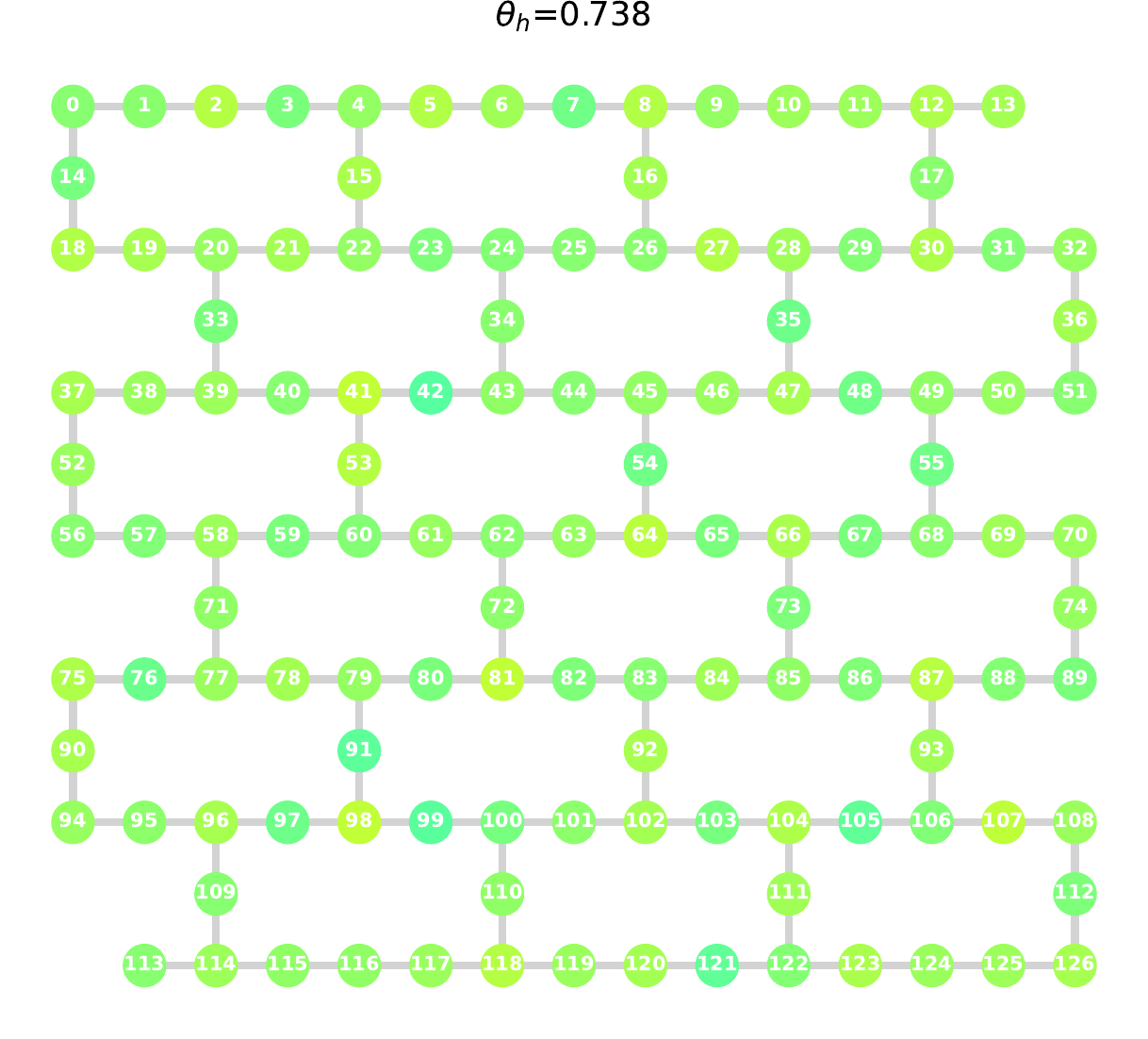}
    \includegraphics[width=0.45\textwidth]{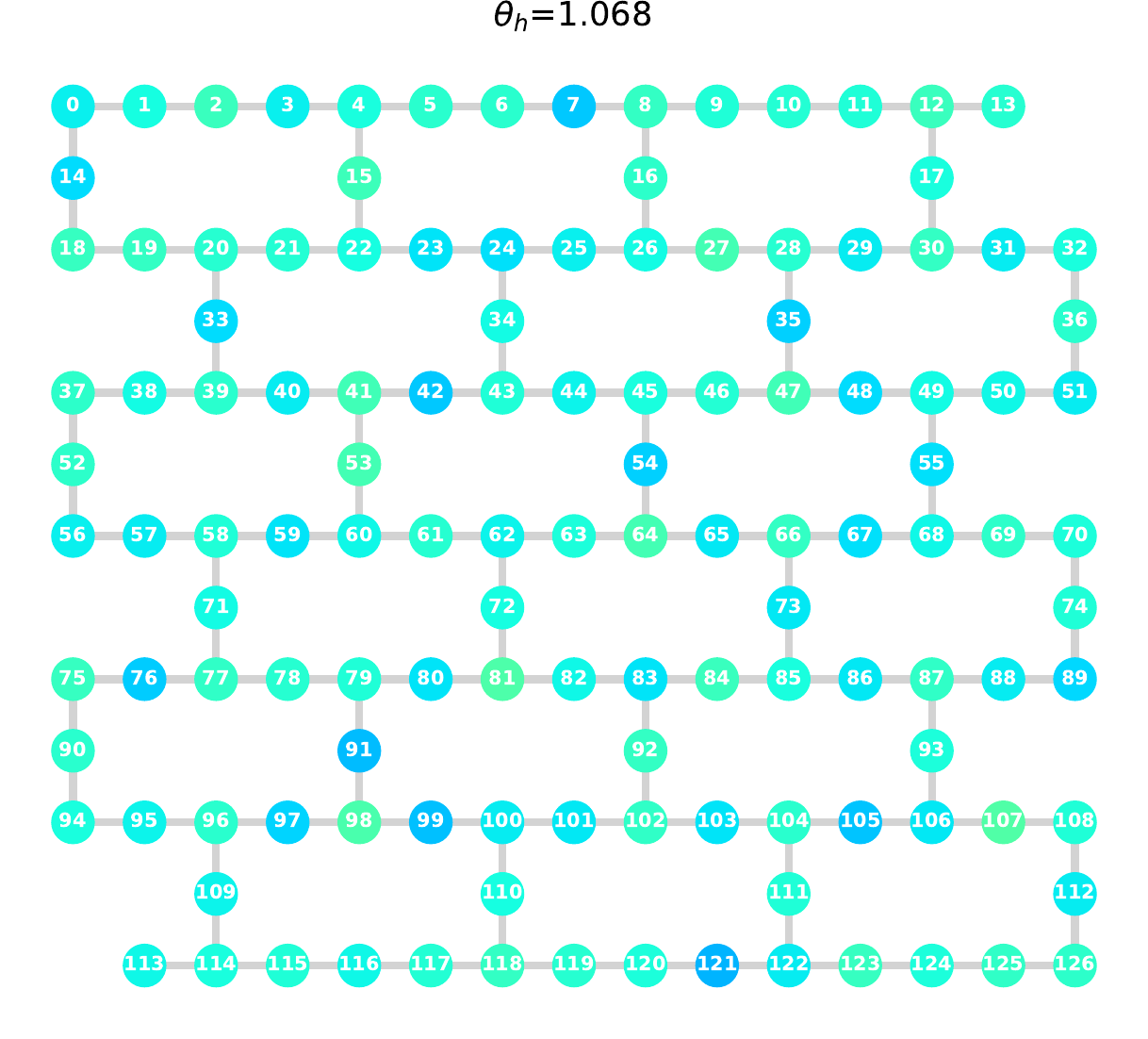}
    \includegraphics[width=0.45\textwidth]{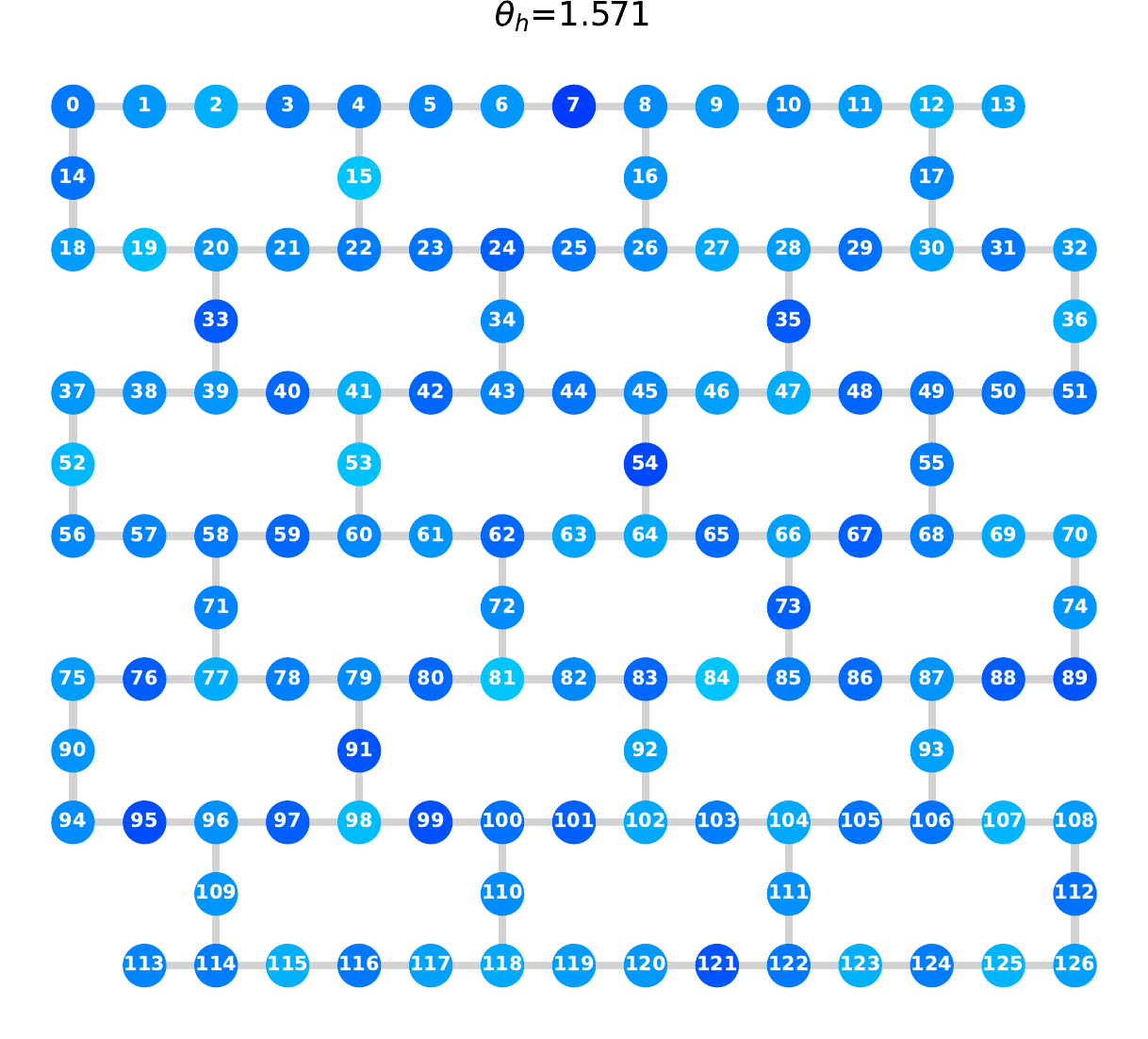}
    \includegraphics[width=0.40\textwidth]{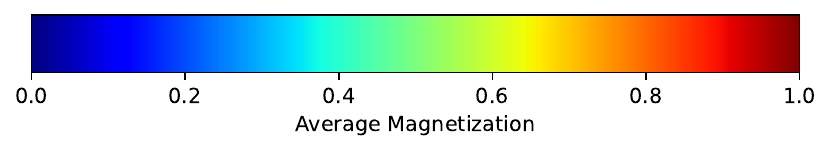}
    \caption{Average single site magnetization (shown by the heatmap below the plots) for all $127$ qubits on the heavy-hex lattice for $6$ intervals of $\theta_h$, from the $126$ time (Trotter) step magnetization data of Figure~\ref{fig:equivalent_magnetization_on_DWave} on \texttt{Advantage\_system6.2} with $J=-0.002$ using reverse quantum annealing. }
    \label{fig:single_site_magnetization_heatmaps}
\end{figure*}

Figure~\ref{fig:single_site_magnetization_heatmaps} shows magnetization heavy-hex lattice heatmaps for varying $\theta_h$ for the magnetization dynamics results from $126$ time step quantum annealing data on \texttt{Advantage\_system6.2} with $J=-0.002$. 

\end{document}